\documentclass[a4paper]{article}
\usepackage{fau1}
\usepackage{hyperref}
\usepackage{graphicx}
\usepackage{amsmath,amssymb,amsfonts,amsthm,latexsym,stmaryrd,physics}
\allowdisplaybreaks[4]
\usepackage{marginnote}
\usepackage{color}
\usepackage{bm}
\usepackage{float}
\usepackage{nicefrac}
\usepackage{microtype} 
\usepackage{booktabs}
\usepackage{verbatim}
\usepackage{setspace}
\usepackage[T1]{fontenc}
\usepackage[utf8]{inputenc}

\hypersetup{
	colorlinks=true
,urlcolor=blue
,anchorcolor=blue
,citecolor=blue
,filecolor=blue
,linkcolor=blue
,menucolor=blue
,pagecolor=blue
,linktocpage=true
,pdfproducer=medialab
}

\def\beq{\begin{equation}}
\def\eeq{\end{equation}}
\newcommand{\bea}{\begin{eqnarray}}
\newcommand{\eea}{\end{eqnarray}}
\def\bi{\begin{itemize}}
\def\ei{\end{itemize}}
\def\ba{\begin{array}}
\def\ea{\end{array}}
\def\bfig{\begin{figure}}
\def\efig{\end{figure}}

\def\C{\mathbb{C}}

\def\ui{\mathrm{i}}
\def\ue{\mathrm{e}}

\def\sgn{\text{sgn}}

\newcommand{\Slc}{\mathrm{SL}(2,\mathbb{C})}

\def\be{\begin{eqnarray}}
\def\ee{\end{eqnarray}}

\newcommand{\cb}{\mathcal B}

\newcommand{\cg}{\mathcal G}

\newcommand{\ck}{\mathcal K}

\newcommand{\cv}{\mathcal V}

\newcommand{\cR}{\mathcal R}
\newcommand{\cI}{\mathcal I}

\newcommand{\g}{\gamma}

\renewcommand{\o}{\omega}

\normalsize

\begin{document}

\title{Analytic Continuation of Spin foam Models}
\author[1,2]{Muxin Han}
\email{hanm(AT)fau.edu}  

\affiliation[1]{Department of Physics, Florida Atlantic University, 777 Glades Road, Boca Raton, FL 33431-0991, USA}

\author[2]{Hongguang Liu} 
\email{hongguang.liu(AT)gravity.fau.de}
\affiliation[2]{Institut f\"ur Quantengravitation, Universit\"at Erlangen-N\"urnberg, Staudtstr. 7/B2, 91058 Erlangen, Germany}

\abstract{
The Lorentzian Engle-Pereira-Rovelli-Livine/Freidel-Krasnov (EPRL/FK) spinfoam model and the Conrady-Hnybida (CH) timelike-surface extension can be expressed in the integral form $\int e^S$. This work studies the analytic continuation of the spinfoam action $S$ to the complexification of the integration domain. Our work extends our knowledge from the real critical points well-studied in the spinfoam large-$j$ asymptotics to general complex critical points of $S$ analytic continued to the complexified domain. The complex critical points satisfying critical equations of the analytic continued $S$. In the large-$j$ regime, the complex critical points give subdominant contributions to the spinfoam amplitude when the real critical points are present. But the contributions from the complex critical points can become dominant when the real critical point are absent. Moreover the contributions from the complex critical points cannot be neglected when the spins $j$ are not large. In this paper, we classify the complex critical points of the spinfoam amplitude, and find a subclass of complex critical points that can be interpreted as 4-dimensional simplicial geometries. In particular, we identify the complex critical points corresponding to the Riemannian simplicial geometries although we start with the Lorentzian spinfoam model. The contribution from these complex critical points of Riemannian geometry to the spinfoam amplitude give $e^{-S_{Regge}}$ in analogy with the Euclidean path integral, where $S_{Regge}$ is the Riemannian Regge action on simplicial complex.
}

\date{\small\today}

\maketitle

\section{Introduction}
Spin foam models (SFM) arise as a covariant formulation of Loop Quantum Gravity (LQG), for a review, see \cite{Thiemann:2007zz,Rovelli:2014ssa,Ashtekar:2017yom,Han:2005km, Perez:2012wv}. The spinfoam model is defined as a state sum model over certain cellular complex $\mathcal{K}$ which contains vertices $v$, edges $e$, and faces $f$. $\ck$ is dual to a triangulation in 4 dimensions. Each face $f$ are bounded by a cyclic sequence of contiguous edges and each edges $e$ are bounded by two vertices. One of the popular spinfoam models is the Engle-Pereira-Rovelli-Livine/Freidel-Krasnov (EPRL/FK) model and the Conrady-Hnybida (CH) extension. These models using certain boundary gauge choices to weakly impose the simplicity constraint. The EPRL model uses the time gauge which leads $SU(2)$ irreducible representations on boundary states, and correspond to quantum spacelike boundary geometries \cite{Engle:2007wy, Freidel:2007py}. The CH extension extends the model to space gauge which uses $SU(1,1)$ irreducible representations for boundary states, thus have timelike boundary geometries \cite{Conrady:2010kc,Conrady:2010vx,Rennert:2016rfp}. Both the EPRL/FK and CH models can be cast into an integral expression $\int e^S$ with the spinfoam action $S=\sum_f S_f$. The action $S_f$ at each face $f$ can be either the space action or the time action for the triangle dual to $f$ being spacelike or timelike. The space action uses representations of the $SU(2)$ or $SU(1,1)$ discrete series while the time action uses continuous-series representation of $SU(1,1)$ . In these models, a spinfoam can be regraded as a Feynmann diagram with 5-valent vertices. Each vertex corresponds to a quantum 4-simplex, as the building block of the discrete quantum spacetime.

The semiclassical behavior of spinfoam model is determined by its large-$j$ asymptotics. Recently there have been many investigations of large-$j$ behaviour of spinfoams, in particular the asymptotics of EPRL/FK model \cite{Conrady:2008mk,Barrett:2009mw,Barrett:2009gg,Han:2011re,Han:2011rf, Han:2013gna} and the asymptotics of CH extension \cite{Kaminski:2017eew,Liu:2018gfc}. It has been shown that, in large-$j$ asymptotics, the spinfoam amplitude is dominant by the contributions from critical configurations, which corresponds to simplicial geometries on a simplicial complex and gives discrete Regge action as its critical action. The models may contain critical configurations of degenerate simplicial geometries, known as vector geometries, except for the CH model with both spacelike and timelike tetrahedra appearing in every 4-simplex. 

Currently, most of the studies about the spinfoam models focus or rely on the critical configurations inside the real integration domain of the spinfoam amplitude. These results shows the semi-classical behaviour and perturbative effect around these semi-classical configurations corresponds to simplicial geometries. However, for the boundary data does not correspond to simplicial geometry configurations, the behavior of amplitude will be dominated by complex critical points away from the real integration domain \cite{Witten:2010cx,Witten:2010zr,Cristoforetti:2013qaa,Basar:2013eka}. Those contributions from complex critical points have not been extensively studied. The complex critical points will give the "sub-dominate" contributions (e.g. analog of instantons) of the model when real critical points appear, which reflects the non-perturbative behavior. %
At smaller $j$ regime, these "sub-dominate" contributions can become important and determine the behaviour of the model. The recent progress on the Monta-Carlo computation of spinfoams \cite{Han:2020npv} also requests a better understanding of these sub-dominant contributions in order to clarify the behavior at smaller $j$'s. Moreover, a complete analysis of these sub-dominate contribution might be a necessary step towards the understanding of the non-perturbative topological property of the model and the study may give different phases and unveil possible quantum phase transition of the model via resurgent trans-series \cite{Basar:2013eka,Dunne:2016nmc}. 

The spinfoam model can be written as an oscillatory integral of type $\int e^S$ over finite dimensional real integration cycle. According to the Picard-Lefschetz theory, we can deform the original integration cycles to the weighted unions of Lefschetz thimbles, each of which is defined as the union of all steepest descent paths ending at a complex critical point of the analytic continued action \cite{Witten:2010cx,Witten:2010zr,Cristoforetti:2013qaa,Basar:2013eka}. The Picard-Lefschetz theory and Lefschetz thimble have been applied to the spinfoam model and turned out to be important in particular for numerical computations. When we analytic continue the spinfoam action $S$, the critical points of the analytic continued action in general live in the complexification of the integraion domain, and contain both the dominate and sub-dominate critical configurations of the model. %
In this paper, we study the analytic continuation of EPRL/FK model and CH extension, extract the complex critical points (of the analytic continued spinfoam action), and analyze their possible geometrical interpretations.

In the analysis we firstly derive the analytic continuation of spinfoam amplitude in the most general EPRL-CH model, to include both spacelike and timelike tetrahedra and triangles. We then analytic continue the action and derive the analytic continued critical equations, from which we extract the complex critical points. At each vertex $v$, the analytic continued critical equations can be written as two copies of parallel transport equations and closure conditions for simple bivectors, which are rotated by $SO(4,\mathbb{C})$ group elements $(\tilde{g}^{+} \in SL(2,\mathbb{C}),\tilde{g}^{-} \in SL(2,\mathbb{C}))$ respectively:
\begin{align}\label{main_eom}
    \tilde{G}_{ve}^{\pm} B_{vef}^{\pm} (\tilde{G}_{ve}^{\pm})^{-1}=  \tilde{G}_{ve'}^{\pm} B_{ve'f}^{\pm} (\tilde{G}_{ve}^{\pm})^{-1} \, , \qquad 0 = \sum_f j_f (-\ui )^{\frac{1-t_f}{2}} B_{ve'f}^{\pm}
\end{align}
where $\tilde{G}^{-} =(\tilde{g}^{-})^{-1}, \tilde{G}^{+} = \tilde{g}^{+} R_{e} $ with $R_e = \mathbb{I}_2 \,\text{or}\, \ui \sigma_3$ respectively for $SU(2)$ or $SU(1,1)$ gauge fixing. Simple bivectors $B_{vef}^{\pm}$ are given as
\begin{align}
    B_{vef}^{\pm} =t_f \mathfrak{a}_{vef}^{\gamma \pm} \tilde{v}_{ef} B_0 \tilde{v}_{ef}^{-1} (\mathfrak{a}_{vef}^{\gamma \pm})^{-1}
\end{align}
with $\tilde{v}_{ef}$ represents the (complexified) coherent states associated to edge $e$ ($\tilde{v}_{ef}= {v}_{ef}\in H $ for boundary edges. $H$ is $SU(2)$ for the EPRL and $SU(1,1)$ for the CH extension). $B_0 = \sigma_3, t_f =1$ for space action (related to spacelike triangles and discrete-series representations) and $B_0 = \sigma_1, t_f =-1$ for time action (related to timelike triangles and continuous-series representations). $ \mathfrak{a}_{vef}^{\gamma \pm} \in SL(2,\mathbb{C})$ is a group element related to phase space variables depending on Immiriz parameter $\gamma$. $B^{\pm}$ satisfies the following condition:
\begin{align}
    \tr(B^{\pm} \cdot (B^{+} - t_f B^{-})) =  \tr((B^{+} - t_f B^{-}) \cdot (B^{+} - t_f B^{-})) = 0
\end{align}
namely, $B^{\pm}$ differs by a null bivector orthogonal to themselves. As a result, $B^{\pm}$ may have different geometrical interpretations and the cross-simplicity condition will be in general broken.

(\ref{main_eom}) are complex holomorphic polynomial equations of complex variables, where the number of equations equals the number of variables. There always exists complex solutions for generic spinfoam boundary data, e.g. even when the boundary data do not satisfy the closure condition, in contrast to the existing results of critical points in the real integration cycle. However, not all of these complex solutions have geometrical interpretation as the simplicial geometries, since they might not always satisfy the cross-simplicity condition for the bivectors.  

We then identify the subset of complex critical points that have clear geometrical interpretations as simplicial geometries, namely all their corresponding bivectors satisfy cross-simplicity condition. These solutions we identified satisfying $\mathfrak{a}_{vef}^{\gamma \pm} = \mathbb{I}_2$. There are three possible signatures of the simplicial geometry arises from the complex critical points: Riemannian, Lorentzian, and split signature. The analytic continued spinfoam action evaluated at these critical points gives:
\begin{itemize}
    \item Riemannian or split signature critical points 
    \begin{align}
        \tilde{S}[\tilde{X}_0]=&  \sum_{f} J_f (-\ui)^{\frac{1-t_f}{2}}\left((-\ui)^{\frac{1-t_f^{\Delta}}{2}} (\pm \gamma \Theta_f  - \ui ( \Phi^B_f + \mu_f \pi)) \mod (\gamma \pi, \ui \pi) \right) \label{riemann_action_final}\\
             & \qquad \qquad \qquad+\sum_{f \in boundary} J_f(\ui + \gamma) \frac{1+t_f}{2}\frac{\omega_f \pi \mod 2 \pi}{2} \notag
    \end{align}
    \begin{align}
        \tilde{S}[\tilde{X}_0]=&\sum_{f} J_f (-\ui)^{\frac{1-t_f}{2}}  \left((-\ui)^{\frac{1-t_f^{\Delta}}{2}} (\ui (\pm \Theta_f - \Phi^B_f - \mu_f \pi)) \mod (\gamma \pi, \ui \pi) \right)\\
             & \qquad \qquad \qquad+\sum_{f \in boundary} J_f(\ui + \gamma) \frac{1+t_f}{2}\frac{\omega_f \pi \mod 2 \pi}{2} \notag
    \end{align}
    
    \item Lorentzian critical points
        \begin{align}
        \tilde{S}[\tilde{X}_0]=&  \sum_{f} J_f  (-\ui)^{\frac{1-t_f}{2}}\left(\ui^{\frac{1-t^{\Delta}_f}{2}} \left( \pm \ui \gamma \Theta_f    - \ui (\Phi^B_f + \mu_f \pi ) - ( \ui + \gamma) \frac{{\omega}_f^{\Delta} \pi}{2}  \mod (\ui \pi, \gamma \pi) \right)\right)\\
             & \qquad \qquad \qquad+\sum_{f \in boundary} J_f(\ui + \gamma) \frac{1+t_f}{2}\frac{\omega_f \pi \mod 2 \pi}{2}
    \end{align}
        \begin{align}
        \tilde{S}[\tilde{X}_0]=&  \sum_{f} J_f (-\ui)^{\frac{1-t_f}{2}} \left( \ui^{\frac{1-t^{\Delta}_f}{2}} ( \pm \Theta_f  - \ui  (\Phi^B_f + \mu_f \pi ) )  \mod (\ui \pi, \gamma \pi)  \right)\\
             & \qquad \qquad \qquad+\sum_{f \in boundary} J_f(\ui + \gamma) \frac{1+t_f}{2}\frac{\omega_f \pi \mod 2 \pi}{2}
    \end{align}
    
\end{itemize}
where $\Theta_f$ ($\Theta_f^B$ ) are deficit (dihedral) angles for internal faces (boundary faces), $ \Phi^B_f$ are determined by the phase convention of the boundary coherent state, which in principle can be $0$ for certain boundary data, and $ \Phi^B_f=0$ for internal faces. ${\omega}_f \in \{0,1\}$ are parameters distinguish the difference of the time gauge or space gauge in EPRL-CH model where ${\omega}_f=1$ when the gauge fixing are different at edges on $\partial f$ and $\o=0$ otherwise. $\omega^{\Delta}_f \in \{0,1\}$ distinguish the difference geometries between the pair of tetrahedra sharing $f$, where $\omega^{\Delta}_f=1 $ when the boundary tetrahedra have different signature and $\o^\Delta=0$ otherwise. Both $\o$ and $\o^\Delta$ are $0$ for internal faces. The extra $\ui \pi$ and $\gamma \pi$ ambiguity appearing in the critical action coming from the analytic continuation of logarithm function which is multi-valued. Thus the analytic continued action has to be defined on the cover space, in which there are infinitely many critical points associated with the same geometrical interpretation. One can easily recognize that the critical action for Riemannian sub-dominate contributions (\ref{riemann_action_final}) is nothing else but the Wick rotated action of the Regge action up to $(-\ui)^{\frac{1-t_f}{2}} \ui \pi$ and $(-\ui)^{\frac{1-t_f}{2}} \gamma \pi$ ambiguity for both space and time action. Namely their contributions to the spinfoam ampltiude are proportional to $e^{-S_{\rm Regge}}$ with  
\begin{align}
    S_{\text{Regge}} = \pm \sum_f A_f \Theta_f
\end{align}
where $A_f :=\gamma J_f$ for both space %
and time action is the area for triangle associated to $f$, and $\pm$ associate to different critical points. $\Theta_f$ is the deficit angle (dihedral angle) when $f$ is an internal (boundary) face. We have analytically continued the spin $J_f\to i J_f$ for the time action to cancel the extra $\ui$ appearing in (\ref{riemann_action_final}). In the case when the tetrahedra contain both timelike and spacelike triangles, this analytical continuation of the spin is required by the closure condition (\ref{main_eom}).%

This paper is organized as follows: In section \ref{sec2}, we give a brief introduction the spinfoam action for EPRL-CH model to fix the notation and derive the analytic continued action. In section \ref{sec3}, we derive and analyze the analytic continued critical equations. The critical equations are reformulated in geometrical form. Then in section \ref{sec4}, we reconstruct geometries from a subset of complex critical points. Finally in section \ref{sec5}, we evaluate the analytic continued action at critical points corresponding to simplicial geometries.

\section{Analytic continuation of the spinfoam amplitude}\label{sec2}
 
The SFM on a simplicial complex $\mathcal{K}$ can be written in the integral representation  \cite{Han:2011re,Conrady:2008ea,Liu:2018gfc,Kaminski:2017eew}
\begin{equation}\label{Z}
    Z({\mathcal{K}})=\sum_{\vec{J}}\prod_f d_{J_f}\int[\mathrm{d} X]\, e^{\sum_f J_f F_f[X]},
\end{equation}
where $f$ are 2-faces in $\mathcal{K}$ colored by half-integer spins $J_f$\footnote{For the amplitude related to timelike triangles in timelike tetrahedra in CH model, $J_f \in \mathbb{Z}/2$ is related to the casimirs of $SU(1,1)$ principle series label $s_f$ by $s_f = -\frac{1}{2}+ \frac{\ui}{2}\sqrt{\frac{4J_f^2}{\gamma^2}-1}$.  }.  %
$d_{J_f}$ labels the choice of the face amplitude:
\begin{align}
    d_{J_f} = \left\{ \begin{array}{cc}
        2 J_f +1 & \qquad  \text{EPRL model}\\
        2 J_f -1 & \qquad  \text{CH model with spacelike triangles in timelike tetrahedra}, J_f \geq 1\\
        1 &  \text{CH model with timelike triangles in timelike tetrahedra}, J_f \geq \gamma/2 
    \end{array} \right. \,.
\end{align}
(\ref{Z}) is a universal expression of SFMs, while different SFMs have different variables $X$ and functions $F_f[X]$ independent of $J_f$. For instance, 

\begin{itemize}

\item Euclidean EPRL model: 
\begin{equation}
X \equiv \left(g^\pm_{ve},\xi_{ef}\right)
\end{equation}
including $(g^+_{ve},g_{ve}^-)\in\mathrm{Spin}(4)$ at each pair of 4-simplex $v$ and 3D tetrahdron $e\subset\partial v$, and $\xi_{ef}\in \mathbb{C}^2$ at each pair of $e$ and $f\subset\partial e$. Each $\xi_{ef}$ is normalized by the Hermitian inner product $\langle\,\cdot\,|\,\cdot\,\rangle$ on $\C^2$. $F_f[X]$ in the exponent is a function of $g^\pm_{ve},\xi_{ef}$ and independent of $J_f$:
\begin{equation}\label{F}
F_f\left[X\right]=\sum_{v, f\subset v}\Big[(1-\gamma)\ln\langle{\xi_{e f}}\big|(g^-_{ve})^{-1}g^-_{ve^\prime}\big|{\xi_{e' f}}\rangle+(1+\gamma) \ln\langle{{\xi_{e f}}\big|(g^+_{ve})^{-1}g^+_{ve'}\big|{\xi_{e' f}}}\rangle\Big].
\end{equation}

\item Lorentzian EPRL model 
\begin{equation}
X\equiv \left(g_{ve},z_{vf},\xi_{ef}\right)
\end{equation}
with $g_{ve}\in\text{SL}(2,\mathbb{C})$, $z_{vf}\in\mathbb{CP}^1$, and $\xi_{ef}\in \mathbb{C}^2$ normalized by the Hermitian inner product. We can equivalently view $\xi_{ef}\in \mathrm{SU(2)}$ since $\xi_{ef}$ corresponds to the SU(2) group element which rotates $(1,0)$ to $\xi_{ef}$. Defining $Z_{vef}=g^\dagger_{ve} z_{vf}$, $F_f[X]$ is written as
\begin{equation}
F_f[X]=\sum_{v, f\subset v}\left( \ln \frac{\left\langle
\xi_{ef},Z_{vef}\right\rangle ^{2}\left\langle Z_{ve^{\prime }f},\xi_{e^{\prime
}f}\right\rangle ^{2}}{\left\langle Z_{vef},Z_{vef}\right\rangle
\left\langle Z_{ve^{\prime }f},Z_{ve^{\prime }f}\right\rangle } - \ui \gamma \ln \frac{\langle Z_{vef},Z_{vef} \rangle}{\langle Z_{ve'f},Z_{ve'f} \rangle}\right)
\end{equation}
with SU(2) invariant inner product $\langle \cdot, \cdot \rangle$.
\item Hnybida-Conrady extension 
\begin{equation}
X\equiv \left(g_{ve},z_{vf},\xi_{ef}^\pm\right)
\end{equation}
with now two different spinors $\xi_{ef}^{\pm} \in \mathbb{C}^2$ which corresponds to the SU(1,1) group element which rotates $\xi_{0}^{+} = (1,0)$ and $\xi_{0}^{-} = (0,1)$  to $\xi_{ef}^{\pm}$. The face action reads
    \begin{equation}
        F_{f}^{\{m\}}[X] = \sum_{v,e \in \partial f}\kappa_{vef} F_{vef}^{m_{ef}}[X]
    \end{equation}
    with
    \begin{align}
         F_{vef}^{\pm}[X]=& -  (\kappa_{vef} - 1) \ln \left( \pm {\langle \xi_{ef}^{\pm},Z_{vef} \rangle} \right) - (\kappa_{vef} + 1) \ln \left( \pm {\langle Z_{vef},\xi_{ef}^{\pm} \rangle}\right) \\
         &- (\ui \gamma - \kappa_{vef} ) \ln \left( \pm \langle Z_{vef},Z_{vef} \rangle \right) 
    \end{align}
    where now $\langle \cdot, \cdot \rangle$ is SU(1,1) invariant inner product and $m_{ef} = \pm1= \langle \xi_{ef}^{\pm},\xi_{ef}^{\pm} \rangle$. $\kappa_{vef} = \pm 1$ defines the direction of simplicial complex satisfying $\kappa_{vef} = -\kappa_{ve'f} = - \kappa_{v'ef}$. Moreover, the integration is restricted to the domain $m_{ef} \langle Z_{vef},Z_{vef} \rangle >0$. All the other variables are the same as EPRL model.
    
\item Hnybida-Conrady extension - timelike triangles $f$ in timelike tetrahedra 
\begin{equation}
X\equiv \left(g_{ve},z_{vf},l_{ef}^{\pm}\right)
\end{equation}
with now again two different spinors $l_{ef}^{\pm} \in \mathbb{C}^2$ which corresponds to the SU(1,1) group element which rotates $l_{0}^{+} = (1,1)$ and $l_{0}^{-} = (1,-1)$  to $l_{ef}^{\pm}$. The face action reads
\begin{equation}
                {F}_{f}^{\{s\}}[X]=\frac{1}{\gamma} \sum_{v, e \subset \partial f} \kappa_{vef} {F}^{s_{vef}}_{vef}[X] 
            \end{equation}
            where $s_{vef} = \pm 1$. ${F}^{s_{vef}}_{vef}[X]$ is given by
            \begin{eqnarray}
                {F}_{vef}^{\pm}[X] =& \gamma \ln \frac{\langle {Z}_{vef}, \, {l}^{\pm}_{ef}\rangle }{ \langle {l}{}^{\pm}_{ef} , \, {Z}_{vef} \rangle  }  \mp \ui \ln \left( \langle {Z}_{vef}, \, {l}^{\pm}_{ef}\rangle  \langle {l}{}^{\pm}_{ef} , \, {Z}_{vef} \rangle  \right) - \ui(1 \mp 1) \ln \langle  {Z}_{vef}, \, {Z}_{vef} \rangle 
            \end{eqnarray}
            with SU(1,1) invariant inner product $\langle \cdot, \cdot \rangle$.

\end{itemize}

\noindent
For both Euclidean and Lorentzian models, the theory have the following gauge transformations
\begin{align}\label{realgauge}
    & z_{vf} \to g_{v} z_{vf} \quad \& \quad g_{ve} \to (g_{v})^{-1 \dagger} g_{ve}, \qquad g_{v} \in \begin{array}{ll} \text{SL}(2, \C) & \text{Lorentzian} \\ \text{Spin}(4) & \text{Euclidean} \end{array} \,, \\
    & \xi_{ef} \to v_{e} \xi_{ef} \quad \& \quad g_{ve} \to (v_{e})^{-1 \dagger} g_{ve}, \qquad v_{e} \in \begin{array}{ll}  \text{SU}(2) & \text{spacelike boundary} \\ \text{SU}(1,1) & \text{timelike boundary} \end{array} \,.
\end{align}
The Lorentzian model has an extra gauge transformation
\begin{equation}
    g_{ve} \to - g_{ve} \,.
\end{equation}

\subsection{Analytic Continuation}

We complexify the integration variables $X$ to complex variables $\tilde{X}$, and analytic continue the integrand in (\ref{Z}) to be the holomorphic function on the space of $\tilde{X}$.

We define the following complexification for integration variables in $X$: The group variables appears in $X$ are complexified as in $\text{SO}(4,\mathbb{C}) \simeq \text{SL}(2,\mathbb{C})_{\mathbb{C}} \simeq \text{Spin}(4)_{\mathbb{C}}$, $\xi_{ef},l_{ef}^\pm$ are complexified via their corresponding group element ${v}_{ef}$ where now becomes in in $\Slc$, and $z_{vf}$ are complexified as in $\C^2$. We look for new critical points of the spinfoam action $\sum_f J_fF_f$ in the space of complex variables.

The complixifications are illustrated below:

\begin{table}[H]
    \centering
    \begin{tabular}{cccc}
    Euclidean  & $\begin{array}{ccc} (g_{ve}^{+},g_{ve}^{-} ) & \to & (\tilde{g}_{ve}^{+}, \tilde{g}_{ve}^{-} ) \\ \text{Spin}(4) & &\text{Spin}(4)_{\mathbb{C}} \end{array}$ &  $\begin{array}{ccc} \xi_{ef} & \to & \tilde{\xi}_{ef} \\ \text{SU}(2) & &\text{SL}(2,\mathbb{C}) \end{array}$ & \\
        Lorentzian  & $\begin{array}{ccc} g_{ve} & \to & \tilde{g}_{ve} \\ {\small \text{SL}(2,\mathbb{C})} & &{ \small \text{SO}(4,\mathbb{C})} \end{array}$ &  $\begin{array}{ccc} \xi_{ef} & \to & \tilde{\xi}_{ef} \\ \text{SU}(2) & &\text{SL}(2,\mathbb{C}) \end{array}$ &  $\begin{array}{ccc} z_{vf} & \to & \tilde{z}_{vf} \\ \mathbb{CP}^1 & &\mathbb{C}^2 \end{array}$ \\
        & $\begin{array}{ccc} g_{ve} & \to & \tilde{g}_{ve} \\ {\small \text{SL}(2,\mathbb{C})} & &{ \small \text{SO}(4,\mathbb{C})} \end{array}$ &  $\begin{array}{ccc} \xi_{ef} & \to & \tilde{\xi}_{ef} \\ \text{SU}(2) & &\text{SL}(2,\mathbb{C}) \end{array}$ &  $\begin{array}{ccc} z_{vf} & \to & \tilde{z}_{vf} \\ \mathbb{CP}^1 & &\mathbb{C}^2 \end{array}$\\
        &  $\begin{array}{ccc} g_{ve} & \to & \tilde{g}_{ve} \\ {\small \text{SL}(2,\mathbb{C})} & &{ \small \text{SO}(4,\mathbb{C})} \end{array}$ &  $\begin{array}{ccc} l_{ef}^{\pm} & \to & \tilde{l}_{ef}^{\pm} \\ \text{SU}(1,1) & &\text{SL}(2,\mathbb{C}) \end{array}$ &  $\begin{array}{ccc} z_{vf} & \to & \tilde{z}_{vf} \\ \mathbb{CP}^1 & &\mathbb{C}^2 \end{array}$
    \end{tabular}
\end{table}
\noindent
where we use $\tilde{\cdot}$ to mark the variables in the space of complex variables.

Some details of complexifying $X$ are given below:
                \begin{itemize}
                    \item Group elements $g \in \text{SL}(2,\mathbb{C})$ or $g \in \text{Spin}(4)$
                    
                    Since $\text{SO}(4,\mathbb{C}) \simeq \text{SL}(2,\mathbb{C})_{\mathbb{C}} \simeq \text{Spin}(4)_{\mathbb{C}}$, we can write $  \tilde{g}_{ve} = (\tilde{g}_{ve}^{+}, \tilde{g}_{ve}^{-} ) \in \text{SO}(4,\mathbb{C})$, where $\tilde{g}_{ve}^{\pm} \in \text{SL}(2,\mathbb{C})$. 
                    For Euclidean model, the complexification is simply defined as
                    \begin{equation}
                        (g_{ve}^{+}, g_{ve}^{-} ) \to (\tilde{g}_{ve}^{+}, \tilde{g}_{ve}^{-} ) \,.
                    \end{equation}
                   For Lorentzian model, we define the complexification as
                    \begin{equation}
                        (g_{ve}, g_{ve}^{\dagger} ) \to (\tilde{g}_{ve}^{+}, \tilde{g}_{ve}^{-} ) \,.
                    \end{equation}
                    Given any $2\times 2$ matrix $x$, the complexification of $g_{ve} x g_{ve}^{\dagger}$ is
                    \begin{equation}
                        g_{ve} x g_{ve}^{\dagger} \to \tilde{g}_{ve}^{+} x \tilde{g}_{ve}^{-}
                    \end{equation}
                    for Lorentzian model while for Euclidean model this simply implies $g_{ve}^{\pm} x g_{ve}^{\pm \dagger} = g_{ve}^{\pm} x (g_{ve}^{\pm})^{-1} \to  \tilde{g}_{ve}^{\pm} x (\tilde{g}_{ve}^{-\pm})^{-1}$.

                    \item Normalized spinors $\xi, l \in \mathbb{C}^2$
                    
                    According to the definition,
                    \begin{equation}\label{def_xixi}
                        \begin{split}
                           &  \xi = v \xi_0, \; v \in \text{SU}(2) \,,\\
                             & \xi^{\pm} = v \xi^{\pm}_0, \; l^{\pm} = v l_0^{\pm}, \; v \in \text{SU}(1,1) \,,
                        \end{split}
                    \end{equation}
                    where $\xi_0,\xi^\pm_0$ are reference spinors $\xi_0,\xi^+_0=(1,0)^t$, $\xi^-_0=(0,1)^t$, $l^\pm_0=(1,\pm1)^t$. The complexifications of $\xi,\xi^\pm$ are equivalent to the complexifications of $\text{SU}(2)$ and $\text{SU}(1,1)$ group variables $v$:
                    \begin{equation}
                        \begin{split}
                            v, v^{\dagger} \to \tilde{v} , \tilde{v}' \in \text{SL}(2,\mathbb{C}) \, , \qquad  v \in \text{SU}(2) \ \text{or} \ \text{SU}(1,1) \,.
                        \end{split}
                    \end{equation}
                    Here $\tilde{v} , \tilde{v}'$ are related to each other. Indeed, $\tilde{v} , \tilde{v}'$ can be expressed by complexifying the parametrization of group elements, where we consider the complex conjugation of a complex parameter $a$ as an independent variable, e.g. $a \to a, \bar{a} \to \tilde{a}$ where $a,\tilde{a}$ are independent complex parameters, see below:
                    \begin{equation}
                        \begin{split}
                            &v = \frac{1}{\sqrt{\bar{a} a \pm \bar{b} b}} \left( \begin{array}{cc} a & \mp \bar{b}\\ b & \bar{a} \end{array} \right) \;\; \to \;\; \tilde{v} = \frac{1}{\sqrt{\tilde{a} a \pm \tilde{b} b}} \left( \begin{array}{cc} a & \mp \tilde{b} \\ b & \tilde{a} \end{array} \right), \;\; \\
                            &v^{\dagger} = \frac{1}{\sqrt{\bar{a} a \pm \bar{b} b}} \left( \begin{array}{cc} \bar{a} & \bar{b} \\ \mp b & a \end{array} \right) \;\; \to \;\; \tilde{v}' = \frac{1}{\sqrt{\tilde{a} a \mp \tilde{b} b}} \left( \begin{array}{cc} \tilde{a} & \tilde{b} \\ \mp b & a  \end{array} \right), \;\; \\
                        \end{split}
                    \end{equation}
                    where $a,b , \tilde{a},\tilde{b} \in \mathbb{C}$, the minus sign in the square-root corresponds to $v,v^\dagger\in {\rm SU}(1,1)$. Note that
                    \begin{equation}
                        \tilde{v} \eta \tilde{v}' = v \eta v^{\dagger} = \eta, \qquad \eta = \text{DiagonalMatrix}[1,\pm 1]
                    \end{equation}
                    where $\det \eta=1$ corresponds to SU(2) and $\det\eta=-1$ corresponds to SU(1,1). The exact form of spinors can be read from (\ref{def_xixi}), for example, $\tilde{\xi}$ and $\tilde{\xi} {}'=\xi_0^t\tilde{v}'{}$ are given by
                    \begin{equation}\label{eq:para_xi}
                        \begin{split}
                            &\xi = \frac{1}{\sqrt{\bar{a} a + \bar{b} b}} \left( \begin{array}{l} a \\ b \end{array} \right) \;\; \to \;\; \tilde{\xi} = \frac{1}{\sqrt{\tilde{a} a + \tilde{b} b}} \left( \begin{array}{l} a \\ b \end{array} \right), \\
                            &\xi^{\dagger} = \frac{1}{\sqrt{\bar{a} a + \bar{b} b}} \left(  \bar{a} , \bar{b} \right) \;\; \to \;\; \tilde{\xi}' = \frac{1}{\sqrt{\tilde{a} a + \tilde{b} b}} \left( \tilde{a}, \tilde{b} \right) \,.
                        \end{split}
                    \end{equation}
                    This also define the complexification of $J \xi , (J\xi)^\dagger$ as
                    \begin{equation}\label{eq:para_jxi}
                        \begin{split}
                            &J \xi = \frac{1}{\sqrt{\bar{a} a + \bar{b} b}} \left( \begin{array}{l} - \bar{b} \\ \bar{a} \end{array} \right) \;\; \to \;\; \widetilde{J\xi} = \frac{1}{\sqrt{\tilde{a} a + \tilde{b} b}} \left( \begin{array}{l} -\tilde{b} \\ \tilde{a} \end{array} \right), \\
                            &(J\xi)^{\dagger} = \frac{1}{\sqrt{\bar{a} a + \bar{b} b}} \left( -b , a \right) \;\; \to \;\; \widetilde{J \xi} ^\prime = \frac{1}{\sqrt{\tilde{a} a + \tilde{b} b}} \left( -b, a \right), 
                        \end{split}
                    \end{equation}
                    $\tilde{\xi}^{\pm}$, $\tilde{\xi}^{\pm} {}'$ and $\tilde{l}^{\pm}, \tilde{l}^{\pm}{}'$ are defined similarly. Note that $\tilde{\xi}$ and $\widetilde{J\xi}$ are linearly independent since there does not exist SL$(2,\mathbb{C})$ group element $\tilde{v}$ such that $\tilde{v} \xi_0 = \alpha \tilde{v} J \xi_0$. Thus $\tilde{\xi}$ and $\widetilde{J\xi}$ form a basis for 2 dimensional spinor space. The same argument hold for pairs of $\tilde{\xi}'$ and $\widetilde{J\xi}'$,  pairs of $\tilde{l}^{\pm}$ and pairs of $\tilde{l}^{\pm}{}'$.
                    
                    In the following, many formulae can unify the treatments of SU(2) $\xi$ and SU(1,1) $\xi^\pm$. In these formulae, we often skip the upper index $\pm$ of $\tilde{\xi}^{\pm}$.

                    \item $\mathbb{CP}^1$ spinors $z$
                    
                    Since $z \in \mathbb{CP}^1$, we can use Gelfand’s choice of the section
                    \begin{equation}\label{para_z_1}
                    z= \left( \begin{array}{l} x \\ 1 \end{array} \right), \;\; x \in \mathbb{C} \,.
                    \end{equation}
                    Under complexification we have
                    \begin{equation}\label{para_z_2}
                        \begin{split}
                            &\bar{z} \to \tilde{z}= \left( \begin{array}{l} \tilde{x} \\ 1 \end{array} \right), \\
                        \end{split}
                    \end{equation}
                    with $\tilde{x} \in \mathbb{C}$ independent of $x$. %

                \end{itemize}
Below we give the analytic continuation of the face action $\tilde{F}[\tilde{X}]$ for specific spinfoam models
\begin{itemize}
\item Euclidean EPRL/FK model: 
\begin{equation}
X \equiv \left(g^\pm_{ve},\xi_{ef}\right) \to \tilde{X} \equiv \left(\tilde{g}^\pm_{ve},\tilde{\xi}_{ef}\right)
\end{equation}
where the analytic continued action for each face now is given by
\begin{equation}\label{Ft_e_a}
\tilde{F}_f\left[\tilde{X}\right]=\sum_{v, f\subset v}\Big[(1-\gamma)\ln \big( \tilde{\xi}'_{e f}(\tilde{g}^-_{ve})^{-1}\tilde{g}^-_{ve'}{\tilde{\xi}_{e' f}}\big)+(1+\gamma) \ln \big( {{\tilde{\xi}'_{e f}}(\tilde{g}^+_{ve})^{-1}\tilde{g}^+_{ve'}{\tilde{\xi}_{e' f}}} \big)\Big].
\end{equation}

\item Lorentzian EPRL model:
\begin{equation}
    X\equiv \left(g_{ve}, g_{ve}^{\dagger}, z_{vf}, z_{vf}^{\dagger}, \xi_{ef}, \xi_{ef}^{\dagger} \right) \to  \tilde{X} \equiv \left(\tilde{g}_{ve}^+, g_{ve}^-, \tilde{z}_{vf}, \tilde{z}'_{vf}, \tilde{\xi}_{ef}, \tilde{\xi}'_{ef} \right) \,.
\end{equation}
We define 
\begin{equation}\label{defZZ'}
    \tilde{Z}_{vef}=\tilde{g}_{ve}^- \tilde{z}_{vf} \qquad \tilde{Z}'_{vef}=\tilde{z}'_{vf} \tilde{g}_{ve}^+  \, ,
\end{equation}
    \begin{itemize}
        \item Spacelike triangles $f$:
            We can unify both the EPRL and CH extension with spacelike triangle with the following $\tilde{F}_f[X]$
            \begin{equation}\label{eq:Ft_sp}
            \tilde{F}_{f}[X]=\sum_{v, e \subset \partial f}\tilde{F}_{vef}[X,\kappa_{vef}] 
            \end{equation}
            with
            \begin{align}
                \tilde{F}_{vef}[X, \kappa_{vef}] =& \kappa_{vef}  \Big[ (1+ \kappa_{vef} \det(\eta_e) )\ln  \left(
                    {\tilde{\xi}'}{}_{ef} \eta_e \tilde{Z}_{vef} \right) \\
                    &+ (\kappa_{vef} \det(\eta_e) -1)  \ln \left(
                        \tilde{Z}'_{vef} \eta_e {\tilde{\xi}}{}_{ef} \right) - (\ui \gamma + \kappa_{vef} \det(\eta_e) )
                        \ln { \tilde{Z}'_{vef} \eta_{e} \tilde{Z}_{vef} } \Big] \nonumber
            \end{align}
            where $\kappa_{vef} = \pm 1$ changes sign when changes $v$ or $e$. This formula of $\tilde{F}_{vef}$ unifies 2 cases: when $e$ is spacelike, $\det(\eta_e)>0$, $\tilde{\xi}$ is the complexification of SU(2) spinor $\xi$; when $e$ is timelike, $\det(\eta_e)<0$, $\tilde{\xi}=\tilde{\xi}^\pm$ is the complexification of SU(1,1) spinor $\xi^\pm$. We often adopt this convention in the following discussion to unify the treatment of $\tilde{\xi}$ and $\tilde{\xi}^\pm$. 

        \item Hnybida-Conrady extension - timelike triangles $f$ in timelike tetrahedra:
             \begin{equation}
                \tilde{F}_{f}^{\{s\}}[X]=\sum_{v, e \subset \partial f} \kappa_{vef} \tilde{F}^{s_{vef}}_{vef}[X] 
            \end{equation}
            where $s_{vef} = \pm 1$ and
            \begin{eqnarray}
                \tilde{F}_{vef}^{\pm}[X] =& \gamma \ln \frac{( \tilde{Z}'_{vef})^t \eta \tilde{l}^{\pm}_{ef} }{{   \tilde{l}'{}^{\pm}_{ef}  \eta \tilde{Z}_{vef} } } \mp \ui \ln \left( {{( ( \tilde{Z}'_{vef})^t \eta \tilde{l}^{\pm}_{ef}) }{ (  \tilde{l}'{}^{\pm}_{ef}  \eta \tilde{Z}_{vef}) }} \right)\\
                & - \ui(1 \mp 1) \ln \left( {\tilde{Z}'_{vef} \eta \tilde{Z}_{vef} } \right) \,. \notag
            \end{eqnarray}
            This defines a series of actions for given sets of $\{ s_{vef} \}$.
    \end{itemize}
\end{itemize}

The analytic continued theory now have the following gauge transformations
\begin{align}\label{complexgauge}
    & \tilde{z}_{vf} \to \tilde{g}^{-}_{v} \tilde{z}_{vf} \quad \& \quad \tilde{g}^{-}_{ve} \to  \tilde{g}^{-}_{ve} (\tilde{g}^{-}_{v})^{-1}, \qquad \tilde{g}^{-}_{v} \in \text{SL}(2, \C) \,, \\
    & \tilde{z}'_{vf} \to \tilde{z}'_{vf}  \tilde{g}^{+}_{v}\quad \& \quad \tilde{g}^{+}_{ve} \to  (\tilde{g}^{+}_{v})^{-1} \tilde{g}^{+}_{ve} , \qquad \tilde{g}^{+}_{v} \in \text{SL}(2, \C) \,,\\
    & \tilde{v}_{ef} \to \tilde{v}_{e} \tilde{v}_{ef}  \quad \& \quad  \tilde{g}^{-}_{ve} \to \tilde{v}_{e} (\tilde{g}^{-}_{v})^{-1} \& \quad  \tilde{g}^{+}_{ve} \to \tilde{v}'_{e}\tilde{g}^{+}_{ve}, \qquad v_{e} \in \text{SL}(2, \C), \; \tilde{v}'_{e} \eta_{e} \tilde{v}_{e} = \eta_{e} \,.
\end{align}
There is still a discrete gauge transformation the analytic continued spinfoam action satisfied:
\begin{align}
    & \tilde{g}^{-}_{ve} \to - \tilde{g}^{-}_{ve} \quad \& \quad  \tilde{g}^{+}_{ve} \to - \tilde{g}^{+}_{ve} \, . \label{complexgauge3}
\end{align}

\section{Semi-Classical analysis of the amplitude}\label{sec3}

We may write the analytic continued action as
\begin{equation}
    \tilde{S} = \lambda \left( \sum_{f} j_f \tilde{F}^{\gamma}_f[\tilde{X}] \right)
\end{equation}
where $J_f = \lambda j_f$. The LQG area spectrum $\text{Ar}_f=8\pi \g \ell_P^2\sqrt{J_f(J_f+1)}$ suggests that $\lambda \to\infty$ should correspond to the $\ell_P \to 0$ while fixing the area $\mathrm{Ar}_f$. Thus the semi-classical limit of the amplitude is given by the asymptotic analysis of the path integral in the $\lambda \to\infty$ limit. In addition to the real critical points which has been studied in the literature, here we focus on the complex critical points emergent from the analytic continuation of the action. The complex saddles give subdominant contributions to the amplitude when the boundary data allow the amplitude to have real saddles. When the boundary data forbids the amplitude to have any real saddle, the contributions from the complex saddles may become dominant to the amplitude. The critical points (critical point) of the analytic continued action are given as the solutions to the equations of motion: 
\begin{equation}
    \delta_{\tilde{g}^{+}}  \tilde{S} = \delta_{\tilde{g}^{-}}  \tilde{S} =\delta_{\tilde{v}}  \tilde{S} =\delta_{\tilde{z}}  \tilde{S} =0 \,.
\end{equation}
We will identify all possible critical point of analytic continued spinfoam action $ \tilde{S}$ on the complexified domain of $\tilde{X}$. We will concentrate on the analysis of Lorentzian model here, while a simple analysis for Euclidean model is given in Appendix \ref{app_Euc}. 

\subsection{Critical equations for Lorentzian Theory}
First of all, from the definition of $\tilde{Z}$ and $\tilde{Z}'$ given in \eqref{defZZ'}
\begin{equation}
    \tilde{Z}_{vef}=\tilde{g}_{ve}^- \tilde{z}_{vf}, \qquad \tilde{Z}'_{vef}=\tilde{z}'_{vf} \tilde{g}_{ve}^+ \,,
\end{equation}
we have the following constraints
\begin{align}\label{eq:sppa_de}
    &(\tilde{g}_{ve}^-)^{-1} \tilde{Z}_{vef}= (\tilde{g}_{ve'}^-)^{-1} \tilde{Z}_{ve'f} &   \tilde{Z}'_{vef} (\tilde{g}_{ve}^+)^{-1} =\tilde{Z}'_{ve'f} (\tilde{g}_{ve'}^+)^{-1}
\end{align}
These constraints hold for both spacelike and timelike faces. The following derives all the equation of motions by variation of the action.
\subsubsection{Space action}
With parametrization (\ref{para_z_2}) of $\tilde{z}$, the variation of spinor variables $\tilde{z}$ can be decomposed as the variation respect to $x$ and $\tilde{x}$ under our parametrization of $z$ given in (\ref{para_z_1}-\ref{para_z_2}). From the analytic continued face action $\tilde{F}$ (\ref{eq:Ft_sp}), 
the variation respects to $z_{vf}$ and $\tilde{z}_{vf}$ leads to 

the following sets of equations 
\begin{align}\label{eq:pat_s}
    & 0 =- (\ui \gamma - 1)\sum_{e \subset \partial f} \kappa_{vef} \mathfrak{\chi}'_{vef} \eta_{e} \tilde{g}_{ve}^{-} \, ,
   & 0= - (\ui \gamma + 1)\sum_{e \subset \partial f} \kappa_{vef} \tilde{g}_{ve}^{+} \eta_{e} \mathfrak{\chi}_{vef}  \, ,
  \end{align}
  with
\begin{align}
    \mathfrak{\chi}'_{vef} =&\frac{ \ui \gamma + \kappa_{ef} \det(\eta_e)}{\ui \gamma - 1} \frac{ \tilde{Z}'_{vef}  }{\tilde{Z}'_{vef}  \tilde{Z}_{vef}}  - \frac{ \det(\eta_{e}) \kappa_{ef} +1}{\ui \gamma - 1} \frac{ \tilde{\xi}'_{ef}   }{\tilde{\xi}'_{ef}\eta_{e} \tilde{Z}_{vef}} \, , \\
    \mathfrak{\chi}_{vef} =&\frac{ \ui \gamma + \kappa_{ef} \det(\eta_e)}{\ui \gamma + 1} \frac{ \tilde{Z}_{vef} }{\tilde{Z}'_{vef} \eta_e\tilde{Z}_{vef}}  - \frac{ \det(\eta_{e}) \kappa_{ef} - 1}{\ui \gamma + 1} \frac{ \tilde{\xi}_{ef}}{ \tilde{Z}'_{vef} \eta_{e} \tilde{\xi}_{ef} } \,, 
  \end{align}
  where $\kappa_{vef} = \pm 1$ flips its sign for changing $e$ to $e'$ with given face $f$.
  
For the variation respect to $\tilde{g}^{\pm} \in \text{SL}(2,\mathbb{C})$, we introduce the small perturbation of $\tilde{g}^{\pm}$ as $\tilde{g}^{\epsilon \pm} = \tilde{g}^{\pm} \ue^{\epsilon \cdot \vec{\sigma}}$ with infinitesimal $\epsilon \in \mathbb{C}$. The variation to $\tilde{g}^{ \pm}$ then becomes the derivation respect to $\epsilon$ evaluated at $\epsilon =0$, which leads to 
\begin{align}
& 0 =- (\ui \gamma - 1) \sum_{f: e \subset \partial f} \kappa_{ef}  \mathfrak{\chi}'_{vef} \sigma_i  \tilde{Z}_{vef} \, , & 0 =- (\ui \gamma + 1) \sum_{f: e \subset \partial f} \kappa_{ef}  \tilde{Z}'_{vef} \sigma_i \mathfrak{\chi}_{vef} \, , \label{eq:clo_s}
\end{align}
where we use the fact that $\eta_e (\sigma)^i = \pm \sigma_i$. 
  
We also derive the variation respect to bulk $\tilde{\xi}$ and $\tilde{\xi}'$. According to the parametrization (\ref{eq:para_xi}), the equations of motion are given by variations respect to $a,b,\tilde{a},\tilde{b}$, which imply

\begin{align}
    &0= \sum_{v \subset \partial e}\kappa_{vef} \Big[ -\kappa_{vef} \tilde{\xi}'_{ef} +(\kappa_{vef} \det(\eta_e)-1) \frac{
                        \tilde{Z}'_{vef}}{\tilde{Z}'_{vef} \eta_e {\tilde{\xi}}{}^{\pm}_{ef}} \Big] \,,\\
    &0= \sum_{v \subset \partial e}\kappa_{vef} \Big[ -\kappa_{vef} \tilde{\xi}_{ef} +(\kappa_{vef} \det(\eta_e)+1) \frac{
                        \tilde{Z}_{vef} }{\tilde{\xi}'_{ef}\eta_{e} \tilde{Z}_{vef}} \Big] \,.
\end{align}
The solution are given by
\begin{equation}\label{eq:dv_spo}
    \tilde{Z}_{vef}(+) \propto_{\C} \tilde{\xi}_{ef}, \qquad \tilde{Z}'_{vef}(-) \propto_{\C} \tilde{\xi}'_{ef} \,,
\end{equation}
where $(\pm)$ correspond to $\kappa_{vef} = \pm \det \eta_e$. 

The set of equations (\ref{eq:pat_s} - \ref{eq:dv_spo}) are equations of motion for general analytic continued space action.

\subsubsection{Time action}
Similar to the space action case, the variation of spinor variables $\tilde{z}$ can be decomposed as the variation respect to $x$ and $\tilde{x}$. From the analytic continued face action $\tilde{F}$ (\ref{eq:Ft_sp}), 
the variation respects to $z_{vf}$ and $\tilde{z}_{vf}$ leads to 
the following equations  
\begin{align}\label{eq:pat_t}
    0 = \ui (\ui \gamma -1)\sum_{e \subset \partial f} \kappa_{vef} \, \chi'^{s_{vef}}_{vef} \eta_e  \tilde{g}_{ve}^{-} \, , \qquad 0 =-\ui (\ui \gamma +1)\sum_{f: e \subset \partial f} \kappa_{vef} \, \tilde{g}_{ve}^{+} \eta_e \chi_{vef}^{s_{vef}} \,, 
\end{align}
with
\begin{align}
    \chi'^{s_{vef}}_{vef}= \frac{\ui \gamma -s_{vef}}{\ui \gamma -1} \frac{\tilde{l}'{}^{\pm}_{ef}  }{  \tilde{l}'{}^{\pm}_{ef}  \eta \tilde{Z}_{vef} }- \frac{1-s_{vef}}{\ui \gamma -1} \ln \frac{\tilde{Z}'_{vef} }{\tilde{Z}'_{vef} \eta \tilde{Z}_{vef} } \, , \\
     \chi_{vef}^{s_{vef}}  = \frac{\ui \gamma + s_{vef}}{\ui \gamma + 1} \frac{ \tilde{l}{}^{\pm}_{ef}  }{  \tilde{Z}_{vef} \eta \tilde{l}{}^{\pm}_{ef}  }+ \frac{1-s_{vef}}{\ui \gamma + 1} \ln \frac{ \tilde{Z}_{vef}}{\tilde{Z}'_{vef} \eta \tilde{Z}_{vef} } \,,
\end{align}
where again $\kappa_{ef} = \pm 1$ flips its sign for changing $e$ to $e'$ with given face $f$.
  
The variation respect to $\tilde{g}^{\pm} \in \text{SL}(2,\mathbb{C})$ leads to 
\begin{align}
& 0 =\ui (\ui \gamma -1) \sum_{f: e \subset \partial f} \kappa_{ef}  \mathfrak{\chi}'_{vef} \eta \sigma_i  \tilde{Z}_{vef} \, , & 0 =-\ui (\ui \gamma +1) \sum_{f: e \subset \partial f} \kappa_{ef}  \tilde{Z}'_{vef} \sigma_i \eta \mathfrak{\chi}_{vef} \, . \label{eq:clo_t}
\end{align}
The variation respect to bulk $\tilde{l}^{\pm}$ and $\tilde{l}'^{\pm}$ leads to
\begin{align}
    0 = \delta_{\{a,b,\tilde{a},\tilde{b}\}} \tilde{F}_{vef}^{s_{vef}} - \tilde{F}_{v'ef}^{s_{v'ef}} \,,
\end{align}
with
\begin{align}
    \delta_{a} \tilde{F}_{vef}^{\pm} = \frac{\pm \ui \tilde{a}}{\sqrt{a \tilde{a}- b \tilde{b}}} + \frac{\gamma \mp \ui}{\sqrt{2}} \left(  \frac{ \tilde{Z}'_{vef} \eta \xi_0}{ \tilde{Z}'_{vef} \eta \tilde{l}^{\pm}_{ef} } \right) + \frac{-\gamma \mp \ui}{\sqrt{2}} \left( \pm \frac{ J \xi_0  \eta \tilde{Z}_{vef}  }{{   \tilde{l}'{}^{\pm}_{ef}  \eta \tilde{Z}_{vef} } } \right) \,,\\
    \delta_{b} \tilde{F}_{vef}^{\pm} =\frac{\mp \ui \tilde{b}}{\sqrt{a \tilde{a}- b \tilde{b}}} + \frac{\gamma \mp \ui}{\sqrt{2}} \left(   \frac{ \tilde{Z}'_{vef} \eta J\xi_0}{ \tilde{Z}'_{vef} \eta \tilde{l}^{\pm}_{ef} } \right) + \frac{-\gamma \mp \ui}{\sqrt{2}} \left( \pm \frac{ \xi_0  \eta \tilde{Z}_{vef}  }{{   \tilde{l}'{}^{\pm}_{ef}  \eta \tilde{Z}_{vef} } } \right) \,,\\
    \delta_{\tilde{a}} \tilde{F}_{vef}^{\pm} =\frac{\pm \ui a}{\sqrt{a \tilde{a}- b \tilde{b}}}+\frac{\gamma \mp \ui}{\sqrt{2}} \left( \pm \frac{ \tilde{Z}'_{vef} \eta J \xi_0}{ \tilde{Z}'_{vef} \eta \tilde{l}^{\pm}_{ef} } \right) + \frac{-\gamma \mp \ui}{\sqrt{2}} \left(\frac{ \xi_0  \eta \tilde{Z}_{vef}  }{{   \tilde{l}'{}^{\pm}_{ef}  \eta \tilde{Z}_{vef} } } \right) \,,\\
    \delta_{\tilde{b}} \tilde{F}_{vef}^{\pm} =\frac{\mp \ui b}{\sqrt{a \tilde{a}- b \tilde{b}}} +\frac{\gamma \mp \ui}{\sqrt{2}} \left(  \pm \frac{ \tilde{Z}'_{vef} \eta \xi_0}{ \tilde{Z}'_{vef} \eta \tilde{l}^{\pm}_{ef} } \right) + \frac{-\gamma \mp \ui}{\sqrt{2}} \left( \frac{ J\xi_0  \eta \tilde{Z}_{vef}  }{{   \tilde{l}'{}^{\pm}_{ef}  \eta \tilde{Z}_{vef} } } \right) \,.
\end{align}
One can show that, after inserting the decomposition of $Z,Z'$ s.t. $\tilde{Z}' = \tilde{l}'{}^{\mp}_{ef} + \alpha'_{vef} \tilde{l}'{}^{\pm}_{ef}  $ and $\tilde{Z}_{vef} = \tilde{l}{}^{\mp}_{ef} + \alpha_{vef} \tilde{l}{}^{\pm}_{ef}  $, the above equation give the following solution:
\begin{align}
 &s_{vef} =s_{v'ef} : \quad  (\ui + s_{vef} \gamma)(\alpha_{v'ef} - \alpha_{vef})  = (\ui - s_{vef} \gamma)(\alpha'_{v'ef} - \alpha'_{vef}) \, , \label{pat_v_t_1} \\
 &s_{vef} =-s_{v'ef} : \;  (\ui + s_{vef} \gamma)\alpha_{vef}  = (\ui - s_{vef} \gamma)\alpha'_{vef}\, , \quad  (\ui + s_{v'ef}  \gamma)\alpha_{v'ef}  = (\ui - s_{v'ef}  \gamma)\alpha'_{v'ef} \, . \label{pat_v_t_2} 
\end{align}

The set of equations (\ref{eq:pat_t} - \ref{pat_v_t_2}) are equations of motion for general analytic continued time actions.
\subsection{Geometric interpretation}
Inspired by (\ref{eq:sppa_de}), (\ref{eq:pat_s}-\ref{eq:clo_s}) and (\ref{eq:pat_t}-\ref{eq:clo_t}), we define the following $2 \cross 2$ matrices for space
and time action
\begin{align}\label{def_x}
&X_{vef}^{-} = Z_{vef} \otimes \chi'_{vef} \eta_{e} \,, &X_{vef}^{+} = \eta_{e} \chi_{vef} \otimes Z'_{vef} \,,
\end{align}
which satisfies
\begin{align}
	\Tr(X^{\pm}\cdot X^{\pm}) =  \Tr(X^{\pm}) = 1 \, .\quad 
\end{align}
Notice that both $X^{\pm}$ are invariant under the transformation $Z \to \lambda Z$ and $Z' \to \lambda' Z'$. Moreover, $X^{+}$ and $X^{-}$ are not totally independent and are related to each other as we shall see lately. . 

We can define a trace-less simple bivector $B^{\pm} \in \mathfrak{sl}(2,\mathbb{C})$ from $X^{\pm}$:
 \begin{align}
	B^{\pm} = X^{\pm} - \frac{1}{2} \mathbb{I}_2 \,, \qquad \Tr(B^{\pm}) = 0
\end{align}
which satisfies
\begin{align}
   |B^{\pm}|^2 : = 2\Tr(B^{\pm}\cdot B^{\pm}) =2 \Tr(X^{\pm} \cdot X^{\pm} -  X^{\pm}  + \mathbb{I}_2/4) = 1 \,.
\end{align}
Thus the bivector $B^{\pm}$ defined above are always simple and timelike \footnote{We define the norm of spin-1/2 bivector as $|B|^2 : = 2\Tr(B \cdot B)$, where $|B|^2 \in  \mathbb{R}$ corresponds to a simple bivector with $|B|^2 > 0 $ the bivector is timelike, $|B|^2 < 0$ is spacelike and $|B|^2 = 0$ is null. The definition generalize to spin-1 representations with $|B|^2 : = \Tr(B \cdot B)$.}.
$B^{\pm}$ can be rewritten as $B^{\pm} = \frac{1}{2} v_c^{\pm i} \sigma_{i}$ where
\begin{align}
	 v_c^{i} = K^i + \ui J^i =  \Tr( B \sigma^{i})  
\end{align}
 with $ v_c^{i} {v_c}_{i} = 2 \Tr( B \cdot B)  =1$. Using $v_c^{i}$, we can induce a map from spin-$\frac{1}{2}$ representation of $B \in \mathfrak{sl}(2,\mathbb{C})$ to spin-1 representation of $B$, where now $B^{IJ} \in \text{SO}(1,3)$ is given as
\begin{align}
  B^{IJ} =  \left( \begin{array}{cccc}
        0 & K^1 & K^2 & K^3   \\
        -K^1 & 0 & J^3 & -J^2   \\
        -K^2 & J^3 & 0 & J^1   \\
        -K^3 & -J^2 & J^1 & 0  
    \end{array} \right)
\end{align}
where $K^i = B^{0i}, J^i =  \epsilon^{ijk} B_{jk}$.
The equation of motion (\ref{eq:sppa_de}), (\ref{eq:pat_s}-\ref{eq:clo_s}) and (\ref{eq:pat_t}-\ref{eq:clo_t}) then can be rewritten as bivector equations contain parallel transport equation:
\begin{align}
    &(\tilde{g}_{ve}^{-})^{-1} B_{vef}^{-} \tilde{g}_{ve}^{-} = (\tilde{g}_{ve'}^{-})^{-1} B_{ve'f}^{-} \tilde{g}_{ve'}^{-} \,, &\tilde{g}_{ve}^{+} B_{vef}^{+} (\tilde{g}_{ve}^{+})^{-1} = \tilde{g}_{ve'}^{+} B_{ve'f}^{+} (\tilde{g}_{ve'}^{+})^{-1} \,, 
\end{align}
and the closure condition:
\begin{align}
    0 = \sum_{f: \text{space action}} j_f \kappa_{ef} B_{vef}^{-} - \ui \sum_{f: \text{time action}} j_f \kappa_{ef} B_{vef}^{-} \,,\\
    0 = \sum_{f: \text{space action}} j_f \kappa_{ef} B_{vef}^{+} + \ui \sum_{f: \text{time action}} j_f \kappa_{ef} B_{vef}^{+} \,,
\end{align}
which holds for both time and space action. The extra $\ui$ appearing in these equations coming from the fact that in our definition, $B$ is always a timelike bivector for both space and time action. These equations then implies that, if we do not complexify the spin $j_f$, after we absorb $i$ into the definition of $B$ for time action, the bivector for time action must have different signatures than space action. We will go back into this lately.

However, the form of the timelike bivectors $B_{vef}^{\pm}$ and their relations to the integration variables are still complicated. Moreover, we also need to impose the parallel transport equation between internal vertices (\ref{eq:dv_spo}) and (\ref{pat_v_t_1}-\ref{pat_v_t_2}) to determine finally the solution. We analyse these bivectors in detail below.

\subsubsection{Space action}

Since pairs of $\tilde{\xi}$ and $\widetilde{J\xi}$ as well as pairs of $\tilde{\xi}'$ and $\widetilde{J\xi}'$ can be regarded as a basis for spinor space, we can make the folllowing decomposition of $Z$ and $Z'$:
\begin{align}
     \tilde{Z}_{vef} \propto_{\mathbb{C}} \mathfrak{z}_{vef}:= \tilde{\xi}_{ef} + \alpha_{vef} \widetilde{J\xi}_{ef}, \qquad \tilde{Z}'_{vef} \propto_{\mathbb{C}} \mathfrak{z}'_{vef} := \tilde{\xi}'_{ef} + \alpha'_{vef} \widetilde{J\xi}'_{ef} \,,
\end{align}
where $\alpha$ is defined as $\alpha_{vef} := \widetilde{J\xi}'_{ef} \eta_e \tilde{Z}_{vef}$.
With the decomposition, the bivectors correspond to space action then can be rewritten as
\begin{align}
   B^{-} =m_{ef} \tilde{\xi}_{ef} \otimes \tilde{\xi}'_{ef} \eta_e -\frac{1}{2} \mathbb{I} +m_{ef} \alpha_{vef} \widetilde{J\xi}_{ef} \otimes \tilde{\xi}'_{ef} \eta_e + E^{-}_{vef} \,,\\
   \eta_{e} B^{+} \eta_{e} =m_{ef} \tilde{\xi}_{ef} \otimes \tilde{\xi}'_{ef} \eta_e -\frac{1}{2} \mathbb{I} + m_{ef} \alpha'_{vef} \tilde{\xi}_{ef} \otimes \widetilde{J\xi}'_{ef} \eta_e + E^{+}_{vef} \,,
\end{align}
where $E^{\pm}$ are given as
\begin{align}
   E^{-}_{vef} =&\frac{ -  \alpha'_{vef} (\gamma-\ui \det \eta_{e} \kappa_{ef})}{(\ui + \gamma)(1 + \det \eta_e \alpha_{vef} \alpha'_{vef})} \Big( \det \eta_e \alpha_{vef} \big(2 m_{ef}\tilde{\xi}_{ef} \otimes \tilde{\xi}'_{ef} \eta_e - I_2+ \alpha_{vef} m_{ef} \widetilde{J\xi}_{ef} \otimes \tilde{\xi}'_{ef} \eta_e  \big) \notag \\
   & \qquad - m_{ef} \tilde{\xi}_{ef} \otimes \widetilde{J\xi}'_{ef} \eta_e \Big)\\
  E^{+}_{vef} =&\frac{- \alpha_{vef} (\gamma-\ui \det \eta_{e} \kappa_{ef})}{(\gamma-\ui )(1 + \det \eta_e \alpha_{vef} \alpha'_{vef})} \Big( \det \eta_e \alpha'_{vef} \big(2 m_{ef} \tilde{\xi}_{ef} \otimes \tilde{\xi}'_{ef} \eta_e - I_2+ \alpha'_{vef} m_{ef} \tilde{\xi}_{ef} \otimes \widetilde{J\xi}'_{ef} \eta_e  \big) \notag\\
  &\qquad - m_{ef} \widetilde{J\xi}_{ef} \otimes \tilde{\xi}'_{ef} \eta_e \Big),
\end{align}
satisfying $\tr(E^{\pm} \cdot E^{\pm})=0, \tr(X^{\pm} \cdot E^{\pm})=0$. $m_{ef}:= \tilde{\xi}'_{ef} \eta_e \tilde{\xi}_{ef}$ is $-1$ when $\tilde{\xi}_{ef}$ is $\tilde{\xi}_{ef}^{-}$, otherwise $m_{ef}=1$.
We check that $B^{\pm}$ are related to each other by the following mapping 
\begin{align}
    \kappa_{ef} \to -\kappa_{ef}, \, \gamma \to -\gamma, \, \alpha_{vef} \leftrightarrow \alpha'_{vef}, \,J\tilde{\xi}_{ef} \otimes \tilde{\xi}'_{ef} \eta_e \leftrightarrow \eta_{e} \tilde{\xi}_{ef} \otimes J\tilde{\xi}'_{ef}
\end{align}
which relate to the fact that $B^{\pm}$ here are related by complex conjugation in the original real domain. Moreover, we can define $M :=X^{-} - \eta_{e} X^{+} \eta_{e} = B^{-} - \eta_{e} B^{+} \eta_{e} $ where
\begin{align}\label{M_def_s}
   M_{vef} =&\frac{1}{(1 + \gamma^2)(1 + \det \eta_e \alpha_{vef} \alpha'_{vef})} \Big( 2 \ui (\det \eta_e \gamma- \ui \kappa_{vef}) \alpha_{vef} \alpha'_{vef} \big(2 m_{ef}\tilde{\xi}_{ef} \otimes \tilde{\xi}'_{ef} \eta_e - I_2) \\
   &+ m_{ef}\alpha_{vef} (\alpha_{vef}  \alpha'_{vef} (1+\ui \gamma) (\det \eta_e+\kappa_{vef})+\ui (\gamma+\ui) (\det \eta_e \kappa_{vef}-1) \widetilde{J\xi}_{ef} \otimes \tilde{\xi}'_{ef} \eta_e  \big) \notag \\
   & + m_{ef} \alpha'_{vef} ( \alpha_{vef}  \alpha'_{vef} (\ui \gamma - 1) (\det \eta_e-\kappa_{vef})-\ui (\gamma-\ui) (\det \eta_e \kappa_{vef}+1)) \tilde{\xi}_{ef} \otimes \widetilde{J\xi}'_{ef} \eta_e \Big) \,. \notag
\end{align}
One can check that
\begin{align}
    \tr(M \cdot M) = \tr(B^{\pm} \cdot M) = 0  .
\end{align}
This relation then implies $B^{-}$ and $\eta_{e} B^{+} \eta_{e}$ differs by a null bivector $M$ orthogonal to them. $M$ is trivial only when both $\alpha$ and $\alpha'$ are zero. As a result, when $B^{-}_{vef}$ at given edge $e$ satisfies the cross simplicity, in general $B^{+}_{vef}$ associated to the same edge will not satisfy it, as cross simplicity conditions
\begin{equation}\label{eq:conditon_simp}
  \forall(f,f'): e \subset \partial (f,f'),\quad \epsilon_{IJKL} B_{vef}^{IJ} B_{vef'}^{KL} =0 
\end{equation}
imposing non-trivial constraints to $M$ thus to $\alpha, \alpha'$.

Since $B_{vef}^{\pm}$ are bivectors satisfying $\tr(B^{\pm} \cdot B^{\pm})=\frac{1}{2}$, we can always define a SL$(2,\mathbb{C})$ group element $\mathfrak{a}_{vef}^\pm$ depending on $\alpha,\alpha'$ such that
\begin{align}
    B_{vef}^{-} =\tilde{v}_{ef} \mathfrak{a}_{vef}^{-} \frac{\sigma_3}{2} (\mathfrak{a}^{-}_{vef})^{-1} (\tilde{v}_{ef})^{-1} \, , \qquad  \eta_{e} B_{vef}^{+} \eta_{e}  =\tilde{v}_{ef} \mathfrak{a}_{vef}^{+} \frac{\sigma_3}{2} (\mathfrak{a}^{+}_{vef})^{-1} (\tilde{v}_{ef})^{-1} \, ,
\end{align}
where $\mathfrak{a}_{vef}^\pm \in \text{SL}(2,\mathbb{C})$ can be defined as
\begin{align}\label{def_of_a_sp}
\mathfrak{a}_{vef}^{-} &= \left(
\begin{array}{cc}
 1 & -\frac{i \alpha_{vef}  ((1+\det \eta_{e} \kappa_{vef}) \alpha_{vef}  \alpha'_{vef} + (1-i \gamma) \det \eta_{e} )}{(\gamma+i) (\alpha_{vef}  \alpha'_{vef} \det \eta_{e}+1)} \\
 \frac{\alpha'_{vef} (\gamma-i \det \eta_{e} \kappa_{vef})}{\gamma+i (\alpha_{vef}  \alpha'_{vef} (\det \eta_{e}+\kappa_{vef})+1)} & \frac{\gamma+i (\alpha_{vef}  \alpha'_{vef} (\det \eta_{e}+\kappa_{vef})+1)}{ (\gamma+i) (\alpha_{vef}  \alpha'_{vef} \det \eta_{e}+1)} \\
\end{array}
\right) \,,\\
  \mathfrak{a}_{vef}^{+} &=  \left(
\begin{array}{cc}
 1 & \frac{\alpha_{vef} (i  \kappa_{vef}- \det \eta_{e} \gamma)}{ (\gamma-i) (\alpha_{vef}  \alpha'_{vef} \det \eta_{e}+1)} \\
  \alpha'_{vef} & \frac{\gamma-i (\alpha_{vef}  \alpha'_{vef} (\det \eta_{e}-\kappa_{vef})+1)}{ (\gamma-i) (\alpha_{vef}  \alpha'_{vef} \det \eta_{e}+1)} \\
\end{array}
\right) \,,
\end{align}
where $\mathfrak{a}_{vef}^{\pm} = \mathbb{I}$ when $\alpha=\alpha'=0$.

The bivectors satisfy the closure condition from which $\{\alpha_{vef}, \alpha_{vef'}\}$ can be solved up to re-scaling. Notice that
\begin{align}
&  (\ui \gamma -1 ) X^{-}_{vef} - (\ui \gamma+1) \eta_{e} X^{+}_{vef} \eta_{e} \\
  =& -2  \tilde{\xi}_{ef} \otimes \tilde{\xi}'_{ef} \eta_e -  {( \kappa_{ef} \det(\eta_{e}) + 1) }( \alpha_{vef} \widetilde{J\xi}_{ef} \otimes \tilde{\xi}'_{ef} \eta_e  ) +  {( \kappa_{ef} \det(\eta_{e}) - 1) }( \alpha'_{vef} \tilde{\xi}_{ef} \otimes \widetilde{J\xi}'_{ef} \eta_e  ) \,. \notag
\end{align}
The closure for $B^{\pm}$ then can be rewritten as the following conditions
\begin{align}
     0 =&\sum_{f} j_f \kappa_{ef} (\ui \gamma -1 ) B^{-}_{vef} - (\ui \gamma+1) \eta_{e} B^{+}_{vef} \eta_{e} = - \sum_{f} j_f \kappa_{vef} ( 2 \tilde{\xi}_{ef} \otimes \tilde{\xi}'_{ef} \eta_e -\mathbb{I}) + \label{new_clo_sp_1}\\
     & \sum_{f} j_f \kappa_{vef} \left( - {( \kappa_{ef} \det(\eta_{e}) + 1) }( \alpha_{vef} \widetilde{J\xi}_{ef} \otimes \tilde{\xi}'_{ef} \eta_e  ) +  {( \kappa_{ef} \det(\eta_{e}) - 1) }( \alpha'_{vef} \tilde{\xi}_{ef} \otimes \widetilde{J\xi}'_{ef} \eta_e  ) \right) \, , \notag \\
       0 =& \sum_{f} j_f \kappa_{vef} \left( B^{-}_{vef} - \eta_{e} B^{+}_{vef} \eta_{e} \right)  = \sum_{f} j_f \kappa_{vef} M_{vef} \, . \label{new_clo_sp_2}
\end{align}
Notice that the second equation are closure condition for null bivectors $M_{vef}$. 

At given edge $e$, since there are only 6 closure conditions, only 3 pairs of $\{\alpha,\alpha'\}$ out of 4 will be fixed. This generates a series of continuous connected solutions $[\mathfrak{a}]_e$, correspond to a continuous deformation of the corresponding bivectors. However, in general not all these solutions $[\mathfrak{a}]_e$ solves the parallel transport equation. Actually $\alpha, \alpha'$ here subject to extra conditions (\ref{eq:sppa_de}) can be viewed as a coordinate change which removes spinor variables $\tilde{z}_{vf}, \tilde{z'}_{vf}$. Thus we have the same number of variables and polynomial critical equations, which in general admits isolated solutions unless the system is degenerate.  {If one carefully counts the d.o.f with parametrization using $\alpha, \alpha'$ and the number of critical equations at each vertex, they are equal: we have in total $2 \times (20+ 4\times 3) = 64$ complex variables for $\alpha, \alpha', g^{\pm}$, and the critical equations contains $2 \times 10 \times (3-1) = 40$ complex bivector equations plus $2 \times 4 \times 3 = 24$ complex closure conditions. }

For the internal edges, from the parallel transport equation between vertices, we have $\alpha_{vef} = 0$ or $\alpha'_{vef} = 0$ for $\kappa_{vef} = \pm \det\eta_{e}$ respectively. As a result, $E^{\mp}_{vef} = 0$ respectively %
and $M_{vef}$ becomes
\begin{align}
   M_{vef} =&\frac{ m_{ef}}{(1 + \gamma^2)} \Big( 
  \alpha_{vef} (\ui (\gamma+\ui) (\det \eta_e \kappa_{vef}-1) \widetilde{J\xi}_{ef} \otimes \tilde{\xi}'_{ef} \eta_e  \big) \notag \\
   & + \alpha'_{vef} (-\ui (\gamma-\ui) (\det \eta_e \kappa_{vef}+1)) \tilde{\xi}_{ef} \otimes \widetilde{J\xi}'_{ef} \eta_e \Big) \,. \notag
\end{align}
As a result, the closure condition given by (\ref{new_clo_sp_1}) becomes
\begin{align}
   0 = \sum_{f} j_f \kappa_{vef} ( 2 \tilde{\xi}_{ef} \otimes \tilde{\xi}'_{ef} \eta_e -\mathbb{I}) \,,
\end{align}
which is independent of $\alpha, \alpha'$ thus constrains internal $\tilde{v}$ to satisfy the closure condition. This is compatible with the argument that the closure constrain in spinfoam models is imposed strongly \cite{Conrady:2009px,Bianchi:2010gc}. The left undetermined $\alpha, \alpha'$ are constrained by (\ref{new_clo_sp_2})
. Notice that, since $\kappa_{vef}$ have opposite sign between the two vertices $v$ and $v'$ associated to the edge $e$, we have $B_{vef} \neq B_{v'ef}$ unless $\alpha = \alpha' =0$. However, one should note that the existence of the $\alpha = \alpha' =0$ solution will be determined finally by solving simultaneously parallel transport equations.

\subsubsection{Time action}
For the time action, a similar analysis can be carried out while now we can expand the bivector using the decomposition $\tilde{Z}' \propto_{\mathbb{C}} \tilde{l}'{}^{\mp}_{ef} + \alpha'_{vef} \tilde{l}'{}^{\pm}_{ef}  $ and $\tilde{Z}_{vef}  \propto_{\mathbb{C}} \tilde{l}{}^{\mp}_{ef} + \alpha_{vef} \tilde{l}{}^{\pm}_{ef}  $, which gives
\begin{align}
    & X_{vef}^{-} %
     =(\tilde{l}{}^{\mp}_{ef} + \alpha_{vef} \tilde{l}{}^{\pm}_{ef}) \otimes \tilde{l}'{}^{\pm}_{ef}  \eta_e +E_{vef}^{-} \,,
    & X_{vef}^{+} %
     =\eta_e  \tilde{l}{}^{\pm}_{ef} \otimes (\tilde{l}'{}^{\mp}_{ef} + \alpha'_{vef} \tilde{l}'{}^{\pm}_{ef})  + E_{vef}^{+} \,,
\end{align}
where now
\begin{align}
    E_{vef}^{-} &= \frac{1}{\alpha_{vef} + \alpha'_{vef}} \Big[ \frac{(1 -s_{vef}) {\alpha_{vef}}^2 }{\ui \gamma - 1} \tilde{l}{}^{\pm}_{ef} \otimes \tilde{l}'{}^{\pm}_{ef}  \eta_e \\
     &\qquad- \frac{(1 - s_{vef})}{\ui \gamma - 1} (-2\alpha_{vef} \tilde{l}{}^{\mp}_{ef} \otimes \tilde{l}'{}^{\pm}_{ef}  \eta_e + \alpha_{vef} I_2+ \tilde{l}{}^{\mp}_{ef} \otimes \tilde{l}'{}^{\mp}_{ef}  \eta_e ) \Big] \,, \nonumber\\
     E_{vef}^{+} &=\frac{1}{\alpha_{vef} + \alpha'_{vef}} \Big[ \frac{( s_{vef} - 1) \alpha'_{vef}{}^2 }{\ui \gamma + 1}  \eta_e \tilde{l}{}^{\pm}_{ef} \otimes \tilde{l}'{}^{\pm}_{ef} \\
    & \qquad + \frac{(1 - s_{vef})}{\ui \gamma + 1} (-2 \alpha'_{vef} \eta_e \tilde{l}{}^{\pm}_{ef} \otimes \tilde{l}'{}^{\mp}_{ef} + \alpha'_{vef} I_2+ \eta_e \tilde{l}{}^{\mp}_{ef} \otimes \tilde{l}'{}^{\mp}_{ef}  ) \Big] \,,  \nonumber
\end{align}
satisfying 
\begin{align}
    \Tr(E_{vef}^{\pm}) =\Tr(E_{vef}^{\pm}.E_{vef}^{\pm}) = 0 \,.
\end{align}
Namely, $E^{\pm}$ is always a null bivector. Notice that we have
\begin{align}\label{M_def_t}
    & M := B^{-} + \eta B^{+} \eta 
    =\frac{1}{\alpha_{vef} + \alpha'_{vef}} \Big[ \Big( \sqrt{\frac{\ui \gamma -s_{vef}}{\ui \gamma - 1}} \alpha_{vef} +\sqrt{\frac{\ui \gamma + s_{vef}}{\ui \gamma + 1}} \alpha'_{vef} \Big)^2 \tilde{l}{}^{\pm}_{ef} \otimes \tilde{l}'{}^{\pm}_{ef}  \eta_e \nonumber\\
     & \qquad + \frac{(1 - s_{vef})}{ \gamma^2 + 1} \Big( 2 \tilde{l}{}^{\mp}_{ef} \otimes \tilde{l}'{}^{\mp}_{ef}  \eta_e - \big( \ui \gamma (\alpha_{vef} + \alpha'_{vef})+ \alpha_{vef} - \alpha'_{vef} \big) \big( 2 \tilde{l}{}^{\mp}_{ef} \otimes \tilde{l}'{}^{\pm}_{ef}  \eta_e - I_2 \big) \Big) \Big] \,, \nonumber
\end{align}
by the fact that $\tilde{l}{}^{\mp}_{ef} \otimes \tilde{l}'{}^{\pm}_{ef}  \eta_e + \tilde{l}{}^{\pm}_{ef} \otimes \tilde{l}'{}^{\mp}_{ef}  \eta_e = I$. One can check that similar to the case of the space action, we have 
\begin{align}
    \tr(M \cdot M) = \tr(B^{\pm} \cdot M) = 0.
\end{align}
When $s=1$ and $\alpha+\alpha'=0$, $M$ is trivial. 

We can define again $\mathfrak{a}_{vef}^\pm \in \text{SL}(2,\mathbb{C})$ :
\begin{align}\label{def_of_a_t}
\mathfrak{a}_{vef}^{-} &= 
\sqrt{\frac{(1+\alpha_{vef})f_a^{-}(\alpha,\alpha')}{1-\ui \gamma}}\left(
\begin{array}{cc}
 1 & 0 \\
 \frac{ \alpha_{vef}}{\alpha_{vef} +1}+\frac{(s_{vef}-1) }{f_a^{-}(\alpha,\alpha')} & \frac{-1}{\alpha_{vef} +1}-\frac{(s_{vef}-1) }{f_a^{-}(\alpha,\alpha')} \\
\end{array}
\right)\,,\\
  \mathfrak{a}_{vef}^{+} &= \sqrt{\frac{(1-\alpha'_{vef})f_a^{+}(\alpha,\alpha')}{1+\ui \gamma}}\left(
\begin{array}{cc}
 1 & 0 \\
 \frac{ \alpha'_{vef}}{\alpha'_{vef} -1}+\frac{(s_{vef}-1) }{f_a^{+}(\alpha,\alpha')} & \frac{-1}{-\alpha'_{vef} +1}+\frac{(s_{vef}-1) }{f_a^{+}(\alpha,\alpha')} \\
\end{array}
\right) \,,
\end{align}
with
\begin{align}
    f_a^{-}(\alpha,\alpha')=1-\alpha'_{vef}+\ui g (\alpha_{vef} +\alpha'_{vef})-(\alpha_{vef} +1) s\\
    f_a^{+}(\alpha,\alpha')=\alpha_{vef}+\ui g (\alpha_{vef} +\alpha'_{vef})+(\alpha'_{vef} -1) s+1
\end{align}
such that the bivectors $B_{ver}^{\pm}$ can be rewritten as
\begin{align}
    B_{vef}^{-} =\tilde{v}_{ef} \mathfrak{a}_{vef}^{-} \frac{\sigma_1}{2} (\mathfrak{a}^{-}_{vef})^{-1} (\tilde{v}_{ef})^{-1} \, , \qquad  \eta_{e} B_{vef}^{+} \eta_{e}  =\tilde{v}_{ef} \mathfrak{a}_{vef}^{+} \frac{\sigma_1}{2} (\mathfrak{a}^{+}_{vef})^{-1} (\tilde{v}_{ef})^{-1} \, ,
\end{align}
Note that when $\alpha=\alpha'=0$ we have $\mathfrak{a}_{vef}^{\pm} = \mathbb{I}$.

As a result, all the argument for space action then follows similarly here, namely $\alpha, \alpha'$ can be solved from the closure condition combining with the parallel transport equation. 
For example, from the fact that
\begin{align}
    (\ui \gamma -1) E^{-} +  (\ui \gamma + 1) \eta E^{+} \eta &=(1 - s_{vef}) \left( (\alpha_{vef} - \alpha'_{vef})  \tilde{l}{}^{\pm}_{ef} \otimes \tilde{l}'{}^{\pm}_{ef}  \eta_e +  2\tilde{l}{}^{\mp}_{ef} \otimes \tilde{l}'{}^{\pm}_{ef}  \eta_e - \mathbb{I}_2 \right) \,,
\end{align}
one of the closure condition can be rewritten as
\begin{align}\label{clo_1_t}
    &\sum_f j_f \kappa_{vef} ((\ui \gamma -1) X^{-} +  (\ui \gamma + 1) \eta X^{+} \eta) \notag\\
    =&\sum_f j_f \kappa_{vef} \left( - 2 s_{vef} \tilde{l}{}^{\mp}_{ef} \otimes \tilde{l}'{}^{\pm}_{ef}  \eta_e  +\mathbb{I}_2 + (\ui \gamma  (\alpha_{vef} + \alpha'_{vef}) -s_{vef} (\alpha_{vef} - \alpha'_{vef} )) \tilde{l}{}^{\pm}_{ef} \otimes \tilde{l}'{}^{\pm}_{ef}  \eta_e \right) \,.
\end{align}
Another closure condition is then given by null closure condition: 
\begin{align}\label{clo_2_t}
  0 = \sum_{f} j_f \kappa_{vef} M_{vef} \,.
\end{align}
Note that, when $s_{vef} =1$, we have $E_{vef}^{\pm} = 0$, the closure conditions becomes
\begin{align}
   0 &= \sum_{f} j_f \kappa_{vef} \left( (\tilde{l}{}^{\mp}_{ef} + \alpha_{vef} \tilde{l}{}^{\pm}_{ef}) \otimes \tilde{l}'{}^{\pm}_{ef}  \eta_e - \frac{1}{2} \mathbb{I}_2 \right), \\
   0 &= \sum_{f} j_f \kappa_{vef} \left( (\tilde{l}{}^{\mp}_{ef} - \alpha'_{vef} \tilde{l}{}^{\pm}_{ef}) \otimes \tilde{l}'{}^{\pm}_{ef}  \eta_e  - \frac{1}{2} \mathbb{I}_2\right) \,.
\end{align}
which are the same set of equations for $\alpha$ and $-\alpha'$ respectively. As a result, in the case when boundary variables at edge $e$ satisfy the closure: $0= \sum_{f} j_f \kappa_{vef} \left( (\tilde{l}{}^{\mp}_{ef} \otimes \tilde{l}'{}^{\pm}_{ef}  \eta_e - \frac{1}{2} \mathbb{I}_2 \right)$, $\alpha$ and $-\alpha'$ differ by only an overall scaling at edge $e$.

For the internal edges, due to (\ref{pat_v_t_1}) and (\ref{pat_v_t_2}), one can check that for all possible $s$, we have $((\ui \gamma -1) X^{-}_{vef} +  (\ui \gamma + 1) \eta X^{+}_{vef}  \eta) = ((\ui \gamma -1) X^{-}_{v'ef}  +  (\ui \gamma + 1) \eta X^{+}_{v'ef}  \eta)$. Thus comparing (\ref{clo_1_t}) at $v$ and $v'$ leads to an equation independent of $\alpha,\alpha'$ which now reads $0= \sum_{f} j_f \kappa_{vef} \left( (\tilde{l}{}^{\mp}_{ef} \otimes \tilde{l}'{}^{\pm}_{ef}  \eta_e - \frac{1}{2} \mathbb{I}_2 \right)$, thus imposing the closure condition to $\tilde{v}_{ef}$. For the case $s_{vef} = s_{v'ef}$, (\ref{clo_1_t}) becomes $0=\sum_f j_f \kappa_{vef} (\ui \gamma  (\alpha_{vef} + \alpha'_{vef}) -s_{vef} (\alpha_{vef} - \alpha'_{vef} )) \tilde{l}{}^{\pm}_{ef} \otimes \tilde{l}'{}^{\pm}_{ef}  \eta_e )$ while this is automatically satisfied for the case $s_{vef} \neq s_{v'ef}$. The left undetermined $\alpha, \alpha'$ are then given by (\ref{clo_2_t}). As a result, this again implies $X_{vef} \neq X_{v'ef}$ in general since $M_{vef} \neq M_{v'ef}$. The possible situation to have $M_{vef} = M_{v'ef}$ is when $M$ is trivial for both $v$ and $v'$, otherwise it will transport nontrivially between $v$ and $v'$. A simple situation for this is given by $s = 1$ for both $v$ and $\alpha + \alpha' = 0$.

For the action composed by both time and space action, the closure condition reads
\begin{align}
   & \sum_{f:spacelike} j_f \kappa_{vef} (1- \ui \gamma ) B_{vef}^{-} + \ui \sum_{f:timelike} j_f \kappa_{vef} (\ui \gamma -1 ) B_{vef}^{-} = 0 \,,\\
    & \sum_{f:spacelike} j_f \kappa_{vef} (-\ui \gamma - 1 ) B_{vef}^{+}  - \ui  \sum_{f:timelike} j_f \kappa_{vef} (\ui \gamma  + 1 )B_{vef}^{+}  = 0  \,,
\end{align}
which implies
\begin{align}
   & \sum_{f:spacelike} j_f \kappa_{vef} B_{vef}^{-} - \ui \sum_{f:timelike} j_f \kappa_{vef} B_{vef}^{-} = 
     \sum_{f:spacelike} j_f \kappa_{vef} B_{vef}^{+}  +  \ui  \sum_{f:timelike} j_f \kappa_{vef} B_{vef}^{+}  = 0  \,.
\end{align}
The compatibility between timelike and spacelike action then requires
\begin{align}
0=&\sum_{f:timelike} j_f \kappa_{vef} \left( B^{-}_{vef} - \eta B^{+}_{vef} \eta \right) - \ui \sum_{f:timelike} j_f \kappa_{vef} \left( B^{-}_{vef} + \eta B^{+}_{vef} \eta \right)\\
=&\sum_{f:spacelike} j_f \kappa_{vef} M_{vef} - \ui \sum_{f:timelike} j_f \kappa_{vef} M_{vef} \,,
\end{align}
which is again a closure condition of null bivectors.

\subsection{Summary}\label{summary_eom}

In summary, the critical point equations are given by two copies of the following bivector equations associated to each vertex $v$:
\begin{align}\label{eq:biveq}
	B^{g}_{vef} &= B^{g}_{ve'f} , \qquad \sum_{f: e \subset \partial f} j_f \kappa_{ef} (\mp \ui)^{\frac{1-t_f}{2}} B^{g \pm}_{vef} =0 \,,
\end{align}
for both space action with $t_f = 1$ and time action with $t_f = -1$. $B^{g}_{vef}$ is defined by
\begin{align}
    	B^{g}_{vef} &:= \tilde{g}_{ve} B_{vef}^{\pm} (\tilde{g}_{ve})^{-1}, \qquad \text{for} \;\; \tilde{g} = (\tilde{g}^{-})^{-1}, \;  \tilde{g}^{+} \,.
\end{align}
The bivectors $B^{\pm}$ satisfying $2 \tr(B^{\pm} \cdot  B^{\pm}) = 1$ and are related to each other by
\begin{align}
    B^{+} = t \, \eta ( B^{-} - M ) \eta \,,
\end{align}
where $M$ are null bivector defined in (\ref{M_def_s}) and (\ref{M_def_t}) satisfying $\tr(M \cdot M)=\tr(M \cdot B) = 0$. The bivector $B^{\pm}$ can also be rewritten as
   \begin{align}\label{B+ina_sum}
    B_{vef}^{-} =\tilde{v}_{ef} \mathfrak{a}_{vef}^{-} B_0 (\mathfrak{a}^{-}_{vef})^{-1} (\tilde{v}_{ef})^{-1} \, , \qquad  \eta_{e} B_{vef}^{+} \eta_{e}  =\tilde{v}_{ef} \mathfrak{a}_{vef}^{+} B_0 (\mathfrak{a}^{+}_{vef})^{-1} (\tilde{v}_{ef})^{-1} \, ,
\end{align}
where $B_0 = \frac{\sigma_3}{2}$ for space action and $B_0 = \frac{\sigma_1}{2}$ for time action, $\mathfrak{a}^{\pm}_{vef} \in \text{SL}(2,\mathbb{C})$ are given by (\ref{def_of_a_sp}) and (\ref{def_of_a_t}) respectively.

For the internal edges $e$ on the complex graph, the variational principle respect to $\tilde{v}_{ef}$ introduce new equations restricts $\mathfrak{a}^{\pm}_{vef}$ and $\tilde{v}_{ef}$, under which the form of $\mathfrak{a}^{\pm}_{vef}$ and $\mathfrak{a}^{\pm}_{v'ef}$ for $v$ and $v'$ associated to the same $e$ are restricted by condition (\ref{eq:dv_spo}) or (\ref{pat_v_t_1}-\ref{pat_v_t_2}) . As a result, one of the closure conditions in (\ref{eq:biveq}) becomes the closure condition for $\tilde{v}_{ef} B_0 (\tilde{v}_{ef})^{-1}$ associated to edge $e$. This is compatible with the fact that the closure constraints in EPRL-CH model are actually imposed strongly \cite{Conrady:2009px,Bianchi:2010gc}.

By using the map $\Phi: \text{SL}(2,\mathbb{C}) \to \text{SO}(3,1)$, we can define 
\begin{equation}
    \text{SO}(3,1) \in R_e = \Phi(\ui \eta_e)=\left\{ \begin{array}{ll} I, & \text{EPRL} \\ \Phi(\ui \sigma_3), & \text{Conrady-Hnybida} \end{array} \right. \, ,
\end{equation}
such that
\begin{equation}
    B^{+}_{vef} = t R_e (B^{-}_{vef} - M_{vef}) (R_e)^{-1} \,.
\end{equation}
We then absorb $R_e$ into definitions of $G^{\pm} = \Phi( \tilde{g}^{\pm})$ as $G^{+} = \Phi( \tilde{g}^{+})R_e$, which leads to the following equations
\begin{align}\label{eq:biveq1}
	B_{f}^{\pm}(v) :=B^{G\pm}_{vef} = B^{G\pm}_{ve'f} , \qquad \sum_{f: e \subset \partial f} j_f \kappa_{ef} (\mp \ui)^{\frac{1-t_f}{2}} B^{\pm}_{f}(v) =0 \,.
\end{align}
with $	B^{\pm}_{f}(v)= (*)^{\frac{1-t}{2}} G^{\pm}_{ve} \left(B_{vef}^{-} + (-1)^{\frac{\pm 1 -1}{2}}  M_{vef} \right) (G^{\pm}_{ve})^{-1}$ for $G = (G^{-})^{-1}, \;  G^{+} R_{e}$.
Clear when $M_{vef} = 0$, $G^{\pm}$ are two possible sets of solutions of above equation.

The parallel transport equations are invariant by the Hodge duality $*$ acting on the bivectors. As a result, we have two possibilities of the geometrical interpretation of the bivectors $B_{f}(v)$. They can be generally interpreted as either timelike bivectors or spacelike bivectors, related by the Hodge duality. However, note that from the closure condition,
the bivector associated to edge $e$ must be defined simultaneously as $B_{vef}$ or $* B_{vef}$. Since $B_{vef}$ is always a timelike bivector in our notation, the corresponding geometrical faces associated to given edge $e$ will be determined up to an overall flip of the signature of the metric associated to these faces. For example, when actions at a given edge $e$ are all space actions, the corresponding geometrical faces can be interpreted as all timelike or all spacelike (Note that however only one set of these signature at $e$ will have a possible geometric explanation as tetrahedron). A special situation is the case where both time and space actions appear at a given edge $e$ (which is the mixed case in \cite{Liu:2018gfc}), the geometric triangles corresponding to the bivector $B$ at edge $e$ will always contain both time and space action. In this case, an extra $\ui$ or Hodge dual for time action in closure condition always appears. As a consequence, an Euclidean signature of the space where these bivectors lie in is possible only when we analytic continue the spin $j$ for time action to $\ui j$. 

In general, as a summary, for each vertex, the solution of the critical equations \eqref{eq:biveq} represents two sets of bivectors subject to closure constraints at each edge $e$ on the complex manifold. There is no simplicial geometric notion for the data associated to each edge. The bivectors lie in a 4D Lorentzian manifold, unless one impose by hand additionally cross simplicity condition (\ref{eq:conditon_simp}).

We shall move to the detailed analysis of this condition in the next section.

\section{Geometrical Interpretation and Reconstruction}\label{sec4}

Since the equations of motion (\ref{eq:biveq1}) contains two sets of equations for bivectors $B_{vef}^{G\pm}$ in 4D Lorentzian space and $\tilde{g}_{ve}^{\pm} \in SL(2,\mathbb{C})$. As a result, we can explain the bivectors $\{ B_{vef}^{G+}, B_{vef}^{G-}\}$ satisfying (\ref{eq:biveq1}) as pair of two geometries in 4D Lorentzian space. The gauge transformations (\ref{complexgauge}) of $SO(1,3)_{\mathbb{C}}$ group elements becomes gauge transformations on $B_{vef}^{G\pm}$ separately. Thus these two geometries can be regarded as independent geometries given by boundary $B_{vef}^{\pm}$ following independent gauge transformations at each vertex. We will summarize all possible geometries appearing in 4D Lorentzian space in this section, and build the link between bivector solutions $B_{vef}^{G\pm}$ and these 4D Lorentzian geometries. The $4$-simplex geometry and degenerate vector geometry will appear as the subsets of all possible geometries correspond to $B_{vef}^{G\pm}$. 

Note the bivectors $\{ B_{vef}^{G \pm}\}$ transform non-trivially between neighboring $v$ and $v'$, the reconstructed geometries can not be glued together unless $B_{vef}^{G \pm} = B_{v'ef}^{G \pm}$. Thus the discussion of this section focuses on the geometrical reconstruction of a single 4-simplex, except for the paragraphs of Eqs. (\eqref{OEE} - \eqref{ERE}) where 4-simplex geometries are glued to form a geometrical triangulation. The boundary geometries in this section means the data of 5 boundary tetrahedra of the 4-simplex.

\subsection{Classification of geometries}

\subsubsection{Non-degenerate simplicial gemometry}

A non-degenerate geometrical 4-simplex up to global scaling is specified by $5$ 4D normals $U_i:=V_i N_i$ where any $4$ of them are linearly independent. Note that the analysis here holds for all signatures of 4D spacetime $M$, not only Lorentzian. The set of $U_i$ satisfy the 4D closure condition:
\begin{align}
    \sum_{i} V_i N_i =\sum_{i} U_i  = 0 \,.
\end{align}
The geometrical 4-simplex is bounded by 3D planes orthogonal to the normals. The 3D boundary is also simplicial, and made by tetrahedra orthogonal to the normals $N_i$. Each $V_i$ is the volume of corresponding boundary tetrahedron. The boundary of these tetrahedron are triangles specified by the bivector
\begin{equation}
    B^{\Delta}_{ij} = V_4 {*( U_{i} \wedge U_{j} )}\,,
\end{equation}
where $V_4$ is the oriented volume of the $4$-simplex given by
\begin{align}
    \frac{1}{V_4} = \frac{1}{5!}\sum_{i,j,k,l} \epsilon_{ijkl}\det[U_i, U_j, U_k, U_l ]
\end{align}
where the orientation of the $4$-simplex is given by the ordering of these 5 normals.
One can check that the bivectors satisfy the following equation from the 4D closure
\begin{equation}
 \forall_{i} \;  \sum_{j, j \neq i} B^{\Delta}_{ij} =0, \qquad N_i \cdot  B^{\Delta}_{ij} = 0 \,.
\end{equation}
This is the closure and linearized simplicity conditions which imply the cross simplicity condition (\ref{eq:conditon_simp}) that results in the simplicial boundary geometry of the 4-simplex. The 3D normal of the triangles in the boundary tetrahedra are given by
\begin{align}
    \vec{n}_{ij} = |B^{\Delta}_{ij}| \frac{ N_{j} - t_i (N_{i} \cdot N_{j}) N_{i} }{ |t_j - t_i (N_{i} \cdot N_{j})^2 | } \,.
\end{align}
The co-frame of the $4$-simplex is specified by 
\begin{align}
    E_{ij}^{I} = \frac{V_4}{3!} \sum_{l,m,n} \epsilon_{ijlmn} \epsilon^{IJKL} U_{lJ} U_{mK} U_{nL} \,,
\end{align}
where $E_{ij}^{I}$ is the vector related to each oriented edge shared by tetrahedra $l,m,n$, as the discretization of the co-tetrad $e_i^I$ of the manifold.
The face bivectors now can be rewritten as
\begin{align}
    B_{ij}^{\Delta} =\frac{1}{3!} \epsilon_{ijlmn} (E_{lm} \wedge E_{ln}) \,.
\end{align}

The shape of the $4$-simplex is determined by it's $10$ edge lengths. This implies that, in order to form a $4$-simplex, the boundary tetrahedra must satisfy the length matching condition (When gluing together boundary tetrahdra to form the $4$ -simplex, the lengths of the common triangle of boundary tetrahedra need to the same. This condition can also be described as shape matching condition). Moreover, %
in order to form a 4-simplex, the oriented volume for the boundary tetrahedra must have the same sign. %
As a result, one has to choose a consistent orientation of the boundary tetrahedra prior to construct the 4-simplex such that their oriented volumes have the same sign.

When the simplicial geometry is composed by several $4$ simpilcies, we can define the co-frame at each $4$-simplex. these co-frames of neighboring $4$ simplices are related to each other by a $SO(M)$ group element $\Omega_I{}^J$ such that
\begin{align}
  \forall_{i \neq j}  \Omega_I{}^J(v',v) E_{ij}(v) = E_{ij}(v') \,, \qquad \Omega_I^J(v',v) N_{e}(v) = N_{e}(v') \label{OEE}
\end{align}
at the shared tetrahedron $t_{e}$ and the group element is determined uniquely by the common edges at the shared tetrahedron $t_{e}$. Notice that, in order to have a consistent orientation on the entire simplicial manifold, for every internal tetrahedron, its orientation seen from different neighboring $4$ simplices must be opposite. When the sign of the oriented volume, $\sgn(V)$, of neighboring $4$ simplices are the same, the above $\Omega_I^J$ is the discrete spin connection. For boundary tetrahedra, the above relation between neighboring co-frames then restricted to boundary symmetry groups $SO(V)$ with $V$ a 3D subspace of $M$.

Simplicial geometries are said to be gauge equivalent if there exists group elements in special orthogonal group  $G_{v} \in SO(M)$ at each vertex $v$ such that the co-frames $\tilde{E}_{ij}(v)$ and ${E}_{ij}(v)$  are related by
\begin{align}
 \forall_{ij}  \tilde{E}_{ij}(v) = G_{v} E_{ij}(v) \,.
\end{align}
The above transformation of co-frames of simplicies is the gauge coordinate transformation which will not change the geometry and orientations. Notice that, for given nondegenerate length data satisfying the length matching condition at each vertex, there are always a geometric 4-simplex up to rotations in the orthogonal group $O(M)$. As a result, there are two non-gauge equivalent geometries related by a reflection:
\begin{align}
 \forall_{ij}  \tilde{E}_{ij}(v) = R_{e_a} E_{ij}(v),\label{ERE}
\end{align}
where $R_{e_a}$ is the reflection with respect to any normalized vector $e_a$. These two geometries then have opposite oriented volume. 

When parametrizing the simplicial geometry in terms of edge lengths and angles, it is manifestly $SO(M)$ invariant. We will see later in the reconstruction that the simplicial geometries appear as the corresponding solutions of the critical point equations. The gauge transformation of $SO(1,3)_{\mathbb{C}}$ is a pair of two $SO(1,3)$ transformations acing on the Lorentzian simplicial geometry, and leaving the geometry invariant.

\subsubsection{Degenerate vector geometry}
A degenerate vector geometry is again specified locally by $10$ faces. However, now these face bivectors $B^{\Delta}_{ij} = -B^{\Delta}_{ji}$ with $i,j \in (1,...,5)$ are all lying in the same three dimensional subspace of the $4$ dimensional Minkowski space, namely,
\begin{equation}
    B^{\Delta}_{ij} = \vec{\cv}^{\Delta}_{ij}  \cdot \vec{\tau} , \quad \vec{\cv}^{\Delta}_{ij} \in \mathbb{R} \,,
\end{equation}
where $\tau^{i}$ represents the generators of $SU(2)$ if the three dimensional subspace is Euclidean or $SU(1,1)$ if the subspace is Lorentzian.
The bivector equations then become vector equations, namely
\begin{equation}
    \vec{\cv}^{\Delta}_{ij}= - \vec{\cv}^{\Delta}_{ji}, \qquad  \forall_{i} \;  \sum_{j, j \neq i} \vec{\cv}^{\Delta}_{ji}=0 \,.
\end{equation}
Thus the geometry is given by $10$ 3D normals by the Minkowski theorem.
The extra simplicial condition for the simplicial geometry are automatically satisfied:
\begin{equation}
   \forall_{i,j} \,  N \cdot  B^{\Delta}_{ij} = 0 \,.
\end{equation}
with $N=(1,0,0,0)$ or $N=(0,0,0,1)$ up to $O(1,3)$ rotations.
Notice that, for a simplicial geometry in 4D Euclidean space or split signature space, since the Hodge duality satisfies $*^2 = 1$, we can always introduce a map on the bivector  $B^{\Delta}_{ij}$ by decomposing it into self dual and anti-self dual part:
\begin{align}
   \Phi^{\pm}: \Lambda^2(M') \to V : \Phi^{\pm}(B^{\Delta}_{ij}) = (* B^{\Delta}_{ij} \pm B^{\Delta}_{ij}) \cdot N =  \vec{\cv}^{\Delta \pm}_{ij} \,,
\end{align}
such that
\begin{align}
    N \cdot \vec{\cv}^{\Delta \pm}_{ij} =(\pm (B^{\Delta}_{ij})_{IJ}N^IN^J + (* B^{\Delta}_{ij})_{IJ}N^IN^J )  = 0 \,.
\end{align}
The inverse map is given by
\begin{align}
    \Phi^{-1}(\vec{\cv}^{\Delta +}_{ij},\vec{\cv}^{\Delta -}_{ij}) =\frac{1}{2} \Big[ ( \vec{\cv}^{\Delta +}_{ij} - \vec{\cv}^{\Delta -}_{ij}) \wedge N + * ( ( \vec{\cv}^{\Delta +}_{ij} + \vec{\cv}^{\Delta -}_{ij}) \wedge N ) \Big] =B^{\Delta}_{ij} \,.
\end{align} 
One can check that, 
\begin{align}
     \Phi^{-1}(\vec{\cv}^{\Delta +}_{ij},\vec{\cv}^{\Delta -}_{ij}) \cdot \Phi^{-1}(\vec{\cv}^{\Delta +}_{ij},\vec{\cv}^{\Delta -}_{ij}) =&\frac{1}{4} \Big[ -t ( \vec{\cv}^{\Delta +}_{ij} - \vec{\cv}^{\Delta -}_{ij})^2 -t (\vec{\cv}^{\Delta +}_{ij} + \vec{\cv}^{\Delta -}_{ij})^2 \Big] \\
    =& - \frac{t}{2} \Big[ ( \vec{\cv}^{\Delta +}_{ij})^2 +  (\vec{\cv}^{\Delta -}_{ij})^2 \Big] \,,\\
     * ( \Phi^{-1}(\vec{\cv}^{\Delta +}_{ij},\vec{\cv}^{\Delta -}_{ij}) ) \cdot \Phi^{-1}(\vec{\cv}^{\Delta +}_{ij},\vec{\cv}^{\Delta -}_{ij}) =&\frac{1}{2} \Big[ -t ( \vec{\cv}^{\Delta +}_{ij} - \vec{\cv}^{\Delta -}_{ij}) \cdot (\vec{\cv}^{\Delta +}_{ij} + \vec{\cv}^{\Delta -}_{ij}) \Big] \\
    =& - \frac{t}{2} \Big[ ( \vec{\cv}^{\Delta +}_{ij})^2 -  (\vec{\cv}^{\Delta -}_{ij})^2 \Big] \,.
\end{align} 
Thus when $( \vec{\cv}^{\Delta +}_{ij})^2 =  (\vec{\cv}^{\Delta -}_{ij})^2$ the bivector $B_{ij}$ is simple and have the same norm specified by the vector up to a signature. As a result, the maps build the correspondence between simplicial geometries in Riemannian or flipped signature space and the vector geometries in their subspace. At given vertex, the flipped signature simplicial geometry and the vector geometries under the maps $\Phi$ clearly have the same boundary geometries, since the boundary bivector are given as $B_{ij} = * ( \vec{\cv}^{\Delta}_{ij} \wedge N )$ which satisfies
\begin{align}
    \Phi^{+}(B_{ij}) = \Phi^{-}(B_{ij}) = \vec{\cv}_{ij} \,.
\end{align}
Notice that, when the original simplicial geometries in Euclidean space or split signature space are degenerate, we have $B^{\Delta}_{ij} =B_{ij}= * ( \vec{\cv}^{\Delta}_{ij} \wedge N )$ up to gauge transformations, such that
\begin{align}
    \Phi^{+}(B^{\Delta}_{ij}) = \Phi^{-}(B^{\Delta}_{ij}) = \vec{\cv}^{\Delta}_{ij} \,.
\end{align}
When $\vec{\cv}^{\Delta +}_{ij} = \vec{\cv}^{\Delta -}_{ij}$, the inverse map gives
\begin{align}
    \Phi(\vec{\cv}^{\Delta}_{ij} ,\vec{\cv}^{\Delta}_{ij} ) = * ( \vec{\cv}^{\Delta}_{ij} \wedge N ) \,.
\end{align}
Namely, non-degenerate $4$-simplex geometries in flipped space are always in one to one correspondence to two non-gauge equivalent vector geometries.

The map also induces a map on transformations with group elements  $G \in SO(4)$ or $G \in SO(2,2)$,
\begin{align}
    \Phi^{\pm}( G B G^{-1}) = \Phi^{\pm}( G ) \vec{\cv}^{\Delta \pm}_{ij}\, , \qquad \Phi^{\pm}( G ) \in O(V)
\end{align}
since it keeps the norm unchanged. As a result, in this case the geometric solution satisfies
$ \Phi^{+}(G) = \Phi^{-}(G) $ if and only if $G N = \pm N$ up to gauge transformations.

\subsubsection{Lorentzian $SO(1,3)$ bivector geometry}
 Generally speaking, the $SO(1,3)$ geometry are specified by $10$ faces whose simple face bivectors $B^{\Delta}_{ij} = -B^{\Delta}_{ji}$ with $i,j \in (1,...,5)$ in the 4D Minkowski space satisfy the closure condition at each $i$:
\begin{equation}
 \forall_{i} \;  \sum_{j, j \neq i} B^{\Delta}_{ij} =0 \,.
\end{equation}
Each $i$ here related to a $SO(1,3)$ boundary geometry composed by $4$ faces with bivectors $B_{ij}, j \neq i$. The simplicial geometries (4-simplex or vector geometries) are a sub class of this geometry where these boundary satisfying further cross simplical constraint, or 
\begin{align}
 \exists N_i, \;\; s.t. \;\; N_i \cdot  B^{\Delta}_{ij} = 0 \,.
\end{align}
This condition actually implies the simplicity to the boundary geometry. In the case when the boundary satisfying closure condition but do not satisfy the cross simplicity constraint, these boundary bivectors do not belong to the same lower dimensional subspace. We call this geometry the $SO(1,3)$ boundary, which does not correspond to a simplicial geometry.

The non-simplicial geometry can be regarded as a composition of two orthogonal vector geometries in a corresponding 3 dimensional Euclidean or Lorentzian subspace, since we can always decompose the bivector as 
\begin{equation}
    B^{\Delta}_{ij} = (\vec{\cv}^{\Delta \cR}_{ij} + \ui \vec{\cv}^{\Delta \cI}_{ij} ) \cdot \vec{\tau} 
\end{equation}
with real 3D vectors $\vec{\cv}^{\Delta \cR}_{ij}$ and $\vec{\cv}^{\Delta \cI}_{ij}$. These vectors satisfy
\begin{equation}
    |\vec{\cv}^{\Delta \cR}_{ij}|^2 - |\vec{\cv}^{\Delta \cI}_{ij}|^2 = | B^{\Delta}_{ij} |^2, \qquad  \vec{\cv}^{\Delta \cR}_{ij} \cdot \vec{\cv}^{\Delta \cI}_{ij} =0 \,,
\end{equation}
where the fact that the face bivector $B^{\Delta}_{ij}$ is simple is encoded in the last equation. The bivector equations then become two vector equations for $\vec{\cv}^{\Delta}=\vec{\cv}^{\Delta \cR}, \vec{\cv}^{\Delta \cI}$
\begin{equation}
    \vec{\cv}^{\Delta}_{ij}= - \vec{\cv}^{\Delta}_{ji}, \qquad  \forall_{i} \;  \sum_{j, j \neq i} \vec{\cv}^{\Delta}_{ji}=0 \,.
\end{equation}
$\{\vec{\cv}^{\Delta \cR}, \vec{\cv}^{\Delta \cI}\}$ can be regarded as the lie algebra element of $\mathfrak{so}(1,3)$ for boost and rotation parts respectively.

We can introduce new bivectors $B^{\Delta \cR}$ and $B^{\Delta \cI}$ defined as
\begin{align}
    B^{\Delta \cR}_{ij} = -  B^{\Delta \cR}_{ji}   = \vec{\cv}^{\Delta \cR}_{ij} \cdot \vec{\tau} , \qquad B^{\Delta \cI}_{ij}= -B^{\Delta \cI}_{ji}  = \vec{\cv}^{\Delta \cI}_{ij} \cdot \vec{\tau} \,
\end{align}
with $\tr(B^{\Delta \cR} \cdot  B^{\Delta \cR}) = 0$. The bivector $B^{\Delta}_{ij} $ is then decomposed as
\begin{align}\label{decom_so13_to2}
    B^{\Delta}_{ij} = B^{\Delta \cR}_{ij} + * B^{\Delta \cI}_{ij} \,,
\end{align}
where both $B^{\Delta \cR}$ and $B^{\Delta \cI}$ satisfy closure condition
\begin{align}\label{clo_for_two}
     \forall_{i} \;  \sum_{j, j \neq i} B^{\Delta \cR}_{ij} = \sum_{j, j \neq i} B^{\Delta \cI}_{ij} =0 \,.
\end{align}
The decomposition (\ref{decom_so13_to2}) are invariant under $SO(1,3)$ transformations for each $i$. As a result, we can always explain the $SO(1,3)$ bivector geometry as the composition of two orthogonal vector geometries \footnote{Here orthogonal means in the 3D subspace, the normals of boundary tetrahedra of these two vector geometries are orthogonal to each other}, related by (\ref{decom_so13_to2}). The geometry is invariant under an overall $SO(1,3)$ rotation which rotates simultaneously two vector geometries. 
Due to (\ref{decom_so13_to2}), the overall $SO(1,3)$ rotation of a single vector geometry is not allowed.

Notice that, $10$ bivectors $B_{ij}^{\Delta} = -B_{ji}^{\Delta}$ are totally determined if the $B_{ij}^{\Delta}$ for the geometries of three boundary tetrahedra are given. This can be seen from the fact that three boundary tetrahedra determine $9$ out of $10$ bivectors, and the only one left needs to satisfy two closure conditions thus is determined uniquely. When the data of three boundary tetrahedra out of five satisfy the closure condition and length matching condition on the gluing triangles, the only geometry it can form is a 4-simplex (or degenerate vector geometry).

\subsection{Geometric condition and solutions}
Clearly by comparing the equations of motion (\ref{eq:biveq1}) with the geometric condition, we see immediately the correspondence between them. More specifically, the bivector solutions $B^{G}$ to the equation of motion corresponds to the geometrical bivectors $B^{\Delta}$ via
\begin{equation}
    B^{\Delta} = r \kappa B^{G}
\end{equation}
where $r = \pm$ related to the orientation and the oriented volume of the geometry. One then can reconstruct geometries from $B^{\Delta}$.
 According to the classification, different geometries are distinguished via their boundary geometries at each vertex. One should keep in mind such boundary geometry is not necessarily a simplicial geometry, unless specified, there will be no simplicial meaning of geometry.
 
 Since the equations of motion (\ref{eq:biveq1}) contains two sets of bivector equations, there will be two 4D geometries reconstructed out from $B^{G\pm}$ respectively at each vertex $v$. As we already argued in Sec. \ref{sec3}, $B^{G\pm}$ may correspond to different geometries. As a result, the two 4D geometries may be in different classes: they can be possible pairs of combinations of non-degenerate Lorentzian simplex, vector geometries and Lorentzian non-simplicial bivector geometry. The pair of geometries reconstructed from $(B^{G
+},B^{G-})$ can be understood as the geometry correspond to $SO(4,\mathbb{C})$ group element which are invariant under $SO(4,\mathbb{C})$ transformations by pairs of $(g^{+},g^{-}) \in SO(4,\mathbb{C})$ respectively. The transformation of the geometry is consistent with the gauge transformations of the analytic continued action given by (\ref{complexgauge}). Moreover, as we shown in Sec. \ref{sec3}, for given edge $e$ the boundary geometries given by $B_{vef}^{\pm}$ and $B_{v'ef}^{\pm}$ may be different. Thus the reconstructed geometries at neighboring vertices may be in different classes.

There is a special case when the boundary geometry correspond to $B^{+}$ are the same with $B^{-}$ up to geometrical gauge transformations. Namely, we will have $\eta_{e} B^{+}_{vef} \eta_{e} = \pm \mathfrak{a}_{ve} B^{-}_{vef} \mathfrak{a}^{-1}_{ve}$ for all $ef$ at a given vertex $v$ for some $\mathfrak{a}_{ve} \in SL(2,\mathbb{C})$. In this case, the pairs of geometries correspond to $B^{G\pm}$ are equivalent to each other up to reflections and $SO(1,3)$ gauge transformations. As we derived in Sec. \ref{sec3}, a simple situation for this is $\alpha =\alpha' = 0$ for space action and $s_{vef} = 1$ as well as $\alpha + \alpha' = 0$ for time action. This seems to be the only possible case to have a same boundary geometry and remove the $v$ dependence for internal vertices since the matrix transform from $\eta_{e} B^{+}_{vef} \eta_{e}$ to $B^{-}_{vef}$ which is $(\mathfrak{a}^{-}_{vef})(\mathfrak{a}^{+}_{vef})^{-1}$ dependents non-trivially on $v$ and $f$.

\subsubsection{Non-simplicial $SO(1,3)$ boundary}
From  (\ref{eq:conditon_simp}), it is clear that the the cross simplicity condition is invariant under the action of group element $g_{ve}^{\pm}$ on boundary bivectors $B_{vef}^{\pm}$ for given edge $e$. This reflects the fact that geometrically the shape of the boundary geometry is invariant under overall $SO(1,3)$ gauge transformations. As a result, the appearance of non-simplicial boundary is determined by $B_{vef}^{\pm}$. From definition (\ref{B+ina_sum}), for boundary edges, since $\tilde{v}_{ef} = v_{ef} \in H \subset SL(2,\mathbb{C})$ are not complexified, the existence of non-simplicial geometry for the boundary edge clearly implies one must have non-trivial solutions of $\alpha, \alpha'$ at edge $v$. This is the case, for example, when the boundary data does not satisfy the closure condition. The existence of $\alpha, \alpha'$ then opens the possibilities to have non-trivial solutions as complex critical point which contribute to the leading order critical action with $\Re(\tilde{S}) < 0$ for the analytic continued action $\tilde{S}$.

For the internal faces, due to the analytical continuation of $\xi$ and $\xi'$, it is not necessarily to have $\alpha, \alpha'$ non-vanishing for a non-simplicial boundary.

\subsubsection{Simplicial boundary}

When the boundary satisfies the cross simplicity constraints, the critical equations are exactly two copies of the equations of motion derived in the original real EPRL-CH model (\cite{Barrett:2009gg,Barrett:2009mw,Barrett:2010ex,Kaminski:2017eew,Liu:2018gfc}), whose solutions corresponding to $4$-simplices or degenerate vector geometries, as described in previous section. We briefly summarize the result here. For the detailed reconstruction of geometry from the solution, we refer to (\cite{Barrett:2009gg,Barrett:2009mw,Barrett:2010ex,Kaminski:2017eew,Liu:2018gfc}).

Since the boundary geometries are simplicial, they correspond to tetrahedra in a 3D subspace. As a result, we can reconstruct lengths of all the tetrahedra at given vertex $v$. Here we will only concentrate on the case when boundary data satisfies the length matching condition and non-degenerate. When it does not satisfy the length matching condition or is degenerate, there will be no solution or only one set of vector geometry solutions exist for each copy of the geometric equations of motion. 

According to the geometric interpretation and reconstruction theorem of EPRL-CH model, we have the following $2$ possibilities at a given vertex determined by their boundaries, which can be described by the signature of length gram matrix contains all boundary lengths at each vertex:
\begin{itemize}
    \item Boundary corresponds to Lorentzian signature signature geometry.
    
     Notice that, for given solution of bivectors $B_{ef}(v)$ satisfying equation of motions, one can reconstruct uniquely up to a sign $s_{ve} = \pm 1$ the normals $ N_{e}(v)$ which satisfying $ B_{ef}(v)  \cdot N_{e}(v)= 0$.
     these normals are given by $N_{e}(v) = G_{ve} u_{e} $ and they are non-degenerate in this case.  The sign $s_{ve}$ here related to the inversion gauge transformations $G_{ve} \to - G_{ve}$. Using the normals, one can shown that the bivectors can be rewritten as
     \begin{align}
         B_{f(e,e')}(v) = \lambda * (N_e \wedge N_{e'})
     \end{align}
     with $\lambda \in \mathbb{R}$.
     
    Compare with the normals and bivectors for geometric $4$ simplicies, we see their relation to geometrical normals $N_{e}^{\Delta}(v)$ and bivectors $B_{ef}^{\Delta}(v)$ of some simplicial geometry are given as
    \begin{align}
        N_{e}(v) = (-1)^{s_{ve}} N_{e}^{\Delta}(v)\,, \qquad B_{f}(v) = {r_{v}} B_{f}^{\Delta}(v) \,.
    \end{align}
    Thus these solutions correspond to geometrical Lorentzian $4$-simplices, which are bounded by 3D planes orthogonal to the normals. Notice that, the existence of $4$-simplex geometry implies that the boundary geometries at each vertex satifying length (shape) matching and orientation matching, otherwise the critical equations have no solution.
    
    From the fact that $N_{e}(v) = G_{ve} u_{e}$, we then have
     \begin{align}
        G_{ve}= G_{ve}^{\Delta} I^{s_{ve}} (I R_{u_e})^{s_v} \,,
    \end{align}
    which implies
     \begin{align}
       \forall_{e : v \subset \partial e} \det G_{ve}^{\Delta} = r_{v} \,,
    \end{align}
    where $r_{v} = \pm 1$ is the Plebanski orientation of the geometric simplcies. Clearly at each vertex, if the boundary satisfies the length matching condition and orientation matching condition, there exists two solutions for given boundary $\tilde{v}_{ef}$, which relates to $4$ simplicies up to the Plebanski orientation. We denote these two solutions as $G$ and $G'$, they are related by the following relation 
    \begin{equation}\label{solution_relation}
        G'_{ve} = R_{e_{\alpha}} G_{ve} R_{u_{e}}
    \end{equation}
   up to geometrical gauge transformations which corresponds to the reflection of geometries. In terms of spin-$\frac{1}{2}$ representation, one can show that these two solutions are related by $g'=J^{-1} g J=g^{-1 \dagger}$.

   In the case when the two boundary geometries correspond to $B^{\pm}$ are the same (in the case $\alpha = \alpha' =0$ for space action and $\alpha + \alpha' = 0$ for time action) , 
   the two copies of equations of motion coincide with each other:
\begin{align}\label{coincide_bdy_eq}
    B^{G}_{ef} : = G_{ve} B_{ef} G_{ve}^{-1} = G_{ve'} B_{ef} G_{ve'}^{-1}, \qquad 0 = \sum_{f} j_f \kappa_{f} B^{G}_{ef} 
\end{align}
with $ G_{ve} = (G_{ve}^{-})^{-1}, \,  G_{ve}^{+} R_e$. As a result, $G_{ve}^{\pm}$ are the two possible solutions of the same sets of geometric equations of motion up to a possible rotation $R_e$.
    As a result, we then have $4$ possibilities for $\tilde{G}=(\tilde{G}^{+}, \tilde{G}^{-})$ at each vertex: $\tilde{G}=(\tilde{G}^{+}, \tilde{G}^{-})$: $\tilde{G}=(G R_e, (G)^{-1})$, $\tilde{G}=(G R_e, (G')^{-1})$ and $\tilde{G}=(G' R_e, (G)^{-1})$, $\tilde{G}=(G' R_e, (G')^{-1})$ for two non-gauge equivalent geometrical solutions $G$ and $G'$ of (\ref{coincide_bdy_eq}). 
    
     \item Boundary corresponds to Riemannian or split signature geometry.
    
    In these cases, the solutions $\{ g\}$ are in the subgroup of $\text{SL}(2,\mathbb{C})$, which is the stabilizer group for some given normal $u$ of the boundary geometry, namely $g \in SU(2)$ for $u = e_0$ and $g \in SU(1,1)$ for $u=e_3$. 
       
            we will have two non-gauge equivalent sets of vector geometry solutions for given boundary bivectors $B_{vef}$, which we denotes as $(\cv_f(v),\cv'_f(v))$. We have $0 = u \cdot \cv_f(v) = u \cdot \cv'_f(v)$ with $u=e_0$ or $u=e_3$ correspondingly. $(\cv_f(v),\cv'_f(v))$ correspond to a Riemannian or Split signature $4$-simplex by the map
            \begin{align}
                 \cb_f(v) = \Phi^{-1}(\cv^{+}_f(v),\cv^{-}_f(v)) \,.\,
            \end{align}
             The reconstruction then follows exactly the same procedure for the non-degenerate Lorentzian simplicial case, with two sets of geometrical simplicies solutions $\cg, \cg' $ related to the vector geometry solutions by the induced map:
\begin{align}
    \cg = \Phi^{-1}(G_f(v),G'_f(v)), \qquad \cg' = \Phi^{-1}(G'_f(v),G_f(v)).
\end{align}

In the case when the two boundary geometries given by $B^{\pm}$ are the same, the two copies of equations of motion are coincide with each other. Thus we have $4$ possibilities for $\tilde{G}=(\tilde{G}^{+}, \tilde{G}^{-})$ again: $\tilde{G}=(G R_e, (G)^{-1})$, $\tilde{G}=(G R_e, (G')^{-1})$ and $\tilde{G}=(G' R_e, (G)^{-1})$, $\tilde{G}=(G' R_e, (G')^{-1})$ with $ $ two non-gauge equivalent sets of vector geometry solutions for boundary $B^{\pm}$. 
            
\end{itemize}
   
   Note that, these solutions will reduce to the usual real solution of EPRL-CH model when we restrict $\tilde{g}$ to $\tilde{g} = (\tilde{g}^{+}, \tilde{g}^{-}) = (g, g^{\dagger})$, and restrict $\tilde{v}$ as the stabilizer group compatible with $\eta_{e}$ appears in the action. The solution in such case can be seen from parallel transport equations and their complex conjugation:
        \begin{equation}
            g_{ve} B_{ef} g_{ve}^{-1} = g_{ve'} B_{e'f} g_{ve'}^{-1}, \qquad \;\;\;  (g_{ve})^{-1 \dagger} (B_{ef})^{\dagger} g_{ve}^{\dagger} = (g_{ve'})^{-1\dagger} (B_{e'f})^{\dagger} g_{ve'}^{\dagger}.
        \end{equation}
        With the fact $\ui^{\frac{1+t_f}{2}} B_{ef} \in \mathfrak{su}(2)$ or $\ui^{\frac{1+t_f}{2}} B_{ef} \in \mathfrak{su}(1,1)$ up to gauge transformations, we have $(B_{ef})^{\dagger} = t_f \eta_{e} B_{ef} \eta_{e}$. Then 
        \begin{equation}
        (g_{ve})^{-1 \dagger} R_{e} B_{ef}  (R_{e})^{-1} g_{ve}^{\dagger} = (g_{ve'})^{-1\dagger} R_{e'} B_{e'f}  (R_{e'})^{-1} g_{ve'}^{\dagger}.
        \end{equation}
        When there is only one solution, this directly implies $(g_{ve})^{-1 \dagger} R_{e} = g_{ve}$, thus $g \in SU(2)$ or $SU(1,1)$, and the solution corresponds to vector geometry. When there are two solutions, in the non-degenerate case since we have $(g_{ve})^{-1 \dagger} R_{e} \neq  g_{ve}$, this means $(g_{ve})^{-1 \dagger} R_{e}$ is another solution for the critical equations, which corresponds to the solution with opposite Plebanski orientation from reconstruction. This is the so-called Parity transformed solution in \cite{Han:2013gna,Barrett:2010ex} and the above relation confirms the fact that there exists two solutions for non-degenerate case, which are related by $g'=J^{-1} g J=g^{-1 \dagger}$. One can then identify solution  $\tilde{G}=(G R_e, (G')^{-1})$ and $\tilde{G}=(G' R_e, (G)^{-1})$ as the real critical point of EPRL-CH model, which leads to $\Re S = 0$. 

\section{Evaluation of the Amplitude}\label{sec5}
One can check that, by inserting the decomposition of $Z$, the function $\tilde{F}$ can be expressed by
\begin{align}
    &\tilde{F}_f[\tilde{X}_0]=\kappa_{f} \sum_{v: f\subset v} \Big[ \theta'_{e'vef}-{ \theta^m_{e'vef}}  + \ui \gamma  \left( { \theta'_{e'vef}} + { \theta^m_{e'vef}} \right)  + f(\alpha,\alpha') \Big] & \text{ space action} \\
   & \tilde{F}_f[\tilde{X}_0]=\kappa_{f} \sum_{v: f\subset v} \Big[  \gamma (\theta_{e'vef}- \theta'_{e'vef}) + \ui (\theta_{e'vef} + \theta'_{e'vef})  + f(\alpha,\alpha') \Big] & \text{time action}
\end{align}
with
\begin{align}
    f(\alpha,\alpha') := \left\{ \begin{array}{cc}
         \ui \gamma \ln \frac{1 + \alpha_{ve'f} \alpha'_{ve'f} }{1 + \alpha_{ve'f} \alpha'_{ve'f}}  +  \ln \frac{(1 + \alpha_{ve'f} \alpha'_{ve'f} )^{-\det \eta_{e'}}}{(1 + \alpha_{ve'f} \alpha'_{ve'f})^{\det \eta_e}}  &  \text{ space action} \\
         \ui \ln \frac{(\alpha_{ve'f}+\alpha'_{ve'f})^{1-s_{ve'}}}{(\alpha_{vef}+\alpha'_{vef})^{1-s_{ve}}} & \text{time action}
    \end{array} \right. \, ,
\end{align}
which is a function depends on $\alpha, \alpha' \in \mathbb{C}$. $\theta^{m}_{e'vef} = \ln m_{ef}m_{e'f} + \theta_{e'vef}$ is a term related to the action. When summing over vertices, the term $\ln m_{ef}m_{e'f} $ in internal faces will cancel with each other thus becomes a pure boundary term.
Here $\theta_{e'vef}$ and $\theta'_{e'vef}$ are defined as
\begin{align}
    &\theta^{m}_{e'vef} = \ln m_{ef}m_{e'f} + \theta_{e'vef} \ln \frac{\zeta_{ve'f}}{\zeta_{vef}}  &\theta'_{e'vef} = \ln \frac{\zeta'_{ve'f}}{\zeta'_{vef}} \,.
\end{align}

At critical configurations,
$\zeta_{vef}, \zeta'_{ve'f} \in \mathbb{C} $ are some complex numbers determined by
\begin{align}\label{space_sol_para}
    Z_{vef}^0 =  {\zeta}_{vef} (\tilde{\xi}_{ef}^0 + \alpha_{vef} J\tilde{\xi}_{ef}^0) \, , \qquad Z'{}^0_{vef} =  {\zeta}'_{vef} ( \tilde{\xi}'{}_{ef}^0+\alpha'_{vef} J\tilde{\xi}_{ef}^0)
\end{align}
for space action and 
\begin{align}\label{time_sol_para}
    Z_{vef}^0 =  \zeta_{vef} (\tilde{l}^{\mp 0}_{ef} +\alpha_{vef} \tilde{l}^{\pm 0}_{ef} ), , \qquad Z'{}_{vef}^0 =  \zeta'_{vef} (\tilde{l}'{}^{\mp 0}_{ef} +\alpha'_{vef} \tilde{l}'{}^{\pm 0}_{ef} )
\end{align}
for time action on critical solutions $Z^0, Z'{}^0 \in X_0$. Since for space action the parallel transport equation implies either $\alpha_{vef} = 0$ or $\alpha'_{vef} = 0$ for internal edges, thus $ f(\alpha,\alpha') $ only involves $\alpha$ and $\alpha'$ at the boundary. Since $\alpha$ and $\alpha'$ can be directly solved via equations of motion, the task is then to determine $\zeta$ and $\zeta'$, which related to loop holonomies along the face.

\subsection{Determine values from EoMs}

From the parallel transport equation for space and time action, we have
\begin{align}
	\tilde{\chi}'^{-}_{vef} \eta_{e}  \tilde{g}_{ve}^{-}  (\tilde{g}_{ve'}^{-})^{-1} =\frac{ \tilde{\zeta}_{vef} } { \tilde{\zeta}_{ve'f}  } \tilde{\chi}'^{-}_{ve'f} \eta_{e'}\, ,  \qquad \tilde{g}_{ve'}^{-}  (\tilde{g}_{ve}^{-})^{-1} \tilde{\mathfrak{z}}_{vef}  = \frac{ \tilde{\zeta}_{ve'f} } { \tilde{\zeta}_{vef}  } \tilde{\mathfrak{z}}_{ve'f} \,, \label{sec5spa1}\\
	\tilde{\mathfrak{z}}'_{vef} (\tilde{g}_{ve}^{+})^{-1}  \tilde{g}_{ve'}^{+} = \frac{ \tilde{\zeta}'_{ve'f} } { \tilde{\zeta}'_{vef}  } \tilde{\mathfrak{z}}'_{ve'f} \, ,  \qquad (\tilde{g}_{ve'}^{+})^{-1}  \tilde{g}_{ve}^{+} \eta_{e} \tilde{\chi}^{+}_{vef}  =\frac{ \tilde{\zeta}'_{vef} } { \tilde{\zeta}'_{ve'f}  } \eta_{e'} \tilde{\chi}^{+}_{ve'f}  \label{sec5spa2}\,,
\end{align}
where we define $Z = \zeta \tilde{\mathfrak{z}}$ and $Z' = \zeta' \tilde{\mathfrak{z}'}$.
The equations can be rewritten as
\begin{align}
	\tilde{g}_{ve'}^{-}  (\tilde{g}_{ve}^{-})^{-1} J  (\tilde{\chi}'^{-}_{vef} \eta_{e})^{\dagger}   =  \frac{ \tilde{\zeta}_{vef} } { \tilde{\zeta}_{ve'f}  }  J (\tilde{\chi}'^{-}_{ve'f} \eta_{e'})^{\dagger}\, ,  \qquad \tilde{g}_{ve'}^{-}  (\tilde{g}_{ve}^{-})^{-1} \tilde{\mathfrak{z}}_{vef}  = \frac{ \tilde{\zeta}_{ve'f} } { \tilde{\zeta}_{vef}  } \tilde{\mathfrak{z}}_{ve'f}  \,,\\
	 (\tilde{g}_{ve'}^{+})^{-1}  \tilde{g}_{ve}^{+} J (\tilde{\mathfrak{z}}'_{vef})^{\dagger} = \frac{ \tilde{\zeta}'_{ve'f} } { \tilde{\zeta}'_{vef}  } J (\tilde{\mathfrak{z}}'_{ve'f})^{\dagger} \, ,  \qquad (\tilde{g}_{ve'}^{+})^{-1}  \tilde{g}_{ve}^{+} \eta_{e} \tilde{\chi}^{+}_{vef}  =  \frac{ \tilde{\zeta}'_{vef} } { \tilde{\zeta}'_{ve'f}  } \eta_{e'} \tilde{\chi}^{+}_{ve'f}  \,,
\end{align}
where we use $J^{-1} g J = g^{-1\dagger}$ for any $\text{SL}(2,\mathbb{C})$ group element $g$.

Using $X$ defined by (\ref{def_x}), we have
\begin{align}
   & X_{ef}^{-} J  (\tilde{\chi}'_{vef} \eta_{e})^{\dagger} = \tilde{\mathfrak{z}}_{vef} \otimes \tilde{\chi}'_{vef} \eta_{e} J  (\tilde{\chi}'_{ef} \eta_{e})^{\dagger} =0\, ,%
   & X_{ef}^{-} \tilde{\mathfrak{z}}_{vef} = \tilde{\mathfrak{z}}_{vef} \otimes \tilde{\chi}'_{vef} \eta_{e} \tilde{\mathfrak{z}}_{vef} = \tilde{\mathfrak{z}}_{vef} \,,
\end{align}
where we use the fact %
$ a^t J a^{t\dagger} = 0$ for arbitrary spinor $a$ and $\tr(X) = 1$. From the definition of bivectors $B = X - \frac{1}{2} I$, we then have
\begin{align}
   2 B_{vef}^{-} J  (\tilde{\chi}'_{ef} \eta_{e})^{\dagger} = - J  (\tilde{\chi}'_{ef} \eta_{e})^{\dagger} \, , \qquad
   2 B_{vef}^{-} \tilde{\mathfrak{z}}_{vef} = \tilde{\mathfrak{z}}_{vef} \,.
\end{align}
Similar argument also holds for $B^{+}$ which leads to
\begin{align}\label{G_theta_internal}
   2 B_{vef}^{+} J  (\tilde{\mathfrak{z}}'_{vef} )^{\dagger} = - J  (\tilde{\mathfrak{z}}'_{vef} )^{\dagger} \, , \qquad
   2 B_{vef}^{+} \eta_{e} \tilde{\chi}_{vef} = \eta_{e} \tilde{\chi}_{vef}  \,.
\end{align}
If we introduce a group element related to boundary variables such that
\begin{align}
    \mathfrak{v}_{ee'}: m_{e'f} \mathfrak{v}_{ee'} \tilde{\xi}_{e'f} \otimes \tilde{\xi}'_{e'f} \eta_{e'} (\mathfrak{v}_{ee'})^{-1}= m_{ef}\tilde{\xi}_{ef} \otimes \tilde{\xi}'_{ef} \eta_{e} \,,
\end{align}
we then have
\begin{align}
   \tilde{\mathfrak{a}}_{vef} \mathfrak{v}_{ee'} (\mathfrak{a}_{ve'f})^{-1} B^{-}_{ve'f} \mathfrak{a}_{ve'f} (\mathfrak{v}_{ee'})^{-1} (\mathfrak{a}_{vef})^{-1} = B^{-}_{vef} \,,
\end{align}
where $\tilde{\mathfrak{a}}_{vef}$ here are related to ${\mathfrak{a}}_{vef}$ defined in (\ref{def_of_a_sp}) and (\ref{def_of_a_t}) by $\tilde{\mathfrak{a}}_{vef}  =\tilde{v}_{ef} {\mathfrak{a}}_{vef} ({\tilde{v}_{ef}})^{-1} $.
Note that since $\tilde{v} \eta \tilde{v}' = \eta$, we have
\begin{align}
    \mathfrak{v}_{ee'} = \tilde{v}_{ef} R_{e}^{-1} (-\ui \sigma_2)^{\frac{1-m_{ef}}{2}} (\ui \sigma_2)^{\frac{1-m_{e'f}}{2}} R_{e'} (\tilde{v}_{e'f})^{-1} \,.
\end{align}
Thus
\begin{align}\label{av_b}
   \mathfrak{a}_{vef} R_{e} \mathfrak{v}_{ee'} (R_{e'})^{-1} (\mathfrak{a}_{ve'f})^{-1} B^{+}_{ve'f} \mathfrak{a}_{ve'f} (\mathfrak{v}_{ee'})^{-1} (\mathfrak{a}_{vef})^{-1} = B^{+}_{vef} \,.
\end{align}
And one can check that,
\begin{align}
    &m_{e'f} \mathfrak{v}_{ee'} J  (\tilde{\xi}'_{e'f} \eta_{e'})^{\dagger}
    =m_{e'f}m_{ef}m_{ef} (\ui)^{\frac{1-\det \eta_{e}}{2}} (-\ui)^{\frac{1-\det \eta_{e'}}{2}} (\tilde{\chi}'_{ef} \eta_{e })^{\dagger} \,,\\
    &\mathfrak{v}_{ee'} \tilde{\xi}_{e'f}
    =m_{e'f} m_{ef} (-\ui)^{\frac{1-\det \eta_{e}}{2}} (\ui)^{\frac{1-\det \eta_{e'}}{2}}  \tilde{\xi}_{ef} \,,\\
    &R_{e} \mathfrak{v}_{ee'} R_{e'}^{-1} J  (\tilde{\xi}'_{e'f})^{\dagger} 
    =J (\tilde{\xi}'_{ef})^{\dagger} \,, \\
     &m_{e'f} R_{e} \mathfrak{v}_{ee'}R_{e'}^{-1} \eta_{e'} \tilde{\xi}_{e'f} 
     = m_{ef} \eta_{e}  \tilde{\xi}_{ef} \,.
\end{align}
In the case when the face contains only one vertex, this then implies
\begin{align}
   &\mathfrak{a}^{-}_{vef} \mathfrak{v}_{ee'} (\mathfrak{a}^{-}_{ve'f})^{-1} \tilde{g}_{ve'}^{-}  (\tilde{g}_{ve}^{-})^{-1} = m_{e'f}m_{ef} \ue^{(2 \theta_{vf} + \ui \pi \omega_{f}) B^{-}_{vef}} \, , \\
   &\mathfrak{a}^{+}_{vef} R_{e} \mathfrak{v}_{ee'} (R_{e'})^{-1} (\mathfrak{a}^{+}_{ve'f})^{-1} (\tilde{g}_{ve'}^{+})^{-1}  \tilde{g}_{ve}^{+} = \ue^{-2 \theta'_{vf} B^{+}_{vef}}
\end{align}
where $\omega_{f} := \frac{|\det \eta_{e'} - \det \eta_{e}|}{2} \in \{0,1\}$ and takes $1$ when $\det \eta_{e'} \neq \det \eta_{e}$, otherwise $\omega=0$.
Then $\theta$ and $\theta'$ can be expressed as
\begin{align}\label{theta_value}
	\theta_{vf}  &= \log \left[  \Tr(m_{e'f} m_{ef} \mathfrak{a}^{-}_{vef} \mathfrak{v}_{ee'} (\mathfrak{a}^{-}_{ve'f})^{-1} \tilde{g}_{ve'}^{-} (\tilde{g}_{ve}^{-})^{-1} X^{-}_{vef} ) \right] - \frac{\ui \omega_{f} \pi}{2}\, ,\\
	\theta'_{vf}  &= -t_f  \log \left[ \Tr( \mathfrak{a}^{+}_{vef} R_{e} \mathfrak{v}_{ee'} R_{e'}^{-1} (\mathfrak{a}^{+}_{ve'f})^{-1} (\tilde{g}_{ve'}^{+})^{-1} \tilde{g}_{ve}^{+} X^{+}_{vef}) \right] \,,
\end{align} 
where the $\log m_{e'f}m_{ef} $ term in $\theta_{vf}$ will cancel exactly the same term appears in the definition of $\theta^m_{vf}$, leading a critical action that independent of $m$. As a result, we can safely remove the $\log m_{e'f}m_{ef} $ terms in all the expressions for simplicity.

The analysis can be generalised to faces containing internal edges, where we can define the following group element
\begin{align}
    G_f^{-}(e_1,e_0):= \mathfrak{a}^{-}_{v_1e_1f} \left( \prod_{v \in \partial f} (\mathfrak{a}^{-}_{ve'f})^{-1} \tilde{g}_{ve'}^{-}  (\tilde{g}_{ve}^{-})^{-1} (\mathfrak{a}^{-}_{vef}) \right) (\mathfrak{a}^{-}_{v_0e_0f})^{-1} \,, \label{eqGfall1}\\
    G_f^{+}(e_1,e_0):=\mathfrak{a}^{+}_{v_1e_1f} \left( \prod_{v \in \partial f} (\mathfrak{a}^{+}_{ve'f})^{-1} (\tilde{g}_{ve'}^{+})^{-1}  (\tilde{g}_{ve}^{+}) (\mathfrak{a}^{+}_{vef})  \right)  (\mathfrak{a}^{+}_{v_0e_0f})^{-1} \label{eqGfall2}
\end{align}
for boundary faces.
For internal faces the definition is the same with identifying $e_1,e_0$ as the same edge.
The above equations \eqref{theta_value} still valid for faces containing internal edges by replacing $(\tilde{g}_{ve'}^{+})^{-1} \tilde{g}_{ve}^{+}$ and $\tilde{g}_{ve'}^{-}(\tilde{g}_{ve}^{-})^{-1}$ by $G_f^{\pm}$correspondingly, and now $\omega_f := \frac{|\det \eta_{e_1} - \det \eta_{e_0}|}{2} \in \{0,1\}$ only contains contribution from boundary edges, thus $\omega = 0$ for internal faces.

\subsection{Special cases: 4D simplicial geometry}
As we derived before, when the geometry forms $4$-simplex, we have the cross simplicity being satisfied. We will restrict our study to the case where we have $\alpha = \alpha' = 0$ for space action and $\alpha = -\alpha'$ for time action and independent of vertex $v$, namely $B^{\pm}$ correspond to the same boundary geometry. The general equations (\ref{eqGfall1}-\ref{eqGfall2}) in previous section then becomes

\begin{align}
    G^{-}_f = \prod_{v \in \partial_f} \tilde{g}_{ve'}^{-}  (\tilde{g}_{ve}^{-})^{-1}, \qquad G^{+}_f = \prod_{v \in \partial_f} (\tilde{g}_{ve'}^{+})^{-1}  \tilde{g}_{ve}^{+}
\end{align}
with

\begin{align}\label{G_theta_internal}
	G_f(e)^{-} = \ue^{2 \sum_{v} \theta_{vf} B_{ef}^{-}  } \, , \qquad G_f(e)^{+} = \ue^{- 2 \sum_{v} \theta'_{vf} B_{ef}^{+}  } =  (R_{e})^{-1}\ue^{- 2 t_f \sum_{v} \theta'_{vf} B_{ef}^{-}  } R_{e}
\end{align} 
for internal faces.
Then $\theta$ and $\theta'$ can be expressed as
\begin{align}
	\sum_{v} \theta_{vf}  = \log \left[ \Tr( G_f(e)^{-} X_{ef}) \right]\, , \qquad \sum_{v} \theta'_{vf}  = -  t_f \log \left[\Tr( (R_{e})^{-1}  G_f(e)^{+} R_{e} X_{ef}) \right]
\end{align} 
where $R_{e} G_f(e)^{+} R_{e}$ is given by 
\begin{align}
    (R_{e})^{-1} G_f(e)^{+} R_{e}=: G_f(e)^{R+} = \prod_{v \in \partial_f} (\tilde{g}_{ve'}^{+} R_{e'})^{-1}  \tilde{g}_{ve}^{+} R_e \,.
\end{align}

For boundary faces, we have
\begin{align}\label{G_theta_boundary}
&\mathfrak{v}_{ee'} G^{-}_f(e',e)  = \ue^{2 \sum_{v} \theta_{vf} B_{ef}^{-} + \ui \pi \omega_f  B_{ef}^{-}  }, \qquad  \mathfrak{v}_{ee'} R_{e'}^{-1} G^{+}_f(e',e)  R_{e}
	=  \ue^{-2 t_f \sum_{v} \theta'_{vf} B^{-}_{ef} } \,,  \nonumber
\end{align} 
where we use the fact that $\sigma_3 \ue^{\epsilon_i \sigma^i} \sigma_3 = \ue^{\epsilon_i \sigma_3 \sigma^i \sigma_3 }$ and here we can again introduce
\begin{align}
    (R_{e'})^{-1} G_f(e',e)^{+} R_{e}=: G_f(e',e)^{R+} = \prod_{v \in \partial_f} (\tilde{g}_{ve'}^{+} R_{e'})^{-1}  \tilde{g}_{ve}^{+} R_e \,.
\end{align}
Note that for time action $\alpha = -\alpha'$ may appear on the boundary edges as shown in (\ref{av_b}). Here we make a redefination of $\tilde{v}_{ef}$ and $\tilde{v}'_{ef}$ to absorb $\tilde{\mathfrak{a}}_{ef}$ and $\tilde{\mathfrak{a}}_{e'f}$ appear on the boundary edge.

\subsection{Geometrical interpretations}

By the reconstruction theorem, when the critical geometry corresponds to simplicial geometry, there are two solutions available at each vertex which defers by a Plebanski orientation. Suppose at each vertex the solution are given by $G$ and $G'$ correspondingly and satisfy (\ref{solution_relation}), one can show that the loop holonomy $G_f$ along a face which are product of these two solutions are related by
\begin{align}
    G'_f(e) &= \prod_{v \in \partial f} (G'_{ve'})^{-1} G'_{ve} = \prod_{v \in \partial f} R_{u_{e'}} (G_{ve'})^{-1} R_{e_{\alpha}}R_{e_{\alpha}} G_{ve} R_{u_e} = R_{u_{e}} G_f(e) R_{u_{e}} \,, \\
    G'_f(e',e) &= \prod_{v \in \partial f} (G'_{ve''})^{-1} G'_{ve} =I^{\frac{1-\frac{\sgn(u_{e'})}{\sgn(u_{e})}}{2}} \prod_{v \in \partial f} R_{u_{e''}} (G_{ve''})^{-1} R_{e_{\alpha}}R_{e_{\alpha}} G_{ve} R_{u_e} \\
    &= I^{\frac{1-\frac{\sgn(u_{e'})}{\sgn(u_{e})}}{2}} R_{u_{e'}} G_f(e',e) R_{u_{e}} \,.
\end{align}
This analysis holds for Lorentzian, Riemannian or split signature simplicial geometries.
The equation then implies
\begin{align}
   {G}_{ve} (G'_f)^{-1} G_f {G}_{ve}^{-1} &= {G}_{ve}  R_{u_{e}} ({G}_{ve})^{-1} (G_f(e)({G}_{ve})^{-1})^{-1} R_{u_{e}} G_f {G}_{ve}^{-1} \\
   &=R_{N_{e}(v)} R_{N_{e}^p(v)} = \ue^{2 \Theta_{f} \frac{N^{p}_{e} \wedge N_{e}}{|N^{p}_{e} \wedge N_{e}|}} %
\end{align}
for internal face holonomies with $N^P(v) :=   G_{ve} (G_f)^{-1} \cdot u_{e} =  G_{ve} (G_f)^{-1} (G_{ve})^{-1} \cdot N_{e}(v) $ which is the parallel transported vector seen in the reference frame specified by $G_{ve}$. $\Theta_f$ is then the dihedral angle between $N^P(v)$ and $N_{e}(v)$, which is given by $\Theta_f :=\cos^{-1} ( \sgn(|N_{e}(v)|) N^P(v) \cdot N_{e}(v) )$ when the plane span by $N^P(v)$ and $N_{e}(v)$ have signature $(--)$ or $(++)$, and $  \Theta_f := \sgn(N_{e}(v) \cdot N_{e}(v) ) \cosh^{-1}(|N^P(v) \cdot N_{e}(v)|)$ when the plane span by $N^P(v)$ and $N_{e}(v)$ have signature $(+-)$.
For boundary faces, similarly we have
\begin{align}
   {G}_{ve} (G'_f)^{-1}(e',e) G_f(e',e) {G}_{ve}^{-1} 
   =I^{\frac{1-\frac{\sgn(u_{e'})}{\sgn(u_{e})}}{2}}R_{N_{e}(v)} R_{N_{e'}^p(v)} = \mathcal{O}^{\frac{1-\frac{\sgn(u_{e'})}{\sgn(u_{e})}}{2}} \ue^{2 \frac{\sgn(u_{e'})}{ \sgn(u_{e})} \Theta_{f} \frac{N^{p}_{e'} \wedge N_{e}}{|N^{p}_{e'} \wedge N_{e}|}}
\end{align}
with now $N^P_{e'}(v) := (G_f G_{ve}^{-1})^{-1} \cdot u_{e'} =  G_{ve} (G_f)^{-1} \cdot u_{e'} $ and $\mathcal{O} = \ue^{\pi *\frac{N^{p}_{e'} \wedge N_{e}}{|N^{p}_{e'} \wedge N_{e}|}}$. $\Theta_f$ is now the dihedral angle between $N^P(e')$ and $N_{e}(e)$. The definition of $\Theta_f$ is the same as internal faces with special cases when $\sgn(N^P(e')) \neq \sgn(N^P(e))$, in which case it is defined as $\Theta_f :=  \sinh^{-1}(N^P(v) \cdot N_{e}(v))$. Note that, for both internal and boundary faces, similar arguments hold for $N'^P(v) :=   G_{ve} (G'_f)^{-1} \cdot u_{e} $ by rewriting above equations based on $G'$, for example,
\begin{align}
  \ue^{2 \Theta_{f} \frac{N^{p}_{e} \wedge N_{e}}{|N^{p}_{e} \wedge N_{e}|}}={G}_{ve} (G'_f)^{-1} G_f {G}_{ve}^{-1} &= {G}_{ve}  (G'_f(e))^{-1} R_{u_{e}} G'_f {G}_{ve}^{-1}{G}_{ve} R_{u_{e}} {G}_{ve}^{-1} \\
   &= R_{N'{}^p(v)_{e}} R_{N_{e}(v)} = \ue^{-2 \Theta'_{f} \frac{N'{}^{p}_{e} \wedge N_{e}}{|N'{}^{p}_{e} \wedge N_{e}|}} %
\end{align}
and we have $\cos(\Theta'_f) :=N'^P(v) \cdot N_e(v) $ where $N'^P(v)$ are in the same plane span by $N^P(v)$ and $N_e$. As a result, $\Theta'_f = -\Theta_f$ which reflects the fact that $G$ and $G'$ differs by the Plebanski orientation.

Notice that, by reconstruction theorem, when the reconstructed geometry admits a consistent orientation and the signature of the $4$-volume $\sgn(V(v))$ of each reconstructed simplex at vertex $v$ along a face is a constant, we have the following equations hold for both $G$ and $G'$ for any co-frame vecotr $E_l$ in the dual triangle othogonal to $N_e$ and $N^P(v)$ or $N^P_{e'}(v)$:
\begin{align}\label{action_on_frame}
    G_{ve}  G^{f}(v) G_{ve}^{-1} E_{l}(v) = (-1)^{\mu_f} E_{l}(v) \,, \qquad  \mu_f  = \sum_{e} \mu_{e} \in \mathbb{R}_{+} \, , \mu_{e} \in \{0,1 \} \,.
\end{align}
For boundary faces, from $ G^{f}(e',e) E_{l}(e) = \mu E_{l}(e')$, we have
\begin{align}
  R_{e'} (-\ui \sigma_2)^{\frac{1-m_{e'f}}{2}} (\tilde{v}_{e'f})^{-1} G^{f}(e',e) \tilde{v}_{ef} (\ui \sigma_2)^{\frac{1-m_{ef}}{2}} R_{e} E^0_{l}(e) = \mu E^0_{l}(e') =  \mu \ue^{\Phi_f^B (*)^{\frac{1+t_f^{\Delta}}{2  }} B_0} E^0_{l}(e)
\end{align}
with $B_0 = \sigma_3$ or $B_0 =\sigma_1$  for space action and time action respectively. We use the fact that both $E^0_{l}(e):=R_e^{-1}(-\ui \sigma_2)^{\frac{1-m_{ef}}{2}} (\tilde{v}_{ef})^{-1} E_l(e) $ and $E^0_{l}(e'):=R_{e'}^{-1}(-\ui \sigma_2)^{\frac{1-m_{ef}}{2}}  (\tilde{v}_{e'f})^{-1} E_l(e') $ are in the plane orthogonal to $B_0$ or $*B_0$. As a result, we have
\begin{align}
   \mathfrak{v}_{ee'} G^{f}(e',e) E_{l}(v) = v_{ef} R_{e} (\ui \sigma_2)^{\frac{1-m_{ef}}{2}}   (-\ui \sigma_2)^{\frac{1-m_{e'f}}{2}} R_{e'} (v_{e'f})^{-1} G^{f}(e',e)  E_{l}(v) = \mu \ue^{\Phi_f^B (*)^{\frac{1+t_f^{\Delta}}{2  }} \frac{B_{ef}}{|B_{ef}|}} E_{l}(e)
\end{align}
which implies
\begin{align}
    G_{ve} \mathfrak{v}_{ee'}  G^{f}(e',e) G_{v e}^{-1} E_{l}(v) = \mu  R_{ee'} E_{l}(v) \,, \qquad R_{ee'} =\ue^{\Phi^B_f (*)^{\frac{1+t_f^{\Delta}}{2  }} \frac{B_f(v)}{|B_f(v)| }} 
\end{align}
for $\Phi_f^B$ some real parameters totally determined by the boundary data. Moreover, one notes that, when the triangles span by $E_l$ are timelike, we have $\mu=0$.

Since $N^P(v) \cdot E_{l} = N^P(v) \cdot E_{l'} = 0$, we have 
\begin{align}
    \frac{N^{p}_{e} \wedge N_{e}}{|N^{p}_{e} \wedge N_{e}|} = r (*)^{\frac{1-t_f^{\Delta}}{2}} \frac{B_f(v)}{|B_f(v)|}
\end{align}
with $r= \pm 1$ is the Plebanski orientation of the reconstructed simplicial geometry, and $r$ related to $\sgn(V(v))$ when the simplicial complex admits a consistent orientation as described before.
For the cases the $\sgn(V(v))$ is not a constant on the reconstructed simplicies, we can perform sub-divisions of the simplicial complex, such that in each sub-complex $\sgn(V(v))$ thus $r$ is a constant. 

As a result, suppose we pick $r= \pm 1$ for solution $G$ and $G'$ respectively, for internal faces, 
\begin{align}\label{eq:geo_connection_int}
     G_{ve} G_f G_{ve}^{-1}= \ue^{ \Theta_{f} (*)^{\frac{1-t_f^{\Delta}}{2}}  \frac{B_f(v)}{|B_f(v)|} + \mu_f  \pi (*)^{\frac{1+t_f^{\Delta}}{2}}  \frac{B_f(v)}{|B_f(v)|}} \, , \\
     G_{ve}  G'_f G_{ve}^{-1}= \ue^{-  \Theta_{f} (*)^{\frac{1-t_f^{\Delta}}{2}} \frac{B_f(v)}{|B_f(v)|} + \mu_f \pi (*)^{\frac{1+t_f^{\Delta}}{2}}  \frac{B_f(v)}{|B_f(v)|}} \,,
\end{align}
which is a rotation (boost) in the plane span by $ (*)^{\frac{1-t_f^{\Delta}}{2}}B_f(v)$ with angle $\Theta_f$. 
For boundary faces, the equation (\ref{action_on_frame}) then determined $G_f$ and $G'_f$ as
\begin{align}
    G_{ve} \tilde{v}_{ee'} G_f G_{ve}^{-1}= \ue^{ \Theta_{f}  (*)^{\frac{1-t_f^{\Delta}}{2}} \frac{B_f(v)}{|B_f(v)|} +  ( \Phi^B_{f} + \mu_f \pi)  (*)^{\frac{1+t_f^{\Delta}}{2}} \frac{B_f(v)}{|B_f(v)|}}\, , \label{eq:geo_connection_bou_1}\\
    G_{ve} \tilde{v}_{ee'} G'_f G_{ve}^{-1}= \ue^{-  \Theta_{f}  (*)^{\frac{1-t_f^{\Delta}}{2}}\frac{B_f(v)}{|B_f(v)|} +  (\Phi^B_{f} + \mu_f \pi - \omega_f^{\Delta} {\pi})   (*)^{\frac{1+t_f^{\Delta}}{2}} \frac{B_f(v)}{|B_f(v)|}}\,, \label{eq:geo_connection_bou_2}
\end{align}
where $\omega^{\Delta}_f=1$ when $\sgn(u_{e'}) \neq \sgn(u_{e})$ for boundary faces, otherwise $\omega^{\Delta}_f=0$.

On the other hand, from (\ref{G_theta_internal}) and (\ref{G_theta_boundary}), we have
\begin{align}
    G_{ve} G_f(e) (G_{ve})^{-1} = \ue^{2 \sum_{v} \theta_{vf} \frac{B_f(v)}{|B_f(v)|}}, \quad G_{ve} v_{ee'}  G_f(e',e) (G_{ve})^{-1} = \ue^{2 \sum_{v} \theta_{vf} \frac{B_f(v)}{|B_f(v)|} - \omega_f \pi * \frac{B_f(v)}{|B_f(v)|}} \,.
\end{align}
 Combine the result we then can determine the value of $\theta$ by relating the solutions of $G^{\pm}$ and geometrical solutions $G$ and $G'$. For example, when $G^{-1} = G$, we have
\begin{align}
    \sum_{v} \Re(\theta_{vf}) = r \frac{\Theta_f}{2} \,, \qquad \sum_{v} \Im(\theta_{vf}) = \frac{\Phi^B_{f}+\sum_{e} \mu_e \pi + \omega_f \frac{\pi}{2}}{2} \,.
\end{align}
Note that here we define both $\Theta_f$ and $\Phi^B_{f}$ to take their principle values, s.t., $\cos^{-1}(x) \in [0,2\pi)$, $\cosh^{-1}(x) \in [0,\infty)$.
The detailed correspondence for simplicial geometries will be built explicitly later. 

For special cases when the critical group elements are in the stabilizer group of normal $u_f$ up to gauge transformations, namely we have vector geometry as the critical geometry, the equations (\ref{G_theta_internal}) and (\ref{G_theta_boundary}) simplifies to 
\begin{align}\label{theta-vector}
    &G_{ve} G_f(e) (G_{ve})^{-1} =\ue^{2 \sum_{v} (-\ui)^{\frac{1+t_f^{\Delta}}{2}} \theta_{vf}  (*)^{\frac{1+t_f^{\Delta}}{2}} \frac{B_{f}(v)}{|B_f(v)|}}%
    = \ue^{2 \sum_{v} (-\ui)^{\frac{1+t_f^{\Delta}}{2}} \theta_{vf} \frac{ \cv_{f} \cdot \vec{\tau} }{|\cv_{f}|} } , \\
    &G_{ve} v_{ee'}  G_f(e',e) (G_{ve})^{-1} =\ue^{ (-\ui)^{\frac{1+t_f^{\Delta}}{2}} (2 \sum_{v}\theta_{vf} + \ui \omega_f \pi ) (*)^{\frac{1+t_f^{\Delta}}{2}} \frac{B_{f}(v)}{|B_f(v)|}}%
    = \ue^{ (-\ui)^{\frac{1+t_f^{\Delta}}{2}} (2 \sum_{v} \theta_{vf} + \ui \omega_f \pi ) \frac{ \cv_{f} \cdot \vec{\tau} }{|\cv_{f}|} } \nonumber
\end{align}
for internal and boundary faces respectively,
where $G'$ and ${G}$ are in the stabilize group of $u_f$, $\vec{\tau} $ are generators of the stabilize group and $(-\ui)^{\frac{1+t_f^{\Delta}}{2}} (2 \sum_{v}\theta_{vf} + \ui \omega_f \pi ) \in \mathbb{R}$ is a real parameter which will be determined later. $t_f^{\Delta}$ is the corresponding signature for the plane orthogonal to both $u_{f}, \cv_{f}$ and we use the fact that $B_f$ is always a timelike plane in our notation. 

Notice that, when there are two gauge in-equivalent vector-geometry solutions available, the solution actually corresponds to $4$-simplicies with Riemannian or split signature. Suppose the two solution are given by $\{G\}$ and $\{G'\}$ and correspond to normal vectors $\cv^{\pm}$ respectively, by using the mapping 
\begin{align}
    \cb_f(v) = \Phi^{-1}(\cv^{+}_f(v),\cv^{-}_f(v)) \,,
\end{align}
and the induced map on group elements
\begin{align}
    \cg = \Phi^{-1}(G_f(v),G'_f(v)), \qquad \cg' = \Phi^{-1}(G'_f(v),G_f(v))
\end{align}
with $\cg \in SO(4)$ for $u = (1,0,0,0)$ and $\cg \in SO(2,2)$ for $u = (0,0,0,1)$.
Following the same analysis as in Lorentzian case, we then have
\begin{align}\label{eq:geo_connection_int_vec}
    \cg_{ve} \cg_f \cg_{ve}^{-1}= \ue^{ \Theta_{f} \frac{\cb_f(v)}{|\cb_f(v)|} + \mu_f \pi * \frac{\cb_f(v)}{|\cb_f(v)|}} \, , \cg_{ve}  \cg'_f \cg_{ve}^{-1}= \ue^{-  \Theta_{f} \frac{\cb_f(v)}{|\cb_f(v)|} + \mu_f \pi * \frac{\cb_f(v)}{|\cb_f(v)|}}
\end{align}
for internal faces and
\begin{align}
    \cg_{ve} \tilde{v}_{ee'} \cg_f \cg_{ve}^{-1}= \ue^{ \Theta_{f} \frac{\cb_f(v)}{|\cb_f(v)|} +  ( \Phi^B_{f} + \mu_f \pi) * \frac{\cb_f(v)}{|\cb_f(v)|}}\, , \label{eq:geo_connection_bou_1_vec}\\
    \cg_{ve} \tilde{v}_{ee'} \cg'_f \cg_{ve}^{-1}= \ue^{-  \Theta_{f} \frac{\cb_f(v)}{|\cb_f(v)|} +  (\Phi^B_{f} + \mu_f \pi)  * \frac{\cb_f(v)}{|B\cb_f(v)|}}\,, \label{eq:geo_connection_bou_2_vec}
\end{align}
for boundary faces. Notice that since 
\begin{align}
     \Phi^{\pm}(\cb_{f}(v)) = \pm \cv^{\pm}_f(v) \,, \qquad \Phi^{\pm}(* \cb_{f}(v)) = \cv^{\pm}_f(v) \,,
\end{align}
above equation recovers (\ref{theta-vector}) for vector geometry solutions $\{G\}$ and $\{G'\}$ with $2 \sum_{v} \theta_{vf} =\ui^{\frac{1+t_f^{\Delta}}{2}} ( \pm \Theta_f + \Phi_f^B + \mu_f \pi ) - \ui \omega_f \pi$, where $\mu =0$ when $t_f^{\Delta} = -1$, since in such case both the plane orthogonal to normals are timelike in the split signature space. $\Theta_f$ is now also given as the angle between $N_{e}^p(v)$ and $N_{e}(v)$ where
\begin{align}
    N_{e}(v) = \cg_{ve} u_{f}\,, \qquad N_{e}^p(v) =\cg_{ve} \cg^{\pm}_f u_{f} \,,
\end{align}
which is the deficit angle along face $f$ in Riemannian or split signature space.

\subsection{Summary and Special Cases}
Now we can relate the above result to different cases to identify the value of $\theta$ and $S$ according to the corresponding critical simplicial geometry. According to previous analysis, when the critical geometry corresponds to non-degenerate simplicial geometry, we have two solutions at each vertex for each set of the equation of motion given by $B^{+}$ or $B^{-}$. As a result, we have four sets of geometrical solutions with two of them correspond to $\tilde{G}^{+}$ and the other two correspond to $\tilde{G}^{+}$ at each vertex. The solution for $\tilde{G}^{\pm}$ may correspond to different geometries.

In the special case when the boundary given by $B^{\pm}$ at each vertex $v$ are the same and does not change at given internal edge $e$ for neighboring vertices $v$ and $v'$ (with $\alpha = \alpha' = 0$ for space action and $\alpha + \alpha' =0$ for time action), the pairs of 4-simplex geometries differ only up to reflection and geometrical gauge transformations since they share the same boundary geometry at each vertex. We then only have two possible sets of geometric solutions correspond to this boundary geometry, denoted as $G,G'$, where the honolomy $G_f$ and $G'_f$ are related to the spin connection compatible with the co-frame specified by the bivector $B_f$ when $\sgn(V)$ is a constant along the face $f$. $\theta$ is then related to the deficit angle between different frame. The solution for $\tilde{G}^{\pm}$ now correspond to the same geometries up to orientation and gauge transformations.

As a result, from the reconstruction, we have $4$ possibilities for solutions $\tilde{G}=(\tilde{G}^{+}, \tilde{G}^{-})$ at each vertex: $\tilde{G}=(\tilde{G}^{+}, \tilde{G}^{-})$: $\tilde{G}=(G R_e, (G)^{-1})$, $\tilde{G}=(G R_e, (G')^{-1})$ and $\tilde{G}=(G' R_e, (G)^{-1})$, $\tilde{G}=(G' R_e, (G')^{-1})$. In the following analysis we assume  $\sgn(V)$ is a constant on the reconstructed simplicial complex. When it is not a constant, we can always make a sub-division of the complex such that in each sub-complex it is a constant. Note that, for the boundary faces whose boundary are boundary of the initial complex, we have $\omega^{\Delta}_f = \omega$. However, for the boundary of sub-divided complexes which contains internal variables of the model, $\omega^{\Delta}_f \neq \omega_f$ is possible.
\subsubsection{4D Riemannian and split signature simplicial geometry}\label{riemann_split_action}
    The solutions correspond to a Riemannian or split signature $4$-simplex at vertex $v$, and in this case, $G,G' \in SO(3)$ for Riemannian and $G,G' \in SO(1,2)$ for split signature. The solution $G, G'$ then subject to (\ref{eq:geo_connection_int}-\ref{eq:geo_connection_bou_2}).
    When $t^{\Delta} = 1$, the corresponding triangles associated to face $f$ in the 4 simplices is spacelike with $\Theta_f \in \pm [0, \pi ) \mod 2 \pi$ is a rotation angle associated to triangle $B_f$ and $\Phi_f^B \in (0, 2\pi )$ corresponds to a phase related to boundary data, while $t^{\Delta} =- 1$ the triangle is timelike with $\Theta_f \in [0, \infty)$, $\Phi_f^B \in [0, \infty )$ and $\mu = 0$. For pure boundary faces of the complex, we must have $\omega^{\Delta}_f = \omega_f = 0$ to have degenerate solutions. For the boundary of sub-divided complexes which are internal variables of the model, only $\omega^{\Delta}_f =0$ is needed. Note that $\omega_f = 0$ when $t_f = -1$.
    Compare with \eqref{theta-vector} we have the following result:
    \begin{itemize}
        \item $\tilde{G}=(G' R_{e}, G^{-1})$
        \begin{align}
             &\sum_v \theta_{vf} = \left(\frac{\ui^{\frac{1+t_f^{\Delta}}{2}}}{2} ( \Theta_f + \Phi^B_f + \mu_f  \pi)  - \frac{1+t_f}{2} \frac{\ui \omega_f \pi}{2} \right)\mod \ui \pi,\\
             &\sum_v \theta'_{vf}=  \frac{\ui^{\frac{1+t_f^{\Delta}}{2}}}{2} t_f ( \Theta_f - \Phi^B_f -\mu_f  \pi)) \mod \pi \ui, \\
             &\tilde{F}_f[\tilde{X}_0]= (-\ui)^{\frac{1-t_f}{2}} \left((-\ui)^{\frac{1-t_f^{\Delta}}{2}} (- \gamma \Theta_f  - \ui ( \Phi^B_f + \mu_f \pi)) \mod (\gamma \pi, \ui \pi) \right) \label{wick_regge}\\
             &\qquad \qquad \qquad+(\ui + \gamma) \frac{1+t_f}{2}\frac{{\omega}_f \pi \mod 2 \pi}{2} \,; \notag
        \end{align}       
        \item $\tilde{G}=(G R_{e}, (G')^{-1})$
        \begin{align}
             &\sum_v \theta_{vf} = \left(\frac{\ui^{\frac{1+t_f^{\Delta}}{2}}}{2} ( - \Theta_f + \Phi^B_f + \mu_f \pi) -\frac{1+t_f}{2} \frac{\ui {\omega}_f\pi}{2} \right)\mod \ui \pi,\\
             &\sum_v \theta'_{vf}=  \frac{\ui^{\frac{1+t_f^{\Delta}}{2}}}{2} t_f ( - \Theta_f - \Phi^B_f -\mu_f \pi) \mod \pi \ui, \\
             &\tilde{F}_f[\tilde{X}_0]= (-\ui)^{\frac{1-t_f}{2}} \left( (-\ui)^{\frac{1-t_f^{\Delta}}{2}} (\gamma \Theta_f  - \ui ( \Phi^B_f + \mu_f \pi)) \mod (\gamma \pi, \ui \pi) \right)\\
             &\qquad \qquad \qquad+(\ui + \gamma) \frac{1+t_f}{2}\frac{{\omega}_f \pi \mod 2\pi}{2} \,;\notag
        \end{align}     
        \item $\tilde{G}=(G R_{e}, (G)^{-1})$
        \begin{align}
             &\sum \theta_{vf} = \left( \frac{\ui^{\frac{1+t_f^{\Delta}}{2}}}{2} ( \Theta_f + \Phi^B_f + \mu_f  \pi) - \frac{1+t_f}{2} \frac{\ui {\omega}_f\pi}{2} \right)\mod \ui \pi, \\
             &\sum \theta'_{vf}=  \frac{\ui^{\frac{1+t_f^{\Delta}}{2}}}{2} t_f ( - \Theta_f - \Phi^B_f- \mu_f \pi) \mod \pi \ui, \\
             &\tilde{F}_f[\tilde{X}_0]= (- \ui)^{\frac{1 - t_f}{2}} \left((- \ui)^{\frac{1-t_f^{\Delta}}{2}} \ui (- \Theta_f - \Phi^B_f  - \mu_f \pi)  \mod (\gamma \pi, \ui \pi) \right) \\
             & \qquad \qquad \qquad+(\ui + \gamma) \frac{1+t_f}{2}\frac{{\omega}_f \pi \mod 2\pi}{2} \,; \notag
        \end{align}  
        \item $\tilde{G}=(G' R_{e}, (G')^{-1})$
        \begin{align}
              &\sum \theta_{vf} = \left(\frac{\ui^{\frac{1+t_f^{\Delta}}{2}}}{2} ( - \Theta_f + \Phi^B_f+ \mu_f  \pi) - \frac{1+t_f}{2} \frac{\ui {\omega}_f\pi}{2}  \right)\mod \ui \pi, \\
              &\sum \theta'_{vf}=  \frac{\ui^{\frac{1+t_f^{\Delta}}{2}}}{2} t_f ( \Theta_f - \Phi^B_f- \mu_f  \pi) \mod \pi \ui, \\
             &\tilde{F}_f[\tilde{X}_0]= (- \ui)^{\frac{1-t_f}{2}}\left( (-\ui)^{\frac{1-t_f^{\Delta}}{2}} \ui ( \Theta_f - \Phi^B_f - \mu_f \pi) \mod (\gamma \pi, \ui \pi) \right)\\
             &\qquad \qquad \qquad+(\ui + \gamma) \frac{1+t_f}{2}\frac{{\omega}_f \pi \mod 2\pi}{2} \,. \notag
        \end{align}   
    \end{itemize}
    Here $f \mod (\gamma \pi, \ui \pi) : = (f \mod \ui \pi) \mod \gamma \pi$.
    Note that the $\gamma \pi$ and $\ui \pi$ ambiguity coming from the fact that the analytic continued action is defined on the cover space due to the analytic continuation of the logarithm. As a result, there are infinitely many critical points on the cover space corresponding to the same geometrical interpretation. 
    
    The original integration path is contained in the case $3$ and $4$ with $t_f = t_f^{\Delta}$ and ${\omega}_f =0 $ with extra requirement that $R_{e} = \mathbb{I}_2$ for Riemannian which implies $g^{+} = (g^{-})^{-1}$ and $R_{e} = \ui \mathbb{\sigma}_3$ for split signature which implies $g^{+} = (g^{-})^{\dagger}$. One can identify them with the degenerate solution of EPRL-CH model shown in \cite{Barrett:2009mw,Han:2011re,Kaminski:2017eew,Liu:2018gfc}. In such case $\theta' = \theta^{\dagger}$ thus the $\tilde{F}_f[\tilde{X}_0]$ is determined up to $2 \pi$, which removes the domain of covering space from analytic continuation. Note that since $j$ can be half integers the total action is determined only up to $\pi$. Some of the $\pi$ ambiguity can be removed by fixing the lift ambiguity according to \cite{Barrett:2009mw,Kaminski:2017eew,Liu:2018gfc}.
    
    By applying the above result to each vertex and summing over the result, due to the cancellation of internal $\Phi_f$ at each vertex, one immediately notice that we have
    \begin{align}
        \Theta_f = \sum_v \Theta_{vf} =\left\{ \begin{array}{ll} \sum_v \theta_{vf}, \quad & t_f^{\Delta} = -1 \\ \sum_v (\pi-\theta_{vf}) , \quad & t_f^{\Delta}  = 1 \end{array} \right. \,,
    \end{align}
    thus $\Theta_f$ is related to the deficit angle $ \epsilon_f $ or boundary deficit angle $\theta_f$ by $ \Theta_f =   \epsilon_f \mod \frac{(1+t_f^{\Delta})}{2} \pi$ or $ \Theta_f =  \theta_f \mod  \frac{(1+t_f^{\Delta})}{2} \pi$ . As a result, we can replace $\Theta_f$ in above equations to $ \epsilon_f $ for internal faces or $ \theta_f $ for boundary. 
    
     Notice that when $t_f^{\Delta} =1$, namely the geometry are Riemannian 4-simplex, the contributions of (\ref{wick_regge}) to the spinfoam ampltiude are proportional to $e^{-S_{\rm Regge}}$ with the Regge action 
\begin{align}
    S_{\text{Regge}} = \pm \sum_f A_f \Theta_f \,,
\end{align}
where $A_f :=\gamma J_f$ for both space %
and time action is the area for triangle associated to $f$. We have analytically continued the spin $J_f\to i J_f$ for the time action to cancel the extra $\ui$ appearing in (\ref{wick_regge}). As we indicated in Section \ref{summary_eom}, in the case when both time and space action appears at a given edge, this analytical continuation of the spin is required by the closure condition given in (\ref{eq:biveq1}).
    
\subsubsection{4D Lorentzian simplicial geometry}\label{lorentzian_action}

The solutions correspond to Lorentzian $4$ simplices at vertices $v$. In such case, $G, G' \in SO(1,3)$. The corresponding spin connection are again given by (\ref{eq:geo_connection_int}-\ref{eq:geo_connection_bou_2})
    with $\Theta_f \in [0,\pi) \, \mod \, 2 \pi$ for timelike triangles of corresponding face $f$ in the 4-simplices with $t^{\Delta}_f = -1$ and $\Theta_f \in \pm [0, \infty)$ for spacelike triagles with $t^{\Delta}_f = 1$. $\Phi_f^B$ is again the angle determined by the boundary. $\mu_f = 0$ when $t^{\Delta}_f =- 1$.
    
    In such case, compare with \eqref{G_theta_boundary}-\eqref{G_theta_internal} we have the following result
    \begin{itemize}
        \item $\tilde{G}=(G' R_{e}, G^{-1})$
        \begin{align}
             &\sum \theta_{vf} = \frac{\ui^{\frac{1-t^{\Delta}_f}{2}} }{2} ( \Theta_f +\ui (\Phi^B_f + \mu_f \pi + {\omega}_f^{\Delta} \pi )) -\frac{1+t_f}{2} \frac{\ui {\omega}_f\pi}{2} \mod \pi \ui,\\
             &\sum \theta'_{vf}=  \frac{1}{2} \ui^{\frac{1-t^{\Delta}_f}{2}} t_f ( \Theta_f - \ui (\Phi^B_f + \mu_f \pi )) \mod \pi \ui, \\
             &\tilde{F}_f[\tilde{X}_0]= (-\ui)^{\frac{1-t_f}{2}}\left( \ui^{\frac{1-t^{\Delta}_f}{2}} \left( \ui \gamma \Theta_f    - \ui (\Phi^B_f + \mu_f \pi ) - ( \ui + \gamma) \frac{{\omega}_f^{\Delta} \pi}{2}\right)  \mod (\ui \pi, \gamma \pi) \right)  \nonumber\\
             & \qquad \qquad \qquad+(\ui + \gamma) \frac{1+t_f}{2}\frac{{\omega}_f \pi \mod 2\pi}{2} \,;
        \end{align}       
        \item $\tilde{G}=(G R_{e}, (G')^{-1})$
        \begin{align}
             &\sum \theta_{vf} = \frac{1}{2} \ui^{\frac{1-t^{\Delta}_f}{2}} ( -\Theta_f +\ui (\Phi^B_f + \mu_f \pi )) -\frac{1+t_f}{2} \frac{\ui {\omega}_f\pi}{2}\mod \pi \ui,\\
             &\sum \theta'_{vf}=  \frac{1}{2} \ui^{\frac{1-t^{\Delta}_f}{2}} t_f ( -\Theta_f - \ui (\Phi^B_f + \mu_f \pi -\omega^{\Delta}_f \pi )) \mod \pi \ui, \\
             &\tilde{F}_f[\tilde{X}_0]= (-\ui)^{\frac{1-t_f}{2}}  \left( \ui^{\frac{1-t^{\Delta}_f}{2}} \left(- \ui \gamma \Theta_f - \ui  (\Phi^B_f + \mu_f \pi ) + ( \ui - \gamma) \frac{{\omega}_f^{\Delta} \pi}{2} \right) \mod (\ui \pi, \gamma \pi) \right)  \nonumber\\
             & \qquad\qquad\qquad+(\ui + \gamma) \frac{1+t_f}{2}\frac{{\omega}_f \pi \mod 2 \pi}{2} \,;
        \end{align} 
        \item $\tilde{G}=(G R_{e}, (G)^{-1})$
        \begin{align}
             &\sum \theta_{vf} = \frac{1}{2} \ui^{\frac{1-t^{\Delta}_f}{2}} ( - \Theta_f +\ui (\Phi^B_f + \mu_f \pi ))-\frac{1+t_f}{2} \frac{\ui {\omega}_f\pi}{2} \mod \pi \ui,\\
             &\sum \theta'_{vf}=  \frac{1}{2} \ui^{\frac{1-t^{\Delta}_f}{2}} t_f ( \Theta_f - \ui (\Phi^B_f + \mu_f \pi )) \mod \pi \ui, \\
             &\tilde{F}_f[\tilde{X}_0]= (- \ui)^{\frac{1-t_f}{2}} \left( \ui^{\frac{1-t^{\Delta}_f}{2}} ( \Theta_f  - \ui  (\Phi^B_f + \mu_f \pi ) )  \mod (\ui \pi, \gamma \pi)  \right) \nonumber \\
             & \qquad\qquad\qquad +(\ui + \gamma) \frac{1+t_f}{2}\frac{{\omega}_f \pi \mod 2 \pi}{2} \,;
        \end{align} 
        \item $\tilde{G}=(G' R_{e}, (G')^{-1})$
        \begin{align}
             &\sum \theta_{vf} = \frac{1}{2} \ui^{\frac{1-t^{\Delta}_f}{2}} ( \Theta_f +\ui (\Phi^B_f + \mu_f \pi  + \omega^{\Delta}_f \pi )) -\frac{1+t_f}{2} \frac{\ui {\omega}_f\pi}{2}\mod \pi \ui,\\
             &\sum \theta'_{vf}=  \frac{1}{2} \ui^{\frac{1-t^{\Delta}_f}{2}} t_f ( -\Theta_f - \ui (\Phi^B_f + \mu_f \pi  + \omega^{\Delta}_f \pi )) \mod \pi \ui, \\
             &\tilde{F}_f[\tilde{X}_0]= (- \ui)^{\frac{1-t_f}{2}} \left( \ui^{\frac{1-t^{\Delta}_f}{2}} (- \Theta_f  - \ui  (\Phi^B_f + \mu_f \pi ) )  \mod (\ui \pi, \gamma \pi)  \right) \nonumber \\
             & \qquad\qquad\qquad +(\ui + \gamma) \frac{1+t_f}{2}\frac{{\omega}_f \pi \mod 2 \pi}{2} \,.
        \end{align} 
    \end{itemize}
    The original integration path is contained in the case $1$ and $2$ with $t_f = t^{\Delta}_f$, ${\omega}_f = \omega^{\Delta}_f$ and $g'R_e = g^{-1 \dagger}$, which corresponds to the parity transformation. One can identify them with the non-degenerate solution shown in \cite{Barrett:2009mw,Han:2011re,Kaminski:2017eew,Liu:2018gfc}. Again in such case $\theta' = \theta^{\dagger}$, and $\tilde{F}_f[\tilde{X}_0]$ is determined up to $2 \pi$ which recovers the original mornodromy, which determine the action up to $\pi$. 
    
    Again by applying the above result to each vertex and summing over the result, we have now
    \begin{align}
        \Theta_f = \sum_v \Theta_{vf} =\left\{ \begin{array}{ll} \sum_v \theta_{vf}, \quad & t_f^{\Delta} = 1 \\ \sum_v (\pi-\theta_{vf}) , \quad & t_f^{\Delta}  = -1 \end{array} \right. \, .
    \end{align}
    As a result,  $ \Theta_f =   \epsilon_f \mod \frac{(1+t_f^{\Delta})}{2} \pi$ or $ \Theta_f =  \theta_f \mod  \frac{(1+t_f^{\Delta})}{2} \pi$, thus we can replace $\Theta_f$ in above equations to $ \epsilon_f $ for internal faces or $ \theta_f $ for boundary.

\section{Discussions and Outlook}\label{sec6}

In this work we study the analytic continuation of the Lorentzian EPRL spinfoam model and the CH extension on 4D simplicial manifold. We then derive the complexified critical equations and find all complex critical points. We also obtain the geometrical correspondence of the complex critical points. Our result is important for understand the subdominant contributions to the large-$j$ spinfoam amplitude when the real critical point is present, and dominant contributions to the amplitude when the real critical point is absent. Our result may also be helpful for studying spinfoam amplitude when $j$ is not very large.
         
There are a few future perspectives from this work: Firstly, we do not take into account the analytic continuation of the Barbero-Immrizi parameter $\g$. The complex critical points with simplicial-geometry interpretations satisfy critical equtations that are independent of $\g$. Thus the effect of possible complex $\g$ may be seen from the critical action with complexified $\g$ and may relate to the Stokes phenomenon.

Secondly, our result should be helpful for improving the Lefschetz-thimble Monte-Carlo computation in \cite{Han:2020npv} at small $j$, because the small-$j$ spinfoam amplitude and correlation functions receive non-negligible contribution from the Lefschetz thimbles of complex critical points. Our work identifies and classifies these complex critical points, thus provide a preparation for the numerical integration on the Lefschetz thimbles.

Lastly, our work propose a realization of Wick rotation in the spinfoam LQG: By the analytic continuation of the Lorentzian model, we identify the complex critical points correspond to Riemanian simplicial geometries, whose contributions to the amplitude behave as $e^{-S_{Regge}}$, similar to the situation in the Euclidean path integral. This provides a possible relation from the spinfoam model to the Euclidean quantum gravity. This relation should be important for applying spinfoams to studies such as the black hole entropy computation and the entanglement entropy computation.

\section*{Acknowledgements}

M.H. receives support from the National Science Foundation through grant PHY-1912278.

	\bibliographystyle{jhep}
	\bibliography{refs}

\providecommand{\href}[2]{#2}\begingroup\raggedright\begin{thebibliography}{10}

\bibitem{Thiemann:2007zz}
T.~Thiemann, {\it {Modern canonical quantum general relativity}},
  \href{http://arxiv.org/abs/gr-qc/0110034}{{\tt gr-qc/0110034}}.

\bibitem{Rovelli:2014ssa}
C.~Rovelli and F.~Vidotto, {\em {Covariant Loop Quantum Gravity}}.
\newblock Cambridge Monographs on Mathematical Physics. Cambridge University
  Press, 2014.

\bibitem{Ashtekar:2017yom}
A.~Ashtekar and J.~Pullin, eds., {\em {Loop Quantum Gravity}}, vol.~4 of {\em
  100 Years of General Relativity}.
\newblock World Scientific, 2017.

\bibitem{Han:2005km}
M.~Han, W.~Huang, and Y.~Ma, {\it {Fundamental structure of loop quantum
  gravity}},  {\em Int. J. Mod. Phys.} {\bf D16} (2007) 1397--1474,
  [\href{http://arxiv.org/abs/gr-qc/0509064}{{\tt gr-qc/0509064}}].

\bibitem{Perez:2012wv}
A.~Perez, {\it {The Spin Foam Approach to Quantum Gravity}},  {\em Living Rev.
  Rel.} {\bf 16} (2013) 3, [\href{http://arxiv.org/abs/1205.2019}{{\tt
  arXiv:1205.2019}}].

\bibitem{Engle:2007wy}
J.~Engle, E.~Livine, R.~Pereira, and C.~Rovelli, {\it {LQG vertex with finite
  Immirzi parameter}},  {\em Nucl. Phys.} {\bf B799} (2008) 136--149,
  [\href{http://arxiv.org/abs/0711.0146}{{\tt arXiv:0711.0146}}].

\bibitem{Freidel:2007py}
L.~Freidel and K.~Krasnov, {\it {A New Spin Foam Model for 4d Gravity}},  {\em
  Class. Quant. Grav.} {\bf 25} (2008) 125018,
  [\href{http://arxiv.org/abs/0708.1595}{{\tt arXiv:0708.1595}}].

\bibitem{Conrady:2010kc}
F.~Conrady and J.~Hnybida, {\it {A spin foam model for general Lorentzian
  4-geometries}},  {\em Class. Quant. Grav.} {\bf 27} (2010) 185011,
  [\href{http://arxiv.org/abs/1002.1959}{{\tt arXiv:1002.1959}}].

\bibitem{Conrady:2010vx}
F.~Conrady, {\it {Spin foams with timelike surfaces}},  {\em Class. Quant.
  Grav.} {\bf 27} (2010) 155014, [\href{http://arxiv.org/abs/1003.5652}{{\tt
  arXiv:1003.5652}}].

\bibitem{Rennert:2016rfp}
J.~Rennert, {\it {Timelike twisted geometries}},  {\em Phys. Rev.} {\bf D95}
  (2017), no.~2 026002, [\href{http://arxiv.org/abs/1611.00441}{{\tt
  arXiv:1611.00441}}].

\bibitem{Conrady:2008mk}
F.~Conrady and L.~Freidel, {\it {On the semiclassical limit of 4d spin foam
  models}},  {\em Phys. Rev.} {\bf D78} (2008) 104023,
  [\href{http://arxiv.org/abs/0809.2280}{{\tt arXiv:0809.2280}}].

\bibitem{Barrett:2009mw}
J.~W. Barrett, R.~J. Dowdall, W.~J. Fairbairn, F.~Hellmann, and R.~Pereira,
  {\it {Lorentzian spin foam amplitudes: Graphical calculus and asymptotics}},
  {\em Class. Quant. Grav.} {\bf 27} (2010) 165009,
  [\href{http://arxiv.org/abs/0907.2440}{{\tt arXiv:0907.2440}}].

\bibitem{Barrett:2009gg}
J.~W. Barrett, R.~J. Dowdall, W.~J. Fairbairn, H.~Gomes, and F.~Hellmann, {\it
  {Asymptotic analysis of the EPRL four-simplex amplitude}},  {\em J. Math.
  Phys.} {\bf 50} (2009) 112504, [\href{http://arxiv.org/abs/0902.1170}{{\tt
  arXiv:0902.1170}}].

\bibitem{Han:2011re}
M.~Han and M.~Zhang, {\it {Asymptotics of Spinfoam Amplitude on Simplicial
  Manifold: Lorentzian Theory}},  {\em Class. Quant. Grav.} {\bf 30} (2013)
  165012, [\href{http://arxiv.org/abs/1109.0499}{{\tt arXiv:1109.0499}}].

\bibitem{Han:2011rf}
M.-X. Han and M.~Zhang, {\it {Asymptotics of Spinfoam Amplitude on Simplicial
  Manifold: Euclidean Theory}},  {\em Class. Quant. Grav.} {\bf 29} (2012)
  165004, [\href{http://arxiv.org/abs/1109.0500}{{\tt arXiv:1109.0500}}].

\bibitem{Han:2013gna}
M.~Han and T.~Krajewski, {\it {Path Integral Representation of Lorentzian
  Spinfoam Model, Asymptotics, and Simplicial Geometries}},  {\em Class. Quant.
  Grav.} {\bf 31} (2014) 015009, [\href{http://arxiv.org/abs/1304.5626}{{\tt
  arXiv:1304.5626}}].

\bibitem{Kaminski:2017eew}
W.~Kaminski, M.~Kisielowski, and H.~Sahlmann, {\it {Asymptotic analysis of the
  EPRL model with timelike tetrahedra}},
  \href{http://arxiv.org/abs/1705.02862}{{\tt arXiv:1705.02862}}.

\bibitem{Liu:2018gfc}
H.~Liu and M.~Han, {\it {Asymptotic analysis of spin foam amplitude with
  timelike triangles}},  {\em Phys. Rev. D} {\bf 99} (2019), no.~8 084040,
  [\href{http://arxiv.org/abs/1810.09042}{{\tt arXiv:1810.09042}}].

\bibitem{Witten:2010cx}
E.~Witten, {\it {Analytic Continuation Of Chern-Simons Theory}},  {\em AMS/IP
  Stud. Adv. Math.} {\bf 50} (2011) 347--446,
  [\href{http://arxiv.org/abs/1001.2933}{{\tt arXiv:1001.2933}}].

\bibitem{Witten:2010zr}
E.~Witten, {\it {A New Look At The Path Integral Of Quantum Mechanics}},
  \href{http://arxiv.org/abs/1009.6032}{{\tt arXiv:1009.6032}}.

\bibitem{Cristoforetti:2013qaa}
M.~Cristoforetti, F.~Di~Renzo, A.~Mukherjee, and L.~Scorzato, {\it {Quantum
  field theories on the Lefschetz thimble}},  {\em PoS} {\bf LATTICE2013}
  (2014) 197, [\href{http://arxiv.org/abs/1312.1052}{{\tt arXiv:1312.1052}}].

\bibitem{Basar:2013eka}
G.~Basar, G.~V. Dunne, and M.~Unsal, {\it {Resurgence theory, ghost-instantons,
  and analytic continuation of path integrals}},  {\em JHEP} {\bf 10} (2013)
  041, [\href{http://arxiv.org/abs/1308.1108}{{\tt arXiv:1308.1108}}].

\bibitem{Han:2020npv}
M.~Han, Z.~Huang, H.~Liu, D.~Qu, and Y.~Wan, {\it {Spinfoam on Lefschetz
  Thimble: Markov Chain Monte-Carlo Computation of Lorentzian Spinfoam
  Propagator}},  \href{http://arxiv.org/abs/2012.11515}{{\tt
  arXiv:2012.11515}}.

\bibitem{Dunne:2016nmc}
G.~V. Dunne and M.~\"Unsal, {\it {New Nonperturbative Methods in Quantum Field
  Theory: From Large-N Orbifold Equivalence to Bions and Resurgence}},  {\em
  Ann. Rev. Nucl. Part. Sci.} {\bf 66} (2016) 245--272,
  [\href{http://arxiv.org/abs/1601.03414}{{\tt arXiv:1601.03414}}].

\bibitem{Conrady:2008ea}
F.~Conrady and L.~Freidel, {\it {Path integral representation of spin foam
  models of 4d gravity}},  {\em Class. Quant. Grav.} {\bf 25} (2008) 245010,
  [\href{http://arxiv.org/abs/0806.4640}{{\tt arXiv:0806.4640}}].

\bibitem{Conrady:2009px}
F.~Conrady and L.~Freidel, {\it {Quantum geometry from phase space reduction}},
   {\em J. Math. Phys.} {\bf 50} (2009) 123510,
  [\href{http://arxiv.org/abs/0902.0351}{{\tt arXiv:0902.0351}}].

\bibitem{Bianchi:2010gc}
E.~Bianchi, P.~Dona, and S.~Speziale, {\it {Polyhedra in loop quantum
  gravity}},  {\em Phys. Rev. D} {\bf 83} (2011) 044035,
  [\href{http://arxiv.org/abs/1009.3402}{{\tt arXiv:1009.3402}}].

\bibitem{Barrett:2010ex}
J.~W. Barrett, R.~J. Dowdall, W.~J. Fairbairn, H.~Gomes, F.~Hellmann, and
  R.~Pereira, {\it {Asymptotics of 4d spin foam models}},  {\em Gen. Rel.
  Grav.} {\bf 43} (2011) 2421--2436,
  [\href{http://arxiv.org/abs/1003.1886}{{\tt arXiv:1003.1886}}].

\end{thebibliography}\endgroup

\appendix 
\section{Critical configurations and action for Euclidean Model}\label{app_Euc}
From the action for Euclidean EPRL model,
\begin{equation}\label{Ft_e_a}
\tilde{F}_f\left[\tilde{X}\right]=\sum_{v, f\subset v}\Big[(1-\gamma)\ln\big( \tilde{\xi}'_{e f}(\tilde{g}^-_{ve})^{-1}\tilde{g}^-_{ve^\prime}{\tilde{\xi}_{e' f}}\big)+(1+\gamma) \ln\big( {\tilde{\xi}'_{e f}}(\tilde{g}^+_{ve})^{-1}\tilde{g}^+_{ve'}{\tilde{\xi}_{e' f}} \big)\Big].
\end{equation}
The variation respect respect to group elements $\tilde{g}^{\pm}$ leads to the following closure condition
\begin{align}
    0 = \sum_{f} j_f \kappa_{ef} \frac{\tilde{\xi}'_{e f}(\tilde{g}^-_{ve})^{-1} \sigma^i \tilde{g}^-_{ve^\prime}{\tilde{\xi}_{e' f}} }{\tilde{\xi}'_{e f}(\tilde{g}^-_{ve})^{-1}\tilde{g}^-_{ve^\prime}{\tilde{\xi}_{e' f}} }\\
    0 = \sum_{f} j_f \kappa_{ef} \frac{ {{\tilde{\xi}'_{e f}}(\tilde{g}^+_{ve})^{-1} \sigma^i \tilde{g}^+_{ve'}{\tilde{\xi}_{e' f}}} }{ {{\tilde{\xi}'_{e f}}(\tilde{g}^+_{ve})^{-1}\tilde{g}^+_{ve'}{\tilde{\xi}_{e' f}}} }
\end{align}
For internal faces, the variation respect to $\xi$ and $\xi'$ becomes the variation respect to the $SL(2,\mathbb{C})$ group element $\tilde{v}$. Since we have $\tilde{v}'\tilde{v} = \mathbb{I}_2$, for $\delta \tilde{v} =  \tilde{v}\vec{\epsilon} \cdot \vec{L}$ , we have $\delta \tilde{v}' = - \vec{\epsilon} \cdot \vec{L} (\tilde{v})^{-1} =  - \vec{\epsilon} \cdot \vec{L} \tilde{v}'$. As a result, we have
\begin{align}
  0= & (1-\gamma)\frac{R_{L}\tilde{\xi}'_{e f}(\tilde{g}^-_{ve})^{-1}\tilde{g}^-_{ve^\prime}{\tilde{\xi}_{e' f}} }{\tilde{\xi}'_{e f}(\tilde{g}^-_{v'e})^{-1}\tilde{g}^-_{ve^\prime}{\tilde{\xi}_{e' f}} } - (1-\gamma)\frac{\tilde{\xi}'_{e'' f}(\tilde{g}^-_{v'e''})^{-1}\tilde{g}^-_{v'e^\prime}{R_{L}\tilde{\xi}_{ef}} }{\tilde{\xi}'_{e'' f}(\tilde{g}^-_{v'e''})^{-1}\tilde{g}^-_{v'e^\prime}{\tilde{\xi}_{e f}} } \\
   & +(1+\gamma) \frac{ R_{L}{{\tilde{\xi}'_{e f}}(\tilde{g}^+_{ve})^{-1}\tilde{g}^+_{ve'}{\tilde{\xi}_{e' f}}} }{ {{\tilde{\xi}'_{e f}}(\tilde{g}^+_{ve})^{-1}\tilde{g}^+_{ve'}{\tilde{\xi}_{e' f}}} }-(1+\gamma)\frac{\tilde{\xi}'_{e'' f}(\tilde{g}^+_{v'e''})^{-1}\tilde{g}^+_{v'e^\prime}{R_{L}\tilde{\xi}_{ef}} }{\tilde{\xi}'_{e'' f}(\tilde{g}^+_{v'e''})^{-1}\tilde{g}^+_{v'e^\prime}{\tilde{\xi}_{e f}} }
\end{align}
where $R_{L} \tilde{\xi} := \tilde{v} \sigma_i \xi_0 $. Since $\sigma_1 \xi_0 \propto \sigma_2 \xi_0 \propto J\xi_0$ and $\sigma_3 \xi_0 = \xi_0$, there are only one non-trivial equation given by
\begin{align}
  0= & (1-\gamma)\frac{\widetilde{J\xi}'_{e f}(\tilde{g}^-_{ve})^{-1}\tilde{g}^-_{ve^\prime}{\tilde{\xi}_{e' f}} }{\tilde{\xi}'_{e f}(\tilde{g}^-_{v'e})^{-1}\tilde{g}^-_{ve^\prime}{\tilde{\xi}_{e' f}} } - (1-\gamma)\frac{\tilde{\xi}'_{e'' f}(\tilde{g}^-_{v'e''})^{-1}\tilde{g}^-_{v'e^\prime}{\widetilde{J\xi}_{ef}} }{\tilde{\xi}'_{e'' f}(\tilde{g}^-_{v'e''})^{-1}\tilde{g}^-_{v'e^\prime}{\tilde{\xi}_{e f}} } \label{euc_para_v}\\
   & +(1+\gamma) \frac{ {{\widetilde{J\xi}'_{e f}}(\tilde{g}^+_{ve})^{-1}\tilde{g}^+_{ve'}{\tilde{\xi}_{e' f}}} }{ {{\tilde{\xi}'_{e f}}(\tilde{g}^+_{ve})^{-1}\tilde{g}^+_{ve'}{\tilde{\xi}_{e' f}}} }-(1+\gamma)\frac{\tilde{\xi}'_{e'' f}(\tilde{g}^+_{v'e''})^{-1}\tilde{g}^+_{v'e^\prime}{\widetilde{J\xi}_{ef}} }{\tilde{\xi}'_{e'' f}(\tilde{g}^+_{v'e''})^{-1}\tilde{g}^+_{v'e^\prime}{\tilde{\xi}_{e f}} } \nonumber
\end{align}

The equations of motion is totally different from the one we obtain in the case of Lorentzian models, but we can still assume for special configurations there are solutions of above equations of motion which satisfies
\begin{align}
    \tilde{g}^{\pm}_{ve^\prime}{\tilde{\xi}_{e' f}} = \ue^{\theta^{\pm}_{e'vef}} \tilde{g}^{\pm}_{ve}{\tilde{\xi}_{e f}} \,, \qquad \tilde{\xi}'_{e' f}(\tilde{g}^{\pm}_{ve'})^{-1} =  \ue^{-\theta^{\pm}_{e'vef}} \tilde{\xi}'_{e f}(\tilde{g}^{\pm}_{ve})^{-1} 
\end{align}
One can check that this ansarz solves \eqref{euc_para_v}. The equation of motion now have the same form as \eqref{eq:biveq1} with bivector $B_f^{\pm}$ defined as
\begin{align}
    B_f^{\pm}(v) := \tilde{g}^{\pm}_{ve} B_{ef}^{\pm} (\tilde{g}^{\pm}_{ve'})^{-1} :=  \tilde{g}^{\pm}_{ve} \left(\tilde{\xi}_{e f} \otimes \tilde{\xi}'_{e' f} - \frac{1}{2}I_2 \right) (\tilde{g}^{\pm}_{ve'})^{-1}
\end{align}
Then the analysis for Lorentzian case follows exactly here. Namely for the (degenerate) simplicial geometry solutions $G,G'$, we have 4 possibilities for solutions $\tilde{G}=(\tilde{G}^{+}, \tilde{G}^{-})$ at each vertex: $\tilde{G}=(\tilde{G}^{+}, \tilde{G}^{-})$: $\tilde{G}=(G, (G)^{-1})$, $\tilde{G}=(G, (G')^{-1})$ and $\tilde{G}=(G', (G)^{-1})$, $\tilde{G}=(G', (G')^{-1})$. 
The critical action associated to each face in this case reads
\begin{equation}\label{Ft_e_a_c_Euc}
\tilde{F}_f\left[\tilde{X}\right]=\sum_{v, f\subset v}\Big[ (1-\gamma)  \theta^{-}_{vf}+(1+\gamma) \theta^{+}_{vf} \Big].
\end{equation}
The parallel transport equations are given by
\begin{align}
\mathfrak{v}_{ee'} G^{\pm}_f(e',e) = \ue^{2 \sum_{v} \theta_{vf}^{\pm} B_{ef}^{-} }
\end{align} 
with $G^{\pm}_f = \prod_{v \in \partial_f} (\tilde{g}_{ve'}^{\pm})^{-1}  \tilde{g}_{ve}^{\pm}$.

We can then get similar result for $\theta^{\pm}$ as in section \ref{riemann_split_action} and \ref{lorentzian_action} by identifying $\theta^{-}$ with $\theta$ and $\theta^{+}$ with $-\theta'$ as well as set ${\omega}_f = 0$. Substitute them to (\ref{Ft_e_a_c_Euc}) gives out the ciritcal action. As a result, we may have the following possibilities:
\begin{itemize}
    \item Riemannian or split signature critical points 
    \begin{align}
        \tilde{S}[\tilde{X}_0]=&  \sum_{f} J_f \left((-\ui)^{\frac{1-t_f^{\Delta}}{2}} \ui (\pm \gamma \Theta_f  +  \Phi^B_f + \mu_f \pi) \mod (\gamma \pi, \ui \pi) \right) 
    \end{align}
    \begin{align}
        \tilde{S}[\tilde{X}_0]=&\sum_{f} J_f \left((-\ui)^{\frac{1-t_f^{\Delta}}{2}} \ui (\pm \Theta_f + \Phi^B_f + \mu_f \pi) \mod (\gamma \pi, \ui \pi) \right)
    \end{align}
    
    \item Lorentzian critical points
        \begin{align}\label{l_regge}
        \tilde{S}[\tilde{X}_0]=&  \sum_{f} J_f \left(\ui^{\frac{1-t^{\Delta}_f}{2}} \left( \pm \gamma \Theta_f  + \ui (\Phi^B_f + \mu_f \pi ) \mod (\ui \pi, \gamma \pi) \right)  + \ui^{\frac{1-t^{\Delta}_f}{2}} \ui ( 1 - \gamma) \frac{{\omega}_f^{\Delta} \pi \mod 2 \pi}{2}  \right)
    \end{align}
    \begin{align}\label{l_regge_1}
        \tilde{S}[\tilde{X}_0]=&  \sum_{f} J_f \left(\ui^{\frac{1-t^{\Delta}_f}{2}} \left( \pm \Theta_f  + \ui (\Phi^B_f + \mu_f \pi ) \mod (\ui \pi, \gamma \pi) \right)  \right)
    \end{align}
\end{itemize}
For the Lorentzian critical points (\ref{l_regge}), their contributions to the spinfoam ampltiude are again proportional to $e^{-S_{\rm Regge}}$ with  
\begin{align}
    S_{\text{Regge}} = \pm \sum_f A_f \Theta_f
\end{align}
the Lorentzian Regge action with $A_f = \gamma J_f$. There is also another subdominant contribution proportional to $e^{-\frac{1}{\gamma} S_{\rm Regge}}$ given by (\ref{l_regge_1})
\end{document}